\documentclass{ucetd}
\pdfoutput=1

\usepackage{amsthm}
\usepackage{amsmath,amssymb,calc}
\usepackage{graphicx}
\usepackage{subfig}
\usepackage[nosort]{cite}
\usepackage{epsfig,psfrag}

\let\non\nonumber

\let\a=\alpha
\let\b=\beta
\let\dd=\delta
\let\e=\epsilon

\let\k=\kappa
\let\l=\lambda
\let\m=\mu
\let\n=\nu

\let\s=\sigma
\let\t=\tau

\let\o=\omega

\newcommand{\La}{\Lambda}
\newcommand{\G}{\Gamma}

\let\S=\Sigma
\let\Th=\Theta

\newcommand{\del}{\partial}
\newcommand{\delbar}{\bar{\partial}}

\newcommand{\bea}{\begin{eqnarray}}
\newcommand{\eea}{\end{eqnarray}}
\newcommand{\be}{\begin{equation}}
\newcommand{\ee}{\end{equation}}
\newcommand{\bma}{\begin{pmatrix}}
\newcommand{\ema}{\end{pmatrix}}
\newcommand{\alt}[1]{\left\{\begin{array}{ll} #1 \end{array}\right.}

\newcommand{\hlf}{\frac{1}{2}}
\newcommand{\hf}{{\textstyle{1\over 2}}}

\def\coeff#1#2{{\textstyle{#1\over #2}}}
\newcommand{\Z}{{\mathbb Z}}
\newcommand{\R}{{\mathbb R}}

\newcommand{\F}{{\mathbb F}}
\newcommand{\PP}{{\mathbb P}}
\newcommand{\CC}{{\mathbb C}}

\newcommand{\cG}{{\cal G}}
\newcommand{\cC}{{\cal C}}
\newcommand{\cA}{{\cal A}}
\newcommand{\cE}{{\cal E}}
\newcommand{\cF}{{\cal F}}
\newcommand{\cJ}{{\cal J}}
\newcommand{\B}{{\cal B}}
\newcommand{\I}{{\cal I}}
\newcommand{\cO}{{\cal O}}
\newcommand{\cR}{{\cal R}}
\newcommand{\cRp}{{\cal R}^{(+)}}

\newcommand{\Om}{\Omega}

\newcommand{\p}{\partial}

\newcommand{\C}[1]{$(\ref{#1})$}

\def\tr{{\rm tr\,}}
\def\Tr{{\rm Tr\,}}

\newcommand{\N}{{\cal N}}
\newlength{\bredde}
\def\slash#1{\settowidth{\bredde}{$#1$}\ifmmode\,\raisebox{.15ex}{/}
\hspace*{-\bredde} #1\else$\,\raisebox{.15ex}{/}\hspace*{-\bredde}
#1$\fi}

\def\Im{{\rm Im ~}}
\def\Re{{\rm Re ~}}

\newcommand{\ena}{\end{eqnarray}}
\newcommand{\beqa}{\begin{eqnarray}}
\newcommand{\eeqa}{\end{eqnarray}}

\def\G{\Gamma}

\def\N{{\cal N}}

\def\cD{{\cal D}}

\renewcommand{\b}{\beta}
\newcommand{\g}{\gamma}

\def\d{{\rm d}}

\newcommand{\ibar}{{\bar \imath}}
\newcommand{\jbar}{{\bar \jmath}}
\newcommand{\thbar}{{\bar \theta}}
\newcommand{\Dbar}{{\bar D}}

\newcommand{\labar}{{\bar \lambda}}

\newcommand{\abar}{{\bar \a}}
\newcommand{\bbar}{{\bar \b}}

\newcommand{\mbar}{{\bar m}}
\newcommand{\nbar}{{\bar n}}

\newfont{\goth}{ygoth.tfm scaled 1200}                   

\def\a{\alpha}
\def\b{\beta}
\def\e{\epsilon}
\def\th{\theta}
\def\f{\phi}
\def\g{\gamma}

\def\j{\psi}
\def\k{\kappa}
\def\l{\lambda}
\def\m{\mu}
\def\n{\nu}
\def\o{\omega}

\def\s{\sigma}
\def\t{\tau}

\def\D{\Delta}
\def\F{\Phi}
\def\G{\Gamma}

\def\L{\mathcal{L}}

\def\S{\Sigma}
\def\U{\Upsilon}

 \numberwithin{equation}{section}

\def\1{{(1)}}
\def\2{{(2)}}
\def\3{{(3)}}

\def\1{{\bf 1}}
\def\a{{\alpha}}
\def\da{{\dot \alpha}}

\def\B{{\mathcal B}}

\def\M{{\mathcal M}}

\newcommand\al{{(\alpha)}}
\newcommand\bt{{(\beta)}}
\newcommand\ab{{(\alpha\beta)}}

\newcommand{\hG}{{\hat \G}}
\newcommand{\Jbar}{{\bar J}}

\def\p{\partial}
\def\pb{\bar{\partial}}
\def\SO{\operatorname{SO{}}}
\def\GU{\operatorname{U{}}}
\def\SU{\operatorname{SU{}}}
\def\ff#1#2{{\textstyle\frac{#1}{#2}}}

\def\normd#1{{{:\!#1\!\!:}}}

\def\la{{\langle}}
\def\ra{{\rangle}}

\newcommand\wb{\overline{w}}
\newcommand\zb{\overline{z}}
\newcommand\gammab{\overline{\gamma}}
\newcommand\psib{\overline{\psi}}

\newcommand\chib{{\overline{\chi}}}
\newcommand\lambdab{{\overline{\lambda}}}
\newcommand\vphi{{\varphi}}
\newcommand\vphib{{\overline{\vphi}}}
\newcommand\vtheta{{\vartheta}}
\newcommand\ep{{\epsilon}}

\def\cA{{\cal A}}

\newcommand\Ab{\overline{A}}
\def\tcA{{\tilde\cA}}

\def\cF{{\cal F}}

\def\Jb{{{\overline{J}}}}
\def\cJ{{\cal J}}

\def\JJ{{{\mathbb J}}}
\def\JJb{{{\overline{\JJ}}}}

\def\tint{{{\text{int}}}}

\def\tYuk{{{\text{Yuk}}}}
\def\tct{{{{\text{c.t.}}}}}
\def\teff{{{{\text{eff}}}}}

\title{Heterotic Flux Geometry from Chiral Gauge Dynamics}
\author{Callum Quigley}
\department{Physics}
\division{Physical Sciences}
\degree{Doctor of Philosophy}
\date{August 2013}

\dedication{{\it To Madeleine and the memory of my grandfather, W.W.}}

\begin{document}

\maketitle

\makecopyright
\makededication

\tableofcontents
\listoffigures
\listoftables

\acknowledgments
It is a great pleasure to acknowledge the many people to whom I am grateful.

First and foremost I must thank my advisor, Sav Sethi, for always being willing to share his ideas, time, and knowledge with me. My understanding of string theory has been deeply enriched by the countless hours we have spent together. This work simply would not have been possible without his guidance, support, and expertise. I would also like to thank Sav for his help in refining my palette for single malt scotch.

Very special thanks to Ilarion Melnikov and Mark Stern, with whom I collaborated on the projects that formed the basis of this work. Their knowledge was instrumental in developing these ideas, and I am pleased to have gleaned a least a part of that knowledge.

I would like to thank all the other professors at the University of Chicago with whom I have interacted, especially Jeff Harvey, David Kutasov, Emil Martinec, Carlos Wagner, and Paul Wiegmann, for sharing their intuition and deepening my understanding of string theory, quantum field theory, and physics in general. Thanks as well to Henry Frisch and Ilya Gruzberg, for serving on my thesis committee and reading this body of work.

Distinguished thanks go to the advisors from my earlier studies in maths and physics: Jim Colliander, Moshe Rozali, and Kentaro Hori, for their tremendous help in building the foundation that got me here. I have also benefitted from discussions with many other physicists, including Allan Adams, Katrin Becker, Nick Halmagyi, Oleg Lunin, Amanda Peet, Ronen Plesser, and Eric Sharpe.

My time at Chicago would not have been nearly as educational, fun and memorable if not for the many wonderful classmates I have met here. Thanks to Denis Erkal, Adam Nahum, Mike Schmidt, Bhujyo Bhattacharya and the rest of the Titanic Twelve, Andy Royston, Sophia Domokos, Jock McOrist, Patrick Draper, Szilard Farkas, Stephen Green, Sam Gralla, Arun Thalapillil, Dan Herbst, Yeunjin Kim, Gabe Lee, Pierre Gratia, Josh Schiffrin, Wynton Moore, Travis Maxfield, Mikhail Solon and the rest of the Crossfit Crew, for all the stimulating conversations, exciting times, and brutal workouts.

I am grateful to the Department of Physics for financial support through Sachs, McCormick, and Bloomenthal Fellowships, and I also thank the Natural Sciences and Engineering Research Council of Canada for two years of funding through a PGS-D Scholarship.

Finally, deep heartfelt thanks go out to my family, especially my mother, for a lifetime of love and encouragement. Most of all, to my wife Violet, for ten years of constant love and support, and for all the sacrifices she has made so I can pursue this crazy dream, I am eternally grateful.

Thank you all.

\abstract
Chiral gauge theories in two dimensions with (0,2) supersymmetry admit a much broader, and
more interesting, class of vacuum solutions than their better studied (2,2) counterparts. In this thesis,
we will explore some of the possibilities that are offered by this additional freedom by including field-dependent $\th$-angles and FI parameters.
The moduli spaces that will result from this procedure correspond to heterotic string backgrounds with non-trivial $H$-flux and NS-brane sources.
Along the way, a remarkable relationship between (0,2) gauge anomalies and $H$-flux will emerge.

\mainmatter

\chapter{Introduction}\label{ch:intro}

String theory remains to date our best hope of unifying all the fundamental building blocks of Nature into a single fully quantum mechanical framework.\footnote{The material presented in this chapter is well established and can be found in any standard textbook, such as~\cite{Green:1987sp,Green:1987mn,Polchinski:1998rq,Polchinski:1998rr}. The material of Section~\ref{sec:het}\ is reviewed in~\cite{Becker:2007zj}.} The basic ingredients of the Standard Model, namely chiral fermions coupled to non-Abelian gauge fields, are present, but perhaps most striking is the appearance of an interacting massless spin-two particle: the graviton. String theory is therefore a quantum theory of gravity! In fact, string theory is the only known way to couple matter to gravity in a way consistent with quantum mechanics.

Beyond these crucial basic features, required of any fundamental theory of Nature, string theory also predicts the existence of several, independently motivated, concepts that lie beyond the Standard Model: these include supersymmetry, gauge-unification, branes, and extra-dimensions. This last idea has enjoyed an interesting change in attitude over the years. Initially, the fact that superstring theory is only consistent in ten dimensions of spacetime (nine space and one time) was considered an obstacle of the theory that needed to be explained away. However, we now realize that many of string theories greatest achievements, such as  the counting of black hole microstates and the AdS/CFT correspondence, stem directly from the existence of these extra dimensions. Furthermore, we have come to understand that the geometry and topology of the internal space has important consequences for the four-dimensional physics that we observe; for example, Yukawa couplings are determined by certain topological invariants associated with the internal dimensions.

During the course of its development, several unexpected discoveries have reshaped the very notion of what we think string theory is. These include the appearance of extended non-perturbative objects (D-branes), the web of dualities linking all known perturbative string theories as different limits of a single unifying \textit{eleven}-dimensional theory (M-theory), the exact equivalence of string/gravitational theories with gauge theories in lower dimensions (AdS/CFT).
However, despite more than half a century of progress, there is still no definitive answer to the question, `\textit{What is string theory?}' The main difficulty stems from the lack of a complete non-perturbative, background independent definition of the theory.
The strong coupling dynamics of string theory has many descriptions: other strings, the same strings, membranes, D-branes, matrix models, gauge theories, and many more depending on the circumstances. Even worse, we have no knowledge of what the correct degrees of freedom are at intermediate couplings.
Fortunately, the weak coupling description of string theory is well understood and it is just what it sounds like: the theory of one-dimensional vibrating strings. We will restrict ourselves to the perturbative regime and study the dynamics of weakly coupled strings propagating through non-trivial spacetime backgrounds.

\section{Worldsheet descriptions}\label{sec:ws}
To determine what sort of action should govern the dynamics of these strings, it helps to recall how the dynamics of point particles are described.
\subsubsection{Point particles}
Suppose we have a particle of mass $m$ traveling through spacetime, $\M$. If we label the points of spacetime by $X^M$ then the worldline, $\G$, is parameterized by a path $X^M(\t)$, where $\t$ is the proper time as observed by the particle. The set of functions $X^M(\t)$ therefore provide an embedding of the particle's worldline in spacetime:
\be
X:~\G\hookrightarrow\M,
\ee
by mapping each point of $\G$, labeled by $\t$, to a point $X^M(\t)$ in $\M$.
If $\M$ is equipped with a metric $G_{MN}(X)$, then the motion of the particle is governed by the action
\be\label{eq:Spp}
S_{pp}[X] = -m\int_\G d\t\sqrt{- G_{MN}(X)\del_\t X^M(\t)\del_\t X^N(\t)},
\ee
which is nothing more than the proper length of the path $\G$. The classical trajectory of the particle is the one that minimizes the length of $\G$.

The action~\C{eq:Spp}\ has two major limitations: first, it is not well-defined for massless particles, and second, the square-root makes it impractical for quantization.
However, there is another action that is classically equivalent to~\C{eq:Spp}, but avoids these two pitfalls. That action is
\be\label{eq:Spp2}
S_{pp}'[X,g] = -\hlf\int_\G d\t\, \sqrt{-g} \left(g^{-1}(\t)G_{MN}(X)\del_\t X^M(\t)\del_\t X^N(\t)  + m^2\right),
\ee
where we should think of $g(\t)$ as a sort of metric on $\G$. Clearly the $m\rightarrow0$ limit of~\C{eq:Spp2}\ is well-defined, and when $m\neq0$ we can solve the equations of motion for $g(t)$ to recover~\C{eq:Spp}.
From the point of view of $\G$, the action~\C{eq:Spp2}\ is a $0+1$ dimensional field theory for a set of scalar fields $X^M$ coupled to some kind of worldline gravity. We cannot add an Einstein-Hilbert type term for $g$ to the action, since there is no notion of curvature in one-dimension.

\subsubsection{The Polyakov action}
Now we repeat the same reasoning for a one-dimensional string, rather than a point particle. A string sweeps out a two-dimensional worldsheet, $\S$, parameterized by $X^M(\s^1,\s^2)$ so that each point $(\s^1,\s^2)$ of $\S$ gets mapped to some point $X^M$ in spacetime:
\be
X:~\S\hookrightarrow\M.
\ee
The analog of~\C{eq:Spp}\ is the Nambu-Goto action:
\be
S_{NG}[X] = -{1\over2\pi\a'} \int_\S d^2\s\sqrt{-\det \left[G_{MN}(X)\,\del_\m X^M(\s)\del_\n X^N(\s)\right]},
\ee
whose classical solutions minimize the area of $\S$. The pre-factor ${1\over2\pi\a'}$ is the tension of the string (or mass per unit length), in analogy with $m$ for the point particle.
The quantity $\a'$ has the dimensions of $(length)^2$ and, as the only dimensionful parameter in the theory, it sets the overall scale of the theory.
For a perturbative string, the string scale $\ell_s=\sqrt{\alpha'}$ is (at least) several order of magnitude larger than the typical scale of quantum gravity, which is the Plank scale $\ell_{Pl}\sim M_{Pl}^{-1}\sim(10^{19}~GeV)^{-1}$.\footnote{Note that only one of these two scales is actually an input for the theory, since the ratio $\ell_s/\ell_{Pl}$ is fixed by the solutions of the theory.}
As in the point particle case, we may remove the pesky square-root appearing in the Nambu-Goto action by introducing a worldsheet metric $g_{\m\n}$. This leads us to
the Polyakov action,
\be\label{eq:Poly}
S_P[X,g] = -{1\over4\pi\a'}\int_\S d^2\s\sqrt{-g}\,g^{\m\n}(\s) G_{MN}(X)\,\del_\m X^M(\s)\del_\n X^N(\s),
\ee
which is equivalent to $S_{NG}$ after solving for $g_{\m\n}$, and this serves the basic starting point for describing perturbative string theory.

\subsubsection{Einstein-Hilbert action, $g_s$, and the dilaton}
From the worldsheet point of view, the Polyakov action~\C{eq:Poly}\ is that of a collection of scalar fields $X^M$ with non-linear kinetic terms coupled to $1+1$ dimensional gravity.
Unlike the $0+1$ dimensional case, we can write down an Einstein-Hilbert action for the string metric $g$:
\be\label{eq:EH}
S_{EH}[g,\S] = {1\over4\pi}\int_\S d^2\s\sqrt{-g}\ {\cal R}[g],
\ee
where ${\cal R}[g]$ is the Ricci scalar associated with the metric $g$. However, $S_{EH}$ does not generate any kinetic terms for the metric $g$ because ${\cal R}$ is a total derivative in two-dimensions. As shown by Gauss and Bonnet back in the $19^{\rm th}$ century, the action~\C{eq:EH}\ is a topological invariant of $\S$ (independent of $g$) known as the Euler number, $\chi$:\footnote{Here we assume $\S$ is a closed 2-manifold without boundary.}
\be
S_{EH}[\S] = \chi(\S) = 2-2h,
\ee
where $h$ is the number of handles on $\S$. When $\S$ is a sphere, $S^2$, then $h=0$, and when $\S$ is a torus, $T^2$, then $h=1$, while when $\S$ is a double-torus $h=2$, and so on.

If we include $S_{EH}$ with a coefficient $\l$ in the (Euclidean) path integral,
\be
\int[DX,Dg] e^{-S_P[X,g] -\l \chi(\S)} \ldots
\ee
then we only affect the relative weighting of worldsheets with different topologies.
Worldsheets with handles are the string analogs of Feynman diagrams with loops, since emitting and reabsorbing a string has the effect of increasing $h$ by 1. Each such quantum process adds a weighting factor of $e^{2\l}$ to the path integral. Therefore, it is natural to define
\be
g_s = e^\l
\ee
as the string coupling constant, which controls the probability of strings splitting and joining.

It might seem that $g_s$ is a free parameter, labeling different string theories by their coupling strengths, but we will now show that this is not the case.
The idea is to generalize the action~\C{eq:EH}\ by including a coupling of the worldsheet curvature to another background field on $\M$, similar to how $S_P$ depends on the spacetime metric $G_{MN}$. The result is
\be\label{eq:Sdil}
S_\varphi[X,g] = {1\over4\pi}\int_\S d^2\s\sqrt{-g}\, \varphi(X) {\cal R}[g],
\ee
where the scalar field $\varphi$ is known as the dilaton. The action~\C{eq:Sdil}\ is no longer a topological invariant unless
$
\varphi(X) = \varphi_0,
$
for some constant value $\varphi_0$. In this case,
\be
g_s = e^{\varphi_0}
\ee
is not a parameter that distinguishes different string theories, but instead it only labels different backgrounds in the \textit{same} theory.

\subsubsection{$B$-fields and $H$-flux}
There is one last background field we must couple to the string worldsheet which is called the $B$-field. To understand the relation between strings and $B$-fields, it helps to first recall the relation between point particles and gauge fields.

Let us return to our point particle example, either with action~\C{eq:Spp}\ or~\C{eq:Spp2}, and give it an electromagnetic charge $q$. The interaction of this particle with a background electromagnetic potential $A_M$ is given by the pullback (via $X^M$) of the gauge field to the worldline:
\be\label{eq:SA}
S_{A}[X]  = q\int_\G d\t \, \del_\t X^M(\t) A_M(X).
\ee
Under a local gauge transformation,
\be
\dd A_M(X) = -\del_M\La(X),
\ee
the action is invariant\footnote{We will assume the transformation is localized, so that $\La$ vanishes at the end points of $\G$.}:
\be
\dd S_{A} = -q\int_\G d\t\,\del_\t X^M(\t)\del_M \La(X) = -q\int_\G d\t\, \del_\t \La(X) = 0,
\ee
and so is the field strength tensor
\be
F_{MN} = \del_M A_N - \del_N A_M.
\ee

Just as the worldline $\G$ couples naturally to a one-form potential $A$, the worldsheet $\S$ couples naturally couples to a (anti-symmetric) two-form ``gauge potential" $B$:
\be\label{eq:int}
S_{B}[X] = -{1\over4\pi\a'}\int_\S d^2\s \,\e^{\m\n} B_{MN}(X)\del_\m X^M(\s) \del_\n X^N(\s) .
\ee
 The $B$-field also enjoys a kind of gauge symmetry that leaves~\C{eq:int}\ invariant:
\be\label{eq:delB1}
\dd B_{MN} = \del_M \La_N(X) -\del_N \La_M(X),
\ee
except now the gauge parameter $\La_M$ is a one-form. The invariant field strength associated to $B$ is a three-form:
\be
H_{MNP} = \del_M B_{NP} + \del_N B_{PM} + \del_P B_{MN}.
\ee
It is interesting to note that $G_{MN}$ and $B_{MN}$ appear on roughly equal footing in the worldsheet actions~\C{eq:Poly}\ and~\C{eq:int}. This should be contrasted to the point particle case, where~\C{eq:Spp2}\ and~\C{eq:SA}\ are quadratic and linear in $\del_\t X^M$.
This seemingly innocent observation underlies many of the striking differences in how a string `sees' spacetime geometry compared to a point particle.

\subsubsection{Non-linear sigma models}
Together the three actions~\C{eq:Poly},~\C{eq:Sdil}, and~\C{eq:int}, comprise the \textit{non-linear sigma model} for the (closed, bosonic) string:
\be\label{eq:nlsm}
S[X,g] = {-1\over4\pi\a'}\int_\S d^2\s\sqrt{-g}\left\{\left[g^{\m\n}G_{MN}(X)+\e^{\m\n}B_{MN}(X)\right]\del_\m X^M\del_\n X^N +\a'\varphi(X){\cal R}[g]\right\}.
\ee
The background fields $G,B$ and $\varphi$, together with those worldsheet couplings, appear in (nearly\footnote{The only exception is the type I string, which does not contain $B$.}) every description of perturbative string theory. Each of the known superstring theories differ in their field content, both bosonic and fermionic, beyond those given~\C{eq:nlsm}. For example, our interest in this thesis will lie in the heterotic strings, which additionally contain gauge fields $A$ as well as various fermionic fields. We will postpone writing down the worldsheet couplings to the heterotic gauge fields until we require them in Section~\ref{ss:leftmovers}.

NB: in subsequent chapters, we will work on a fixed flat worldsheet, $\S\simeq\R^{1,1}$, with Minkowski metric, $g_{\m\n}=\eta_{\m\n}$, and so the curvature coupling $\varphi{\cal R}$ will be absent.

\section{Spacetime descriptions}\label{sec:het}
A remarkable feature of string theory is its complementary descriptions from the worldsheet and spacetime perspectives. From the spacetime point of view, the different vibrations of a string appear as different particle excitations. The typical mass scale of these excitations is set by the string tension, $\a'$, and so we expect generally all massive string states to have masses of order $M_{Pl}$. There is an important exception to this, which is the set of massless string states. These correspond to small perturbations of the background fields appearing in the sigma-model action~\C{eq:nlsm}, and its generalizations. For example, the graviton $h_{MN}$ is a massless spin-2 particle that is a perturbation of the background spacetime metric $G_{MN}$.

At energies that are small compared to $M_{Pl}$, and curvatures scales that are large compared to $\ell_s$, the dynamics of a string theory are well approximated by a low energy effective field theory for its massless degrees of freedom. These effective descriptions all contain General Relativity coupled to various matter and force fields. When the string theory is supersymmetric, the low energy description contains a ten-dimensional supergravity theory. Corrections beyond the leading supergravity approximation are suppressed by powers of $\a'$.

\subsubsection{Heterotic supergravity}
Our interest is primarily in heterotic string theory, which at low energies is well approximated by ten-dimensional $\N=1$ supergravity coupled to super-Yang-Mills theory with gauge group ${\cal G}=(E_8\times E_8)\ltimes\Z_2$ or $Spin(32)/\Z_2$.\footnote{Colloquially, these groups are often referred to simply as ${\cal G}=E_8\times E_8$ and $SO(32)$.} The restriction on the choice of ${\cal G}$ comes from the requirement that all local anomalies (gauge, gravitational, and mixed) cancel.\footnote{Although the anomalies also cancel for the gauge groups $E_8\times U(1)^{248}$ and $U(1)^{496}$, it has recently been shown that these low energy theories do not admit a proper UV completion~\cite{Adams:2010zy}. These theories lie in the string `swampland'~\cite{Vafa:2005ui}, as opposed to the string landscape, and therefore should not be considered.}
\begin{table}[ht]
\begin{center}
\begin{tabular}{cllc}
\hline\hline
\underline{ Field }     & \underline{ Name  }      & \underline{Representation  }          & \underline{d.o.f.}\\
$G_{MN}$ & metric & traceless symmetric~ tensor & 35\\
$B_{MN}$ &  $B$-field & anti-symmetric tensor & 28\\
$\varphi$ & dilaton & real scalar & 1\\
$A_M^a$ & gauge field & adjoint valued one-form & $8\cdot{\rm dim}\,{\cal G}$\\
\hline\hline
\end{tabular}
\end{center}
\caption{Bosonic field content of  $\N=1$ $d=10$ supergravity and super-Yang-Mills}
\label{tab:bosonic}
\end{table}
The bosonic field content of the theory is listed in Table~\ref{tab:bosonic}, and these degrees of freedom\footnote{In Table~\ref{tab:bosonic}\ we are only counting the number of on-shell degrees of freedom} are governed by the action
\be\label{eq:sugra}
S_{het} ={1\over2\k^2}\int d^{10}x\sqrt{-G}e^{-2\varphi}\left[{\cal R}+4|d\varphi|^2 -\hlf\left|H\right|^2 -{\a'\over4}\left({\rm tr}\left|\cF\right|^2-{\rm tr}\left|{\cal R}^{(+)}\right|^2\right)+O(\a'^2)\right],
\ee
where ${\cal R}^{(+)}={\cal R}^{(+)\,AB}_{MN}$ is the curvature two-form computed with the spin-connection $\Om^{(+)}$, which has been twisted by $H$-flux:
\be
\Om^{(\pm)}_M{}^{AB} = \Om_M{}^{AB} \pm \hlf H_M{}^{AB} +O(\a').
\ee
The $H$-flux already includes $O(\a')$ corrections:
\be\label{eq:H}
H = dB +{\a'\over4}\left(CS(\Om^{(+)})-CS(A)\right),
\ee
where $CS$ denotes the Chern-Simons three-form
\be
CS(A) = \tr\left(A\wedge dA -{2\over3}A\wedge A\wedge A\right)
\ee
and similarly for $\Om^{(+)}$. These corrections to $H$ are necessary for the cancelation of spacetime anomalies in the theory~\cite{Green:1984sg}, but we will see them emerge much more naturally from a worldsheet argument in Section~\ref{ss:BHAnom}.
Once we include the $\a'$ suppressed term $CS(\Om^{(+)})$, supersymmetry requires that we also include the other higher derivative interaction $\tr|\cR^{(+)}|^2$.
The Einstein-Hilbert term, ${\cal R}$, in the action is computed using the standard spin connection, $\Om$. Our convention for the norms of $p$-form fields is
\be
|C_p|^2 = {1\over p!}G^{M_1N_1}\ldots G^{M_pN_p} C_{M_1\ldots M_p}C_{N_1\ldots N_p}.
\ee
The equations of motion which follow from this action are
\bea
{\cal R}_{MN}+2 \nabla_M \nabla_N\varphi -{1\over 4} {H}_{MAB} {{H}_N}^{AB} \qquad\qquad\qquad\qquad\qquad\qquad\qquad\qquad\qquad\cr-{\a'\over4}\left[\tr\cF_{MP}\cF_N{}^P-\tr\cR_{MP}^{(+)}\cR_{N}^{(+)P}\right]= O(\alpha'^2), \label{eom0}\\
{\cal R} +4\nabla^2\varphi -4\nabla_M\varphi\nabla^M\varphi - \hlf|H|^2 -{\a'\over4}\left(\tr|\cF|^2-\tr|\cR^{(+)}|^2\right)= O(\a'^2),\\
 d \left( e^{-2\varphi} \star {H}\right) = O(\alpha'^2), \\
e^{2\varphi}d\left(e^{-2\varphi}\star\cF\right) + A\wedge\star\cF-\star\cF\wedge A +\cF\wedge\star H = O(\a'^2),
\eea
where we have used the dilaton equation to simplify the Einstein equation appearing above. In addition to these equations of motion, valid spacetime solutions must also satisfy the modified Bianchi identity:
\be\label{eq:bianchi0}
dH = {\a'\over4}\left(\tr\cR^{(+)}\wedge\cR^{(+)} -\tr\cF\wedge\cF\right),
\ee
which follows from~\C{eq:H}.

Heterotic supergravity also contains a set of Majorana-Weyl fermions, listed in Table~\ref{tab:fermionic}.
\begin{table}[ht]
\begin{center}
\begin{tabular}{cllc}
\hline\hline
\underline{ Field }     & \underline{ Name  }      & \underline{Representation  }          & \underline{d.o.f}\\
$\Psi_{M\a}$ & gravitino & right-chiral vector-spinor & 56\\
$\l_\da$ & dilatino & left-chiral spinor & 8\\
$\chi^a_\a$ & gaugino  & adjoint valued right-chiral spinor & $8\cdot{\rm dim}\,{\cal G}$\\
\hline\hline
\end{tabular}
\end{center}
\caption{Fermionic field content of  $\N=1$ $d=10$ supergravity and super-Yang-Mills}
\label{tab:fermionic}
\end{table}
Note that the theory contains an equal number of bosonic and fermionic degrees of freedom, as required by supersymmetry.
Spacetime supersymmetry is preserved if the variations of these fermions vanishes. To lowest order in $\a'$, the bosonic terms in the Killing spinor equations that must be satisfied are
\bea
\delta \Psi_M  &=& \left(\del_M +{1\over 4} \Om^{(-)}_{MAB} \G^{AB}\right)\epsilon=0, \label{eq:gravvar}\\
\delta \lambda &=& \left(  \del_M \varphi\G^M  -{1\over 12} H_{MNP}\G^{MNP} \right) \epsilon=0, \label{eq:dilvar} \\
\delta\chi &=& -\hlf \cF_{MN}\G^{MN}\epsilon=0 \label{eq:gaugvar}.
\eea
The advantage of these first-order supersymmetry equations is that their solutions automatically satisfy the second-order equations given above, provided we also impose the Bianchi identity~\C{eq:bianchi0}.

\subsubsection{Heterotic compactifications}
We can now consider compactifications of the heterotic string to  phenomenologically relevant spacetimes of the form
\be
\R^{1,3}\times \M,
\ee
where $\M$ is a compact six-dimensional manifold, and see what constraints the equations~\C{eom0}-\C{eq:gaugvar}\ impose on $\M$.
This was the analysis carried out in~\cite{strominger-torsion}\ (see also~\cite{Hull:1986kz}), where it was found that $\M$ must be a complex manifold equipped a nowhere vanishing holomorphic top form, $\Omega$, satisfying
\be\label{eq:holotop}
d\left( e^{-2\varphi} \Omega \right)=0,
\ee
and a Hermitian metric $G_{i\jbar}$, which defines the fundamental two-form
\be
J \equiv i G_{i\jbar}dz^i d\bar z^{\jbar},
\ee
and determines the $H$-flux and dilaton via the relations
\bea
&& H = i(\bar\del-\del)J,\label{eq:HJ} \\
&& d\left(e^{-2\varphi} J\wedge J\right) =0.
\eea
Furthermore, the field strength $\cF$ must satisfy the Hermitian-Yang-Mills equations:
\be\label{eq:HYM}
\cF_{ij} = \cF_{\ibar\jbar} = J^{i\jbar}\cF_{i\jbar} =0.
\ee
The first two equations above imply that $\cF$ takes values in a holomorphic vector bundle ${\cal E}$ over $\M$.\footnote{If $\cE$ has structure group $\cG_\cE$, then $\cE$ breaks the spacetime gauge symmetry down to the centralizer of $\cG_\cE$ in ${\cal G}$; that is to say, the gauge symmetry that survives is the largest group $\cG_{st}$ such that $\cG\supseteq \cG_{st}\times \cG_{\cE}$. For example, when $\cG=E_8\times E_8$ and $\cG_\cE=SU(3)$, $SU(4)$, or $SU(5)$, then $\cG_{st}=E_6\times E_8$, $SO(10)\times E_8$, or $SU(5)\times E_8$, respectively.}
The latter equation is known to be extremely difficult to solve.\footnote{When $\M$ is K\"ahler, the Donaldson-Uhlenbeck-Yau theorem provides a simple criteria for the existence of solutions to~\C{eq:HYM}, which suffices for our purposes. However for a general heterotic solution, in particular when $H\neq0$, there is no such theorem available.} On top of all this, the Bianchi identity
\be\label{eq:bianchi1}
dH  =  {\a'\over4}\left(\tr\cR^{(+)}\wedge\cR^{(+)} -\tr\cF\wedge\cF\right),
\ee
must be satisfied.

The set of relations~\C{eq:holotop}-\C{eq:bianchi1}\ are a set of non-linear first-order ODEs and they are highly intractable.
In fact, in the nearly thirty years since they were first written down, only one non-trivial class of solutions has ever been found~\cite{Dasgupta:1999ss},
(see also~\cite{Becker:2002sx,Goldstein:2002pg,Becker:2003gq,Becker:2003sh,Becker:2003yv,Becker:2005nb,Fu:2006vj,Becker:2006et,Becker:2006xp,Becker:2007ea,Becker:2009zx}),
along with a handful of generalizations of this basic example~\cite{Fu:2008ga,Becker:2008rc,Adams:2009av,Becker:2009df}. This should be contrasted to the much simpler set of constraints obtained by setting the $H$-flux to zero:
\be
H=0\quad\Rightarrow\quad d\Omega = dJ = d\varphi =0,\ \cRp=\cF.
\ee
This system was found in~\cite{Candelas:1985en}, roughly around the same time as the general case, and the solutions are known to be Calabi-Yau (CY) manifolds (i.e. complex, Ricci-flat manifolds with $dJ=0$). At the time only a few CY manifolds were known in three complex dimensions (i.e. six real dimensions), but thirty years later that number has grown astronomically. For example, one method\footnote{This method constructs CYs as hypersurfaces of quasi-homogeneous polynomials in four-dimensional complex weighted projective space.} of constructing CY manifolds is known to generate up to 473,800,776
distinct spaces~\cite{Kreuzer:2000xy}!\footnote{This number is only an upper bound on the number obtained by this construction, since some of these solutions may actually be isomorphic. There is a lower bound of 30,108 solutions, which can be distinguished by topological invariants and so are necessarily distinct.}
This only represents a (small) subset of all known examples, and it is still an open question whether the number of CY manifolds is finite in complex dimension three.

The arduous task of constructing non-trivial solutions to~\C{eq:holotop}-\C{eq:bianchi1}\ is further compounded by the fact that, once found, these are only solutions of supergravity and not the full, $\a'$-corrected, equations of string theory.
The CY solutions are reliable because all the length scales can be made large, and so $\a'$-perturbation theory can be trusted. On the other hand, heterotic solutions with $H\neq0$ will necessarily involve some cycles with sizes of order $O(\a')$.
To see this consider equations~\C{eq:HJ}\ and~\C{eq:bianchi1}, which together imply
\be\label{eq:scale}
2i\del\bar\del J = {\a'\over4}\left(\tr\cR^{(+)}\wedge\cR^{(+)} -\tr\cF\wedge\cF\right).
\ee
Under a rescaling of the metric, $G\rightarrow \l^2G$, the left-hand side would also rescales by $\l^2$, but the right-hand side is invariant. Therefore, those cycles where $dH\neq0$ are frozen at a size set by the only scale appearing in~\C{eq:scale}, namely $O(\a')$.
Note that the class of solutions in~\cite{Dasgupta:1999ss}\ are only trusted because they were derived by duality with known M-theory solutions, and not by directly solving the supergravity equations of motion.
Since solutions with $H$-flux generically contain $\a'$-scale cycles, we cannot trust the supergravity approximation and we must search for a worldsheet description of them.

\section{GLSMs in a nutshell}
In order to find compact solutions supporting $H$-flux, we are forced to examine them from the worldsheet.
However, just because a worldsheet theory is capable of describing $H$-flux solutions does not make it any easier to determine what those solutions are.
The non-linear nature of the sigma model makes explicit computations extremely prohibitive.
Fortunately, there is an alternative to studying the sigma models for $H$-flux solutions directly, which is to consider simpler theories that lie in the same universality class.
The full string solution with $H$-flux may then emerge as the low energy limit of these simpler theories.

The gauged linear sigma model (GLSM) was introduced by Witten in~\cite{Witten:1993yc}\ to study CY manifolds in exactly this manner. Let us review very briefly how these models work; a more detailed discussion will be presented in Chapter~\ref{ch:Review}. The basic idea is to couple charged scalars to gauge fields and to each other by potentials. The relevant part of the action is just
\be
S_{GLSM} = -\int d^2\s\left(\left|D_\m X^M\right|^2 -V(X)\right).
\ee
Generically the minimum of the scalar potential $V(X)$ forces some of the charged scalar fields to acquire vacuum expectation values. This generates masses for the gauge fields via the Higgs mechanism. Classically integrating out the gauge fields at low energies leads to the following replacement:
\be\label{eq:sub}
A_\m \rightarrow A_M(X)\del_\m X^M.
\ee
Restricting to the minimum of $V$ and carrying out this substitution leads to a non-linear action for the fields $X^M$:
\be
\left|D_\m X^M\right|^2\Big|_{V'=0} \rightarrow\ G_{MN}(X)\del_\m X^M\del^\m X^N.
\ee
So we see that the GLSM reduces to a Polyakov-type action at low energies, but what about the $B$-field?

There is another term we can add to the GLSM,
\be\label{eq:Sth}
S_\th = {\th\over2\pi}\int d^2\s\, F_{01} = {\th\over4\pi}\int d^2\s\, \e^{\m\n}F_{\m\n},
\ee
which is the two-dimensional analog of the coupling $\int F\wedge F$ in four dimensions. $S_\th$ does not affect the gauge field's equation of motion~\C{eq:sub}, because it is a total derivative and so only contributes topologically. Under the substitution~\C{eq:sub},
\be
F_{\m\n} \rightarrow  F_{MN}(X) \del_\m X^M\del_\n X^N,
\ee
where $F_{MN} = \del_M A_N - \del_N A_M$, and so at low energies we can identify
\be
B_{MN} = {\th\over4\pi} F_{MN}(X).
\ee
Notice, however, that this $B$-field is closed:
\be
H_{MNP} \simeq \del_{[M}B_{NP]} = {\th\over4\pi} \del_{[M} F_{NP]} = 0,
\ee
and so as it stands the standard GLSM construction is incapable of producing solutions with $H$-flux.

\section{Outline}
The goal of this thesis is to remedy the problem sketched at the end of the previous section. In particular, we seek to generalize the GLSM framework in such a way that their low energy descriptions are compact non-linear models with $H$-flux. The results we present here were reported earlier in a series of papers~\cite{Quigley:2011pv,Quigley:2012gq,Melnikov:2012nm}\ by the author and collaborators. We should mention that the many of the results of the~\cite{Quigley:2011pv}\ appeared simultaneously in~\cite{Blaszczyk:2011ib}.\footnote{See also~\cite{NibbelinkGroot:2010wm,Blaszczyk:2011hs,Ludeling:2012cu}\ for related work.} Also, the models we present here share some features of those in~\cite{Adams:2006kb}, which were further explored in~\cite{Adams:2009av,Adams:2007vp,Adams:2009tt,Adams:2009zg,Adams:2012sh,Israel:2013hna}.

For reasons to be explained, we will work exclusively in the context of $(0,2)$ supersymmetric theories. Chapter~\ref{ch:Review}\ reviews the necessary features of $(0,2)$ models needed to understand the rest of this work. We will briefly explain the structure and importance of $(0,2)$ supersymmetry before developing the language of $(0,2)$ superspace and superfields. We will then explore $(0,2)$ non-linear sigma models, and explain their relevance for heterotic string theory. We will close this review chapter by constructing  $(0,2)$ GLSMs, and show in detail how their low energy dynamics realize non-linear models and heterotic strings solutions.

In Chapter~\ref{ch:Hflux}\ we begin to develop the central theme of this thesis, which the incorporation of $H$-flux into the GLSM framework. The basic idea will be to promote the constant $\th$ parameters of~\C{eq:Sth}\ to field-dependent quantities, $\Th$. $H$ will no longer vanish in the non-linear models, but will instead take the schematic form $H\sim d\Th\wedge F$. The GLSM allows two basic possibilities: either $\Th$ is gauge-invariant, or it can shift as $\dd\Th\sim \La$. We consider examples in both cases. In the former, we find that when $\Th$ is both gauge-invariant and globally defined, then (not very surprisingly) $H$ is trivial in cohomology.
In the latter case, where we allow $\Th$ to shift, there are many interesting effects: chief among these being that the shift of $\Th$ violates the gauge invariance of the action unless we include a compensating violation by a quantum gauge anomaly. These models have quantized $H$, but without further information it is unclear whether this delicate balancing between classical and quantum gauge violations leads to any pathologies in the theory. Another puzzling feature of the non-invariant cases is that they can lead to non-complex target spaces, such as $S^4$, despite the fact that all $(0,2)$ models must be complex. We leave these issues open at this stage in the thesis in order to develop some concepts we will require to resolve them.

We make a slight detour in Chapter~\ref{ch:Neutral}\ to study a class of models that lie midway between those of the previous chapter. In particular, we examine $\Th$ which are gauge-invariant but not globally defined, shifting under some \textit{global} symmetries. These models share many of the interesting features of the gauge-varying models, without the subtleties associated with the anomaly. Since $\Th$ is not globally defined, it turns out that $H$ is quantized here as well. An important realization about these models is that they can be generated on novel branches of the standard $(0,2)$ GLSM's moduli space. These branches arise when a pair of non-chiral fermions become massive and are integrated out, leaving behind the $\Th$ coupling. The other main result of this chapter is that all solutions in this class of models contain explicit magnetic sources for $H$ (NS-branes). We explore several examples, both compact and non-compact.

In Chapter~\ref{ch:Charged}, we return to the models with gauge variant $\Th$, and apply the lessons we learned in the previous chapter. We show that this class of theories can arise on certain branches of the $(0,2)$ GLSMs where chiral fermions with different charges acquire a mass. Integrating them out generates the non-invariant $\Th$ coupling and leaves behind an anomalous spectrum of fermions. The gauge variations of the two effects cancel exactly so as to preserve the gauge symmetry of the original UV theory. The low energy metric of the sigma model is also modified in an important way, developing a singular boundary at finite distance. This is the resolution of the non-complex target manifolds. In the case of $S^4$ the singularity essentially divides the space in two leaving just a 4-ball. Understanding the nature and implications of this singularity are left for future work.

Finally, we summarize our results in Chapter~\ref{ch:conc}, and contemplate future lines of research. A set of Feynman rules for $(0,2)$ supergraphs, needed for computing the effective actions of Chapter~\ref{ch:Charged}, are collected in the Appendix.

\chapter{Review of $(0,2)$ models}\label{ch:Review}
In this chapter, we review the structure of $(0,2)$ supersymmetric field theories in $\nobreak{1+1}$ dimensions. These theories will play a crucial role throughout this thesis, and they provide the framework for all the developments in later chapters. We will also highlight the relationship between $(0,2)$ theories and heterotic strings.
For another nice review of this and related topics, see~\cite{McOrist:2010ae}.

\section{$(0,2)$ supersymmetry} \label{sec:supersymmetry}

\subsection{What is $(0,2)$ supersymmetry?}\label{ss:whatis}
We are all probably used to the fact that in $3+1$ dimensions $CPT$ relates Weyl fermions of opposite chiralities. It does not make sense to consider a theory of left-handed particles without also including the corresponding right-handed anti-particles. Each Weyl spinor, $\psi_\a$, has two complex components leading to a total of four real degrees of freedom.\footnote{$\bar\psi_\da$ does not contribute additional degrees of freedom since it is equivalent to $\psi_\a$ by $CPT$.} The same goes for the fermionic charges that generate supersymmetry: for every $Q_\a$ there is a corresponding $\bar Q_\da$. It is also possible to consider theories with $\N$-extended supersymmetry, meaning there exist multiple pairs of fermionic charges $(Q_\a^A, \bar Q_\da^A)$, where $A=1,\ldots,\N$. Then the amount of supersymmetry in a $3+1$ dimensional theory is specified by the single integer $\N$, and there a total of $4\N$ supercharges: one for each real component of $Q^A_\a$.

This is not the case in $1+1$ dimensions where $CPT$ maps Weyl fermions back to themselves. In such a situation one refers to $\N=(p,q)$ supersymmetry, where $p$ and $q$ are integers labeling the number of left- and right-moving supercharges, respectively~\cite{Hull:1985jv}. While the minimal spinor representation in $3+1$ dimensions is a two-component, complex Weyl spinor\footnote{Equivalently, one can use a four component Majorana (real) spinor representation.}, the smallest spinor representation in $1+1$ dimension is a single, real Grassmann variable. We will use $\pm$ subscripts to denote the chirality of spinors in $d=1+1$, so for example
\bea
\psi_+ \ &=&\  \textrm{right-moving chiral fermion, and}\non\\
\psi_- \ &=&\  \textrm{left-moving chiral fermion.} \non
\eea
Thus, theories with $\N=(p,q)$ supersymmetry are generated by supercharges
\bea
&Q^{A_-}_-&, \quad  A_-=1,\ldots,p, \quad \textrm{and}~\\
&Q^{A_+}_+&, \quad A_+=1,\ldots,q,
\eea
and therefore have a total of only $p+q$ (real) generators.

In particular, $(0,2)$ supersymmetry has no left-moving supercharges and two right-moving supercharges, ($Q^1_+,Q^2_+)$, which are usually combined into the complex combination
\be
Q_+ = {1\over\sqrt{2}}(Q_+^1 +i Q^2_+).
\ee
The complex supercharges satisfy the algebra
\bea\label{eq:algebra}
\{Q_+,Q_+\} = \{\bar Q_+,\bar Q_+\} =0 ,\quad \{Q_+,\bar Q_+\} = 2(H-P)
\eea
where $H$ and $P$ generate translations in time and space, respectively, for the $1+1$ dimensional spacetime.

Every heterotic sigma model has at least $(0,1)$ supersymmetry, but only those with at least $(0,2)$ can produce target spaces that preserve (spacetime) supersymmetry~\cite{Banks:1987cy}.

\subsection{$(0,2)$ superspace}\label{ss:superspace}

Throughout this thesis, we will use the language of $(0,2)$ superspace, which is an incredibly convenient and powerful formalism for working with supersymmetric theories. For one of the earlier references on the topic, see~\cite{Dine:1986by}. The basic idea is to enlarge the dimension of spacetime to include fermionic directions, so that supersymmetry transformations become translations in these extra (anti-commuting) dimensions.

Let us begin by establishing our conventions for $1+1$ dimensional spacetime. One option is to label the coordinates of spacetime by $x^\m$, with $\m=0,1$, and define the metric and Levi-Civita symbol to be
\be
\eta_{\m\n} = \bma -1& \ \\ \ & 1 \ema,\qquad \e_{\m\n} = -\e^{\m\n} = \bma \ &  -1  \\ 1 &\  \ema.
\ee
However, because of the chiral nature of $(0,2)$ supersymmetry, it will often prove more convenient to work with the lightcone coordinates
\be
x^\pm = {x^0\pm x^1\over2},
\ee
which we have normalized so that the lightcone derivatives
\be
\del_\pm = \del_0\pm \del_1
\ee
are simple. In particular, $\del_\pm x^\pm =1$. This simplification comes at the cost of a slightly more complicated metric and $\e$-symbol: in the basis $(x^+,x^-)$, we have
\be
\eta_{\m\n} = \bma \ & -2 \\ -2 & \ \ema,\qquad \e_{\m\n} = \bma \ &  2  \\ -2 &\  \ema.
\ee

To extend our $1+1$ dimensional spacetime to $(0,2)$ superspace, we must include two (anti-commuting) Grassmann coordinates, $\th^+,\thbar^+$. The $(0,2)$ supercharges can now be realized as differential operators on superspace:
\bea
Q_+ = \del_{\th^+} +i\thbar^+\del_+,\qquad \bar Q_+ = -\del_{\bar\th^+} - i\th^+\del_+.
\eea
It is easy to see that the algebra~\C{eq:algebra}\ is satisfied by these operators. We recall that Grassmann differentiation is the same as integration:
\be
\del_{\th^+}\th^+ = \int d\th^+\,\th^+ = 1
\ee
and we define the $(0,2)$ measure for Grassmann integration to be $d^2\th^+ \equiv d\thbar^+ d\th^+$, so that
\be
\int d^2\th^+\, \th^+\thbar^+ =1.
\ee

When constructing supersymmetric actions it also helps to introduce the $(0,2)$ super-derivatives
\be
D_+ = \del_{\th^+} - i\thbar^+\del_+, \qquad \Dbar_+=-\del_{\thbar^+} +i\th^+\del_+,
\ee
which anti-commute with $Q_+,\bar Q_+$, and satisfy an algebra similar to~\C{eq:algebra}:
\be
\{D_+,D_+\} = \{\Dbar_+,\Dbar_+\} = 0, \qquad \{\Dbar_+,D_+\} = 2i\del_+.
\ee

\subsection{$(0,2)$ superfields}\label{ss:superfields}
A function on $(0,2)$ superspace, such as  $Y(x^\pm,\th^+,\thbar^+)$, is called a $(0,2)$ superfield. Because of the anti-commuting nature of Grassmann variables, the Taylor expansion in $\th^+,\thbar^+$ of any superfield  terminates at finite order. For example,
\be\label{eq:gensuper}
Y(x^\pm,\th^+,\thbar^+) = y(x^\pm) + \th^+\psi_+(x^\pm) +\thbar^+ \chi_+(x^\pm) +\th^+\thbar^+ W_+(x^\pm).
\ee
The coefficients of each term in the $\th^+$ expansion are called component fields, and it should be clear from this example that superfields unify component fields of different spin, and in particular combine bosonic and fermionic fields, into a single object. In the example above, the superfield $Y$ contains component fields of spin-0, spin-$\hlf$ and spin-1. A completely general superfield, such as~\C{eq:gensuper}, is always reducible. In order to obtain non-trivial representations of supersymmetry we must impose some constraints on the component fields, in a way compatible with supersymmetry.

\subsubsection{Chiral superfields}
Consider a field $\Phi$ that is annihilated by the $\Dbar_+$ operator:
\be
\Dbar_+\F=0.
\ee
Since $\{\bar D_+,Q_+\}=\{\bar D_+,\bar Q_+\}=0$ this constraint is compatible with supersymmetry. Superfields that satisfy this relation are called chiral superfields, and it is trivial to determine that they have the following component expansion:
\bea
\F = \f+\sqrt{2}\th^+\j_+ - i \th^+\thbar^+\del_+\f, \label{chiral}
\eea
Thus a chiral superfield contains a complex scalar, $\phi$, and a right-moving Weyl fermion $\psi_+$. Note that the top component of $\F$ is a total derivative, so the integral over all of superspace vanishes:
\be\label{eq:chiralact}
\int d^2xd^2\th^+\ \F = -i \int d^2x\,\del_+\f =0.
\ee
Furthermore, it is easy to see that chiral superfields form a ring: that is if $\F_1$ and $\F_2$ are chiral superfields, then so are $\F_1+\F_2$ and $\F_1\F_2$. Therefore, any integrand built solely from chiral fields will also vanish when integrated over superspace, since the top component will always be a total derivative.

To get a non-vanishing result requires integrands built from chiral fields, $\F$, and \textit{anti-chiral} superfields, $\bar\F$, which satisfy the conjugate constraint $D_+\bar\F=0$. Also, since the action must be a Lorentz scalar, and the fermionic measure $d^2\th^+ =d\bar\th^+ d\th^+$ carries spin $+1$, the integrand must have spin $-1$. The simplest possibility turns out to be the action for a single, free, chiral superfield:
\be
S_\F = -{i\over2}\int d^2x d^2\th^+\, \bar\Phi \del_-\Phi = \int d^2x\left(-|\del_\m\f|^2 +i\bar\psi_+\del_-\psi_+\right).
\ee
More general actions for chiral superfields will be considered in the next section.

\subsubsection{Fermi superfields}
The other type of matter multiplet is a left-moving fermionic superfield, $\G_-$, which obeys
\be
\Dbar_+\G_- = \sqrt{2} E,
\ee
where $E$ is some chiral superfield.\footnote{$E$ must be the chiral, since $\bar D_+^2=0$.} The main case of interest will be when $E=E(\Phi)$ is a holomorphic function of the chiral superfields. Then the $\th^+$ expansion of $\G$ is
\be
\G_- = \g_- +\sqrt{2}\th^+F - i\th^+\thbar^+\del_+\g_- - \sqrt{2} \thbar^+ \left(E(\f) + \sqrt{2}\th^+ E'(\f)\psi_+\right).
\ee
Apart from the $\F$ dependence, we see that $\G_-$ contains a left-moving Weyl fermion, $\g_-$, and a complex scalar, $F$. We will see in a moment that $F$ is actually an auxiliary field, and so the only propagating degree of freedom is $\g_-$. This is consistent with the fact that $(0,2)$ supersymmetry does not act on left-movers, and so $\g_-$ should not have any superpartner. The simplest action built from Fermi superfields is
\bea
S_\G &=& -\hlf\int d^2xd^2\th^+\, \bar\G_-\G_- \\
&=& \int d^2x\left( i\bar\g_-\del_+\g_- +|F|^2-|E(\f)|^2 -\bar\g_- E'(\f)\psi_+ - \bar\psi_+ \bar E'(\bar\f)\g_-\right).\non
\eea
We see that $F$ is indeed non-propagating, and its equation of motion in this simple example is just $F=0$.

With the inclusion of Fermi superfields we can now add superpotential couplings to our action. Superpotentials are non-derivative interactions integrated over only half of superspace:
\be\label{eq:superpotgen}
S = \int d^2x d\th^+\,\La_-\Big|_{\thbar^+=0} +c.c.\,,
\ee
where Lorentz invariance requires that $\La_-$ be a left-moving fermion. A straightforward exercise reveals that such a coupling is invariant under $Q_+$ variations, while under $\bar Q_+$ it transforms as
\bea
\dd S &=& \int d^2x d\th^+\left(\bar\e^+\bar Q_+\right)\La_-\Big|_{\thbar^+=0} +c.c. \non\\
&=& \bar \e^+\int d^2x d\th^+\left(\bar D_+ -2i\th^+\del_+ \right)\La_-\Big|_{\thbar^+=0} +c.c. \\
&=& \bar\e^+\int d^2x d\th^+ \bar D_+\La_-\Big|_{\thbar^+=0} +c.c. \non,
\eea
because integrating over $\th^+$ in the second term leaves a total derivative. Therefore, in order for the superpotential coupling~\C{eq:superpotgen}\ to be invariant we conclude that $\La_-$ must be a chiral operator. For example, given a collection of Fermi fields $\G^\a$, with $\Dbar_+\G^\a=\sqrt{2}E^\a(\F)$, we can introduce superpotential couplings
\be\label{eq:super}
S_J = -{1\over \sqrt{2}}\int\d^2x\d\th^+\, \G^\a_- J_\a(\F)\Big|_{\thbar^+=0} + c.c.,
\ee
which are supersymmetric provided
\be
\Dbar_+(\G_-^\a J_\a) = \sqrt{2}E^\a J_\a =0.
\ee
In components, the superpotential couplings give
\be
S_J = -\int d^2x\left(F^\a J_\a(\f) + \g^\a_-J'_\a(\f)\psi_+ +c.c.\right).
\ee
$F$ remains auxiliary, but now its equation of motion becomes $F^\a = \bar J_\a$. After integrating out $F$, the combined fermionic actions yield the scalar potential
\be
V = \sum_\a\left(|E^\a|^2 + |J_\a|^2\right).
\ee

\section{Non-linear sigma models}\label{sec:02NLSM}
In the last section the only interactions we considered were mediated by $E^\a$ and $J_\a$ couplings. Let us now examine a much broader class of theories, which will play an important role throughout this thesis.

\subsection{Sigma model actions}\label{ss:sigmamodels}

To begin, we will restrict ourselves to theories built solely from a collection of chiral superfields, $\F^i$. The most general renormalizable action for chiral superfields is completely specified by a $(1,0)$-form $K = K_i\d\f^i$ with complex conjugate  $K^*=K_\ibar\d\f^\ibar$:
\be \label{eq:K}
S_{\s,1} =-{i\over4}\int d^2x\d^2\th^+\left[K_i(\F,\bar\F) \del_- \F^i - K_{\ibar}(\F,\bar\F) \del_- {\bar \F}^{\ibar} \right].
\ee
The one-form $K$ is the analogue of the K\"ahler potential found in $(2,2)$ theories. The  $(0,2)$ analogue of a K\"ahler transformation is
\be \label{eq:K1}
K(\F,\bar{\F}) \rightarrow K(\F,\bar{\F}) + K'(\F)
\ee
where $ K'(\F)$ is any holomorphic $(1,0)$-form. These transformations are in fact symmetries of the action, since they shift the Lagrangian by a purely chiral term, which integrates to zero as in~\C{eq:chiralact}. Furthermore, a shift in $K$ of the form
\be \label{eq:K2}
K \rightarrow K +i\,\del U,
\ee
for any real-valued function $U$, shifts the Lagrangian by a total derivative and is therefore is also a symmetry of the action. It will become clear that this latter symmetry corresponds to the $B$-field transformation~\C{eq:delB1}.

The component expansion of the action~\C{eq:K}\ is called a non-linear sigma model, and it reads
\be\label{eq:component}
S_{\s,1} =\int d^2x\left[ -\left(\eta^{\m\n}G_{i\jbar} + \e^{\m\n}B_{i\jbar}\right) \del_\m\f^i\del_\n\bar\f^\jbar + i G_{i\jbar}\,\bar\psi^\jbar_+ D_-\psi_+^i\right]
\ee
where
\be\label{eq:covder}
D_-\psi_+^i = \del_-\psi_+^i +\left[\left(\G^i_{jk} -\hf H^i{}_{jk}\right)\del_-\f^j +\left(\G^i_{\jbar k} -\hf H^i{}_{\jbar k}\right)\del_-\f^{\jbar}\right]\psi^k_+,
\ee
and the various couplings appearing in~\C{eq:component}\ and~\C{eq:covder}\ are all determined by $K$:
\be\label{eq:couplings}
G_{i\jbar} = \del_{(i} K_{\jbar)}, \qquad  B_{i\jbar} = \del_{[i} K_{\jbar]},\qquad  H_{\ibar\jbar k} = \del_{k[\jbar}K_{\ibar]}.
\ee
The connection $\G^i_{jk}$ is the usual Levi-Civita  connection associated to the metric $G_{i\jbar}$.
We recognize the bosonic terms in~\C{eq:component}\ as the worldsheet action of a string propagating on a target manifold $\M$, equipped with a metric, $G$, and $B$-field.
$(0,2)$ supersymmetry guarantees that $\M$ is a complex manifold~\cite{Hull:1985jv,Sen:1986mg}, with holomorphic coordinates given by $\phi^i$ and a Hermitian metric $G_{i\jbar}=\del_{(i}K_{\jbar)}$. The metric may be used to construct an associated fundamental form
\be
J = iG_{i\jbar}\,d\f^i d\bar\f^\jbar,
\ee
and $(0,2)$ supersymmetry guarantees that the couplings appearing in~\C{eq:couplings}\ are related by
\be\label{eq:susyreln}
H = dB = i(\bar\del- \del)J.
\ee

This story should be contrasted with the results of $(2,2)$ supersymmetric sigma models, where instead of a one-form $K_i$ the theory is specified by a K\"ahler potential function, $K^{(2,2)}$. The resulting target space, $\M_{(2,2)}$, is called a K\"ahler manifold with a metric
\be
G^{(2,2)}_{i\jbar} = \del_{i\jbar} K^{(2,2)}.
\ee
In such cases, the fundamental form $J$ is closed
\be
dJ^{(2,2)}=0
\ee
and it is called the K\"ahler form associated with the metric. Note that any $(0,2)$ target space will also be K\"ahler whenever $K_i=\del_i K$ for some function $K$ which, given~\C{eq:couplings}, makes it clear that $H$ and $B$ must vanish on K\"ahler manifolds.

\subsubsection{Right-movers}

The fermions $\psi_+^i$ behave as tangent vectors on $\M$, or more precisely they couple to the tangent bundle $T\M$ pulled back to $\S$ via the maps $\f^i$. The appearance of $H$ in the covariant derivatives~\C{eq:covder}\ signify that they are parallel transported not with the standard connection $\G$, but rather the connection \textit{with torsion}\footnote{See~\C{eq:torconn} for an explanation for why $\G^{(+)}$ is defined with relative $-$ sign.}:
\be
\G^{(+)} = \G -\hf H.
\ee
To see that $\psi_+^i$ do indeed transform as sections of the tangent bundle, let us introduce a set of $n$-beins $e^a_i$, $e^{\bar b}_\jbar$  (with inverses $e^i_a$, $e^\jbar_{\bar b}$) on $\M$. The metric is determined by the local frame fields in the usual way
\be
G_{i\jbar}(\f,\bar\f) = \dd_{a\bar b}\, e^a_i(\f,\bar\f) e^{\bar b}_\jbar(\f,\bar\f).
\ee
Defining $\psi_+^a \equiv e^a_i \psi_+^i$, we can rewrite the fermionic part of the sigma model action as
\be \label{eq:psi}
S_\psi =i\int d^2x\ \dd_{a\bar b}\, \bar\psi_+^{\bar b}\left(\del_-\psi_+^a +\left[\Om^{(+)}_i{}^a{}_b \del_-\f^i + \Om^{(+)}_\jbar{}^a{}_b\del_-\f^{\jbar}\right]\psi^b_+\right),
\ee
where $\Om^{(+)}$ is the spin-connection with torsion:
\be
\Om^{(+)}_I{}^a{}_b = \Om_I{}^a{}_b +\hf H_I{}^a{}_b  = e^a_j e^k_b \left(\G^{(+)j}_{I}{}_k - e^j_c \del_I e^c_k\right),\qquad I=i,\ibar
\ee
The fermionic action~\C{eq:psi}\ is then invariant under the combined transformations
\be\label{eq:delpsi}
\dd \psi^a_+ = \Th^a_b(\f,\bar\f) \psi_+^b,\qquad \dd \Om^{(+)}_I{}^a{}_b = -\del_I\Th^a_b(\f,\bar\f) - [\Om^{(+)}_I,\Th(\f,\bar\f)]^a_b.
\ee
We can interpret $\Th^a_b$ as an infinitesimal parameter for local Lorentz transformations on $T\M$. Then $\Om^{(+)}$ has the correct transforms properties for the spin-connection, and $\psi_+^a$ transforms as a tangent vector (i.e. as a section of $T\M$) should.

\subsection{Sigma models with left-movers}\label{ss:leftmovers}
Now that we understand the meaning of the non-linear sigma model for chiral superfields, we can ask what happens when we couple Fermi superfields to it. At the renormalizable level, and ignoring superpotential couplings, the general action for Fermi fields is\footnote{We have omitted terms of the form $(h_{\a\b}\G^\a\G^\b +c.c.)$ since these may be removed by a (non-holomorphic) redefinition of $\G^\a$. If we insisted on $\Dbar_+\G^\a=0$ then we would not be allowed to make such non-holomorphic change of basis, but having $E^\a\neq0$ allows this simplification to be made.}
\be\label{eq:Fermi}
S_{\s,2} = -\hlf \int d^2xd^2\th^+\ h_{\a\bar{\b}}(\F,\bar\F)\bar{\G}^{\bar \b}\G^\a
\ee
The $E^\a$ couplings introduce potential and Yukawa couplings, much like a superpotential which we have omitted, so let us $E^\a=0$ for now as well. Performing the superspace integral in~\C{eq:Fermi}\ gives the component action
\be\label{eq:bundle}
S_{\s,2} = \int d^2x\left[ih_{\a\bar{\b}}\,\bar\g_-^\bbar D_+ \g_-^\a +  \left(\cF_{i\jbar}\right)_{\a\bbar}\bar\psi^\jbar_+\j^i_+\bar\g^\bbar_-\g^\a_- +\L_{aux}\right]
\ee
where
\bea
D_+\g_-^\a &=& \del_+\g^\a_- + \left(A_i\right)^\a_\b \del_+\f^i\g_-^\b, \non\\
\left(A_i\right)^\a_\b &=& h^{\a\bar{\a}}\del_i h_{\b\bar{\a}}, \label{eq:YMfield}\\
\left(\cF_{i\jbar}\right)_{\a\bbar} &=& h_{\b\abar} \left(\dd^\b_\a\del_i\left(A_\jbar\right)^\abar_\bbar - \left(A_i\right)^\b_\a \left(A_\jbar\right)^\abar_\b\right) \non.
\eea
Much like $\psi_+^i$ couples to (the pullback of) the tangent bundle, $T\M$, the couplings in~\C{eq:YMfield}\ make it apparent that $\g^\a_-$ couple to (the pullback of) some holomorphic vector bundle ${\cal E}$ over $\M$, with Hermitian metric $h_{\a\bar\b}$, connection $A_i$, and curvature ${\cal F}_{i\jbar}$. Note that~\C{eq:bundle}\ is invariant under the combined transformations
\be\label{eq:delgam}
\dd \g^\a_- = \La^\a_\b(\f) \g^\b_-,\qquad \dd \left(A_i\right)^\a_\b = -\del_i\La^\a_\b(\f) -[A_i,\La(\f)]^\a_\b,
\ee
where $\La(\f)$ is a \textit{holomorphic} function. These transformations can be interpreted as gauge transformations on ${\cal E}$, just as~\C{eq:delpsi}\ could be interpreted as local Lorentz transformations on $T\M$.

The final term in~\C{eq:bundle}\ is just action for the auxiliary fields
\be
\L_{aux} =  \left(F^\a + \left(A_i\right)^\a_\b\psi^i_+\g_-^\a \right)h_{\a\bbar}\left(\bar{F}^\bbar - \left(A_\jbar\right)^\bbar_{\bar{\e}}\bar\psi^\jbar_+\g_-^{\bar{\e}} \right),
\ee
which vanishes once the $F^\a$ are integrated out.

\subsection{$B$-fields, $H$-flux and anomalies}\label{ss:BHAnom}
While the sigma model action
\be
S_\s = S_{\s,1}+S_{\s,2}
\ee
is classically invariant under the combined field redefinitions~\C{eq:delpsi}\ and~\C{eq:delgam} this is not the case quantum mechanically. The reason is that chiral symmetries such as~\C{eq:delpsi}\ and~\C{eq:delgam}, which act separately on left- and right-moving fermions, are generically not respected by the one-loop effective action. As shown by Fujikawa, such anomalous transformations of the action can be traced to a non-invariance of the path integral measure for the chiral fermions~\cite{Fujikawa:1979ay}. In this particular case, the anomalous shift in the action is given by
\be\label{eq:siganom}
\dd S_\s = {1\over8\pi}\int\d^2x\,\textrm{tr}\left(\Th {\cal R}^{(+)}_{i\jbar} - \La\cF_{i\jbar}\right)\e^{\m\n}\del_\m\f^i\del_\n\bar\f^\jbar,
\ee
where ${\cal R^{(+)}}$ is the curvature two-form associated to $\Om^{(+)}$.

The only way to make~\C{eq:siganom}\ vanish identically is if we can identify $\Th\sim\La$ and ${\cal R}^{(+)}\sim\cF$. This is only possible in the non-chiral case where the left- and right-movers have identical couplings. In particular, this means that $\psi_+$ and $\g_-$ must couple to the (pullback of) the \textit{same} vector bundles, and so
\be
\dd S_\s =0 \quad \Leftrightarrow \quad  {\cal E}\simeq T\M.
\ee
Since such a theory must be left-right symmetric it follows that this choice of ${\cal E}$ leads to $(2,2)$ supersymmetry. This choice is often called the \textit{standard embedding}.

Rather than forcing~\C{eq:siganom}\ to vanish on the nose, we can instead try to cancel it by assigning additional field transformations beyond those of $\Om^{(+)}$ and $A$ appearing in~\C{eq:delpsi}\ and~\C{eq:delgam}. This can be achieved by positing the following transformation for $B$:\footnote{The coefficient of $\dd B$ arises if we canonically normalize the action $S_\s\rightarrow(2\pi\a')^{-1}S_\s$.}
\be\label{eq:delB}
\dd B_{i\jbar} = {\a'\over4}\,\textrm{tr}\left(\Th {\cal R}^{(+)}_{i\jbar} - \La\cF_{i\jbar}\right).
\ee
While this anomalous variation of $B$ keeps $S_\s$ invariant under local Lorentz and gauge transformations of the target space, the field strength $H=dB$ is no longer an invariant tensor. The sigma model anomaly therefore induces a quantum correction to the definition of $H$, namely
\be
H = dB +{\a'\over4}\left(CS(\Om^{(+)}) - CS(A)\right),
\ee
where
\be
CS(A) = \tr\left(A\wedge dA -{2\over3}A\wedge A\wedge A\right)
\ee
is the Chern-Simons three-form for the target space gauge field $A$, and similarly for $\Om^{(+)}$. This quantum corrected $H$ is no longer closed, but instead satisfies the modified Bianchi identity
\be
dH = {\a'\over4}\left(\tr {\cal R}^{(+)}\wedge {\cal R}^{(+)} - \tr\cF\wedge\cF\right).
\ee

\section{$(0,2)$ superconformal models and heterotic strings}\label{sec:stringsol}
The main interest in $(0,2)$ sigma models stems from the fact that they can often be used for supersymmetric compactifications of the heterotic string. Suppose we perform a Kaluza-Klein reduction of the full ten-dimensional theory on a spacetime of the form $\R^{1,3}\times\M_6$, where $\M_6$ is some (real) six-dimensional manifold, to obtain an effective description in the four-dimensional spacetime. Then, under some mild assumptions\footnote{Namely, we must assume that all states carry integer charge under the $U(1)_R$ symmetry of the $(0,2)$ superconformal algebra.}, the four dimensional theory will have at least $\N=1$ supersymmetry if and only if the sigma model for $\M_6$ has $(0,2)$ superconformal symmetry~\cite{Banks:1987cy}.

A detailed discussion of $(0,2)$ superconformal symmetry will take us far beyond our needs in this thesis; for details, refer to the review~\cite{McOrist:2010ae}. The main point that concerns us is that a $(0,2)$ superconformal field theory (SCFT) is one that is $(0,2)$ supersymmetric (obviously) and also conformally invariant. A remarkable property of string theory is that the beta-function equations for the sigma model couplings $G,B,A$ and $\varphi$ (the dilaton), which by definition must vanish for conformal models, are precisely the supergravity equations of motion for those fields:
\bea
{\cal R}_{MN}+2 \nabla_M \nabla_N\varphi -{1\over 4} {H}_{MAB} {{H}_N}^{AB} \qquad\qquad\qquad\qquad\qquad\qquad\qquad\qquad\qquad\cr
-{\a'\over4}\left[\tr\cF_{MP}\cF_N{}^P-\tr\cR_{MP}^{(+)}\cR_{N}^{(+)P}\right] = O(\alpha'^2), \label{eom}\\
{\cal R} +4\nabla^2\varphi -4\nabla_M\varphi\nabla^M\varphi - \hlf|H|^2 -{\a'\over4}\left(\tr|\cF|^2-\tr|\cR^{(+)}|^2\right)= O(\a'),\\
 d \left( e^{-2\varphi} \star {H}\right) = O(\alpha'^2), \\
e^{2\varphi}d\left(e^{-2\varphi}\star\cF\right) + A\wedge\star\cF-\star\cF\wedge A +\cF\wedge\star H = O(\a'^2),
\eea
For K\"ahler metrics, which necessarily have $H=0$ and $\varphi$ constant, the condition for a supersymmetric Minkowski solution is Ricci-flatness:
\be\label{eq:ricci}
{\cal R}_{MN}=0.
\ee
For K\"ahler metrics,~\C{eq:ricci}\ is equivalent to the Monge-Amp\`ere equation
\be\label{eq:monge-ampere}
{\cal R} \equiv \del\bar\del\log\det\left( G\right) = 0,
\ee
where the target space metric $G$ is expressed in holomorphic coordinates.
For backgrounds with NS-flux, the conditions are more involved because of the $H$-field and associated varying dilaton. For $(2,2)$ models with flux and varying dilaton, a generalized Monge-Amp\`ere equation constraining the generalized $(2,2)$ K\"ahler potential (which includes semi-chiral fields) was described in~\cite{Hull:2010sn}.

Most $(0,2)$ sigma models are not conformal and will not provide solutions to the heterotic spacetime equations of motion. The heterotic conditions for a spacetime supersymmetric solution were derived from supergravity in~\cite{strominger-torsion}, and considered from a pure spinor perspective in~\cite{Andriot:2009fp}. We would like to understand the local conditions on $K$, analogous to~\C{eq:monge-ampere}, required for a spacetime solution with Minkowski spacetime.\footnote{In a perturbative $\alpha'$ expansion, there are no four-dimensional solutions of de Sitter or anti-de Sitter type, with or without spacetime supersymmetry~\cite{Green:2011cn, Gautason:2012tb}.}

 In addition to the equations of motion, we expect spacetime supersymmetry to be unbroken by the metric, flux and dilaton. It might be broken by the choice of gauge bundle but that is an effect higher order in $\alpha'$. Ignoring the gaugino constraint, spacetime supersymmetry requires the existence of a Killing spinor $\e$ satisfying
\bea
 \label{gravitinovar}\delta \Psi_M  &=& \left(\del_M +{1\over 4} \Om^{(-)}_{MAB} \G^{AB}\right)\epsilon=0, \\
  \label{dilvar} \delta \lambda &=& \left(  \del_M \varphi\G^M  -{1\over 12} H_{MNP}\G^{MNP}  \right) \epsilon=0.
\eea
The first condition~\C{gravitinovar}\ requires $SU(n)$ structure for a complex $n$-dimensional target manifold. This implies the existence of a nowhere vanishing holomorphic top form $\Omega$ satisfying
\be
d\left( e^{-2\varphi} \Omega \right)=0.
\ee
Note that condition~\C{eq:susyreln},
\be\label{eq:susyreln1}
H = i(\delbar-\del)J,
\ee
is automatically satisfied for any model with $(0,2)$ supersymmetry. Spacetime supersymmetry also implies a constraint on $J$:
\be
d\left( e^{-2\varphi}  J^{n-1} \right) =0.
\ee

\subsection{Conditions for heterotic solutions}\label{ss:alternate}

The constraints on the geometry, flux and dilaton of a $(0,2)$ solution can be elegantly encoded in properties of the torsional connection
\be \Om_M^{(-)}=\Om_{M}-\hlf H_M, \ee
with $\Om_M$ the usual spin connection; see, for example, Appendix A of~\cite{Gillard:2003jh}\ or~\cite{Ivanov:2000ai}. Note that the torsional affine connection contains a relative sign:
\be\label{eq:torconn}
\G_{MN}^{(\pm)P} = e^P_A\left(\del_M e^A_N + e^B_N\Om_M^{(\pm)A}{}_B \right) = \G^P_{MN} \mp\hlf H^P{}_{MN}.
\ee
The two Killing spinor equations~\C{gravitinovar}\ and~\C{dilvar}\ imply the existence of an integrable complex structure that is covariantly constant with respect to $\Om^{(-)}$.  A Hermitian manifold satisfying this property is called K\"ahler with torsion (KT). Covariant constancy of the complex structure implies the constraint~\C{eq:susyreln1}, which can be re-written as follows,
\be\label{eq:rewrite}
\G_{i\jbar}^{(-)k} = \G_{\ibar\jbar}^{(-)k}= 0,
\ee
so that $\Om^{(-)}$ has $U(n)$ holonomy.
The gravitino equation~\C{gravitinovar}\ implies that the holonomy of $\Om^{(-)}$ is actually in $SU(n)$ rather than $U(n)$. This holds iff  ${\cal R}^{(-)} = d\o^{(-)}=0$, where
\be
\o_i^{(-)} = i\G_{ij}^{(-)j} - i\G_{i\jbar}^{(-)\jbar} = 2i G^{j\bar{k}}\del_j G_{i\bar{k}} - i G^{j\bar{k}}\del_i G_{j\bar{k}}
\ee
is the connection on the canonical bundle induced by $\Om^{(-)}$. This is a natural torsional generalization of a Calabi-Yau space. Condition~\C{eq:rewrite}, which is a rewriting of~\C{eq:susyreln1}, follows automatically from $(0,2)$ superspace whether the model is conformal or not.  Imposing conformal invariance requires $SU(n)$ structure.

To solve the dilaton supersymmetry constraint~\C{dilvar}, it is useful to introduce the Lee form of a KT manifold defined by,
\be \xi = -2i\delbar^\dagger J, \label{definexi}\ee
where $\delbar^\dagger$ is the adjoint of $\delbar$.\footnote{There is a factor of $2$ in~\C{definexi}\ because the Lee form appears in the modification of the K\"ahler identities for a non-K\"ahler space. See page 307 of~\cite{Demailly_2009}.}
The components of $\xi$ are determined in terms of $G$,
\be
\xi_i = iH_{ij\bar{k}}J^{j\bar{k}} = G^{j\bar{k}}\left(\del_i G_{j\bar{k}} - \del_j G_{i\bar{k}}\right).
\ee
 In terms of the Lee form, the dilatino equation~\C{dilvar}\ becomes,
\be \label{xitriv}
\xi =2\partial\varphi,
\ee
with $\varphi$ real. KT manifolds with exact Lee forms are conformally balanced. An explicit check that conformally balanced KT manifolds with $SU(n)$ structure solve the supergravity equations~\C{eom}\ can be found~\cite{Ivanov:2000ai}.

One might ask under what conditions $SU(n)$ structure implies a solution of the dilaton constraint. To relate the two constraints, note that
\be\label{omegaminus}
\o^{(-)} = i(\del-\delbar)\log\det G -2i\xi + 2i\bar{\xi}.
\ee
The condition of $SU(n)$ structure then requires,
\be\label{flatness}
 \del\xi - \delbar\bar{\xi} + \delbar\xi -\del\bar{\xi} +\del\delbar\log\det G=0.
\ee
This condition is the generalization of the Monge-Amp\`ere equation~\C{eq:monge-ampere}\ to KT manifolds. Following~\cite{strominger-torsion}, we can examine the $(0,2)$ part of this equation which implies
\be
\delbar \bar\xi = 0.
\ee
At least on a space with $h^{(0,1)}=0$, we can conclude that $\bar\xi = 2\delbar \varphi$ for some complex $\varphi$. It remains to show that $\varphi$ can be chosen real. It is not unreasonable to expect this to be true in fairly general circumstances for compact manifolds.\footnote{It might be possible to show this for compact spaces with $h^{(0,1)}=0$ by modifying the argument of~\cite{strominger-torsion}, where a simply-connected space is assumed. There are two complications that need to be addressed. First: on a non-K\"ahler space, $\sum_{p+q=n} h^{p,q} \geq b_{n}$ (see, for example~\cite{Becker:2003yv}) so simply-connected is not sufficient to guarantee exactness of the Lee form; however, assuming $h^{(0,1)}=0$ is good enough for $\delbar$ triviality. The second complication is that the $\square_{\partial}$ and $\square_{\bar\partial}$ Laplacians differ by linear differential operators that depend on $\xi$ (see~\cite{Demailly_2009}). This complicates the original proof of~\cite{strominger-torsion}\ that ${\rm Im}(\varphi)$ is constant on a compact space. } Our examples will be both non-compact and non-simply-connected so we will need to examine what can be said about the Lee form in each case.

When~\C{xitriv}\ is satisfied with a real $\varphi$, we can rewrite the generalized Monge-Amp\`ere equation~\C{flatness}\ as follows:
\be
{\cal R}^{(-)} = \del\bar\del\log\left(e^{-4\varphi}\det G\right) = 0.
\ee
In summary, a KT manifold with $SU(n)$ structure and a (de Rham) exact Lee form provides a supersymmetric heterotic string solution.

\subsection{A counter-example: $S^3\times S^1$}
\label{ss:S3xS1}

The $SU(2)\times U(1)$ WZW models are a well studied family of conformal field theories associated with a compact non-K\"ahler manifold. See, for example,~\cite{Spindel:1988nh, Spindel:1988sr, Rocek:1991vk}.  The target space is $S^3\times S^1$, which we can view as $E\rightarrow \PP^1$, where $E$ is a torus constructed from the Hopf fiber of $S^3\cong S^1\rightarrow S^2$ together with the free circle. See, for example,~\cite{Witten:2005px}. The family of conformal field theories is labeled by the amount of integer $H$-flux threading the $S^3$.

Let $(\f,\th)$ be coordinates on the Hopf fiber and the free circle, respectively, and let $z=\f+i\th$ be a complex combination parameterizing $E$.  Note that $z$ is not a complex coordinate on $S^3\times S^1$ because the complex structure operator maps $d\th$ to $d\phi+A$, where $A$ is the potential for the K\"ahler form on $\PP^1$: $dA=J_{FS}$. In these coordinates, the fundamental form for the space is given by
\be
J = J_{FS} +i(dz+A)\wedge(d\bar{z}+A).
\ee
The $H$-flux threading the $S^3$ takes the form
\be
H = \hlf\left( dz+A\right)\wedge J_{FS} + c.c.,
\ee
where we have assumed one unit of flux.
A non-linear sigma model with this target space is a perfectly good CFT; however, this is not an admissible string background because there is no well-defined dilaton. In particular, the Lee form
\be
\xi + {\bar \xi} = {i\over 2}\left( d\bar{z} + A\right) - {i\over2}\left(dz+A\right) = d\th
\ee
is not exact. Only when we replace $S^1 = {\mathbb R}/{\mathbb Z}$ by its cover ${\mathbb R}$ does $\th$ becomes a globally defined function. With this replacement, it make sense to identify $\th\sim2\varphi$.

\section{Gauged linear sigma models}\label{sec:GLSM}
The gauged linear sigma model (GLSM), introduced by Witten in~\cite{Witten:1993yc}, is a powerful tool for studying non-linear sigma models.
The basic idea is to construct supersymmetric gauge theories whose low-energy dynamics are described by complicated sigma models.
For some of the early developments and applications of GLSMs with $(0,2)$ supersymmetry, see~\cite{Distler:1993mk,Distler:1995mi,Blumenhagen:1995tt,Distler:1996tj}.
An advantage of this construction is that the CFTs that emerge from $(0,2)$ GLSMs are not destabilized by worldsheet instanton effects~\cite{Silverstein:1995re,Basu:2003bq,Beasley:2003fx}, while no such guarantee exists for the generic non-linear model.
Furthermore, many interesting though difficult to compute quantities of the non-linear models can be evaluated in the much simpler gauge theories.
For a highly incomplete sampling of some of the calculations that have been performed, see~\cite{Adams:2003zy,Katz:2004nn,Guffin:2007mp,McOrist:2007kp,McOrist:2008ji,Kreuzer:2010ph,Melnikov:2010sa,McOrist:2011bn,Benini:2013nda}.
Before we can explore this framework we must introduce supersymmetric gauge fields and their couplings to matter superfields.

\subsection{$(0,2)$ vector superfields}\label{ss:vectors}
For a general $U(1)^n$ abelian gauge theory, we require a pair $(0,2)$  gauge superfields $A^a$ and $V_-^a$ for each abelian factor, $a=1,\ldots,n$. Let us restrict to $n=1$ for now. Under a super-gauge transformation, the vector superfields transform as follows,
\bea
\dd A = {i}(\bar{\La} - \La)/2, \qquad \dd V_-  - \del_-(\La + \bar{\La})/2,
\eea
where the gauge parameter $\La$ is a chiral superfield: $\Dbar_+ \La=0$.  In Wess-Zumino gauge\footnote{Wess-Zumino gauge is a partial fixing of the super-gauge symmetry, which uses the component fields $\Im\!(\La)$ and $\psi_{\La,+}$ of $\La$ to eliminate the lowest components of $A$. Standard $U(1)$ gauge transformations, with gauge parameter $\Re\!(\La)$, are still a residual symmetry in this gauge.} the components of the gauge superfields are
\bea
A &=& \th^+\thbar^+ A_+, \\
V_- &=& A_- - 2i\th^+\labar_- -2i\thbar^+\l_- + 2\th^+\thbar^+ D,
\eea
where $A_\pm = A_0 \pm A_1$ are the components of the gauge field. We note that the vector multiplet contains a gauge field, $A_\pm$, two left-moving gauginos, $\l_-,\labar_-$, and a scalar field $D$ that will turn out to be auxiliary.

For a field of charge $Q$, we denote the gauge covariant derivative by
\be
\cD_\pm = \del_\pm + i Q A_\pm.
\ee
We must also introduce the supersymmetric gauge covariant derivative,
\be\label{eq:nabla}
\nabla_- = \del_- +  Q (\del_-A +i V_-),
\ee
which contains $\cD_-$ as its lowest component. The field strength is contained in the gauge-invariant Fermi multiplet defined as follows:
\be
\Upsilon_- =[\bar{D}_+,\nabla_-] =  \Dbar_+(\del_- A + i V_-) = -2\big(\l_- - i\th^+(D-iF_{01}) - i\th^+\thbar^+\del_+\l_-\big).
\ee
Kinetic terms for the gauge field are given by
\be \label{eq:LU}
S_\Upsilon = -{1\over8e^2}\int\d^2x\d^2\th^+\, \bar{\Upsilon}_-\Upsilon_- = {1\over e^2}\int\d^2x\left(\hlf F_{01}^2 + i\labar_-\del_+\l_- + \hlf D^2\right).
\ee
As claimed earlier, we see that $D$ is indeed non-propagating. Note that in $d=1+1$, the gauge coupling $e^2$ has mass dimension 2, and so all gauge interactions (even Abelian ones) are free in the ultraviolet, but strongly coupled in the infrared.

Since $\Upsilon_-$ is a Fermi superfield annihilated by $\bar D_+$ (and gauge-invariant) we are free to introduce a superpotential coupling for it. The simplest possibility involves just a constant coefficient
\be\label{eq:t}
t=ir + {\th\over2\pi},
\ee
where $r$ is called the Fayet-Iliopoulos parameter, and $\th$ is called the $\th$-angle of the gauge theory, for reasons that will become apparent in a moment. We will refer to the complex combination~\C{eq:t}\ as the \textit{complexified FI parameter}.
The action for this FI coupling is
\be \label{eq:LFI}
S_{FI}={t\over4}\int\d^2x\d\th^+\, \U_-\Big|_{\thbar^+=0} + c.c. = \int\d^2x\left(-rD + {\th\over2\pi}F_{01}\right).
\ee
More general superpotential couplings for $\U_-$ are possible, and in fact their study will be the main focus of this thesis.

\subsubsection{Quantization of $F$ and the $\th$-angle}
The $\th$-angle term in~\C{eq:LFI}\ is a total derivative:
\be
F_{01} = \hf \e^{\m\n}F_{\m\n} =\e^{\m\n}\del_\m A_\n,
\ee
and one might naively suspect that it integrates to zero, but this is not always the case. Indeed, the integral over $F$ reduces to a boundary integral
\be
{1\over2\pi}\int_\S F = {1\over2\pi}\oint_{S^1_\infty} A,
\ee
where $S^1_\infty$ is, roughly speaking, the boundary circle of $\S$ at infinity. Now, the gauge field $A$ takes values in $U(1)\simeq S^1$, and so we have a homotopy mapping problem:
\be
A: \ S^1_\infty \longmapsto U(1).
\ee
Maps of this type are characterized by number $n$ called the \textit{winding number}, which counts the number of times the $U(1)$ circle is wrapped as one goes around the $S^1_\infty$. Since $n$ is an integer, it cannot vary smoothly as one  continuously deforms the map $A$, therefore $n$ is a topological invariant. This leads to the quantization of $F$:
\be
S_\th = {\th\over2\pi} \int d^2x \, F_{01} = n\th.
\ee
The value of $\th$ can have important consequences on the dynamics of the gauge theory. However, since the path-integral only depends on $\th$ via the combination
\be
e^{iS_\th} = e^{in\th},
\ee
we see that the physics is invariant under the identification of $\th \sim \th+2\pi$. Thus, $\th$ behaves like a periodic angle.

\subsubsection{Gauge-invariant interactions}
Returning now to the GLSMs, the form of the supersymmetric covariant derivative $\nabla_-$, in~\C{eq:nabla}, ensures that if $\F$ transforms according to
\be
\F\rightarrow e^{iQ\La}\F,
\ee
then so does $\nabla_-\F$:
\be
\nabla_-\F\rightarrow \left(\nabla_- -iQ\del_-\La\right)e^{iQ\La}\F= e^{iQ\La}(\nabla_-\F).
\ee
The standard kinetic terms for charged chirals in $(0,2)$ GLSMs are
\bea \label{eq:LPhi}
S_\F &=& -{i\over2}\sum_i\int\d^2x\d^2\th^+\ \bar{\F}^i e^{2Q_iA} \nabla_- \F^i, \\
&=& \int\d^2x\left(-\big|\cD_\mu \f^i\big|^2 + \bar{\psi}_+i\cD_-\psi_+^i - \sqrt{2}iQ_i \bar{\f}^i\l_-\j^i_+ + \sqrt{2}iQ_i\f^i\bar{\j}_+^i\labar_- + Q_i \big|\f^i\big|^2\right). \non
\eea
Fermi superfields are treated similarly. If we make the standard assumption that $E$ is a holomorphic function of the $\F^i$, then the standard kinetic terms for the Fermi fields are
\bea \label{eq:LLa}
S_\G &=& -\hlf\sum_\a\int\d^2x\d^2\th^+\,  \bar{\G}^\a e^{2Q_\a A}\G^\a, \\
&=& \int\d^2x\left(i\bar{\g}_-^\a\cD_+\g_-^\a + \big|F^\a\big|^2 - \big|E^\a\big|^2 - \bar{\g}^\a_-\del_i E^\a \j_+^i - \bar{\j}_+^i \del_\ibar \bar{E}^\a \g^\a_-\right). \non
\eea
It is also possible to add a superpotential to the theory, so long as the couplings are gauge-invariant. Since these does not add anything new beyond the uncharged case~\C{eq:super}\ studied earlier, we will not write them down explicitly here. In the absence of any superpotential couplings, the action consisting of the terms \C{eq:LU}, \C{eq:LFI}, \C{eq:LPhi} and \C{eq:LLa} comprises the standard $(0,2)$ GLSM. In particular, after integrating out the auxiliary fields, the scalar potential of such theory is given by
\be\label{eq:scalarpot}
V = \sum_a{1\over e_a^2}D^aD^a +\sum_\a |E^\a(\f)|^2,
\ee
where we have allowed for a multiple $U(1)$ gauge factors, and the solutions for the auxiliary fields $D^a$ is
\be\label{eq:D}
D^a = -e_a^2\left(\sum_iQ^a_i|\f^i|^2 -r^a\right).
\ee

\subsection{NLSMs from GLSMs}
\label{ss:classicalgeometry}
Now we come to the real utility of the GLSM, which is that we can use it to engineer non-linear sigma model as its effective low energy description. The idea is that the gauge fields typically acquire masses via the Higgs mechanism, and when we integrate them out at low energies we induce a non-trivial metric on the target space of the theory.

The classical infra-red limit of a  $U(1)^n$ GLSM corresponds to sending $e_a\rightarrow\infty$, since these gauge couplings are  dimensionful quantities. In this limit, formally the $\U_-^a$ kinetic terms disappear resulting in the simple on-shell bosonic action,
\be\label{eq:Sbos}
S_{bos} = - \left|D_\mu\f^i\right|^2 +{\th^a\over2\pi}F_{01}^a - V(\phi),
\ee
where the scalar potential is given in~\C{eq:scalarpot}. Let us once again consider the case where all $E^\a$ are zero and assume $N$ fields $\phi^i$. The vacuum manifold of this theory is then
\be
\M = \left\{\f^i\in\CC^N\,\Big|\, \sum_iQ^a_i|\f^i|^2 = r^a\right\}/U(1)^n.
\ee
Such a space is sometimes called a toric variety, realized as the symplectic quotient $\CC^N//U(1)^n$ with moment maps $D^a$. $\M$ is $(N-n)$ complex dimensional, since each $D$-term constraint ($D^a=0$) and $U(1)$ quotient together remove one complex degree of freedom.

In principle we could continue to work in components and solve the (now algebraic) equations of motion for the gauge fields $A^a_\m$, stemming from~\C{eq:Sbos}, to find the low energy metric induced on $\M$. However this procedure has drawbacks since the resulting metric is not manifestly Hermitian, and one must use the constraints $D^a=0$ to massage it into the appropriate form. Since Hermiticity of the metric is a consequence of $(0,2)$ supersymmetry, a more efficient approach is to work directly in $(0,2)$ superspace to obtain the $K_i$ that define the sigma model, and then compute quantities like the $G_{i\jbar}$.

We will begin by returning to the GLSM action in $(0,2)$ superspace. Since we are primarily interested in the target space geometries, and not the vector bundles over them, we will ignore the Fermi superfields (even though they will be crucial in canceling the sigma model anomalies of Section~\ref{ss:BHAnom}). In the $e_a^2\rightarrow\infty$ limit, the $\U^a_-$ kinetic terms decouple, but the FI superpotential remains, leaving
\be
S = {1\over4}\int\d^2x\left(-i\int\d^2\th^+\sum_i \bar{\F}^ie^{2Q_i\cdot A}\nabla_-\F^i   + \int\d\th^+\, t^a \Upsilon_-^a \right)+c.c.\,.
\ee
By writing the FI superpotential as a standard superspace integral,
\be
\int\d^2x\d\th^+ \U_- = \int\d^2x\d\th^+ \bar D_+(\del_-A+iV_-) = \int d^2x\d^2\th^+(\del_-A+iV_-)
\ee
we can cast the low energy gauge theory action in the form
\bea
S &=& -{1\over2}\int\d^2x\d^2\th^+\, \left[\left({i\over2}\sum_i\bar{\F}^i e^{2Q_i\cdot A} \del_-\F^i - c.c. \right) \right.\\
&&\qquad\qquad\qquad\qquad -\left. {\th^a\over2\pi}\del_-A^a +  \left(\sum_i Q^a_i|\F^i|^2e^{2Q_i\cdot A} -r^a  \right)V_-^a\right]\non.
\eea
Even though the $\th$ term is a total derivative we will keep it anyways, since it can have important effects when $\S$ has a non-trivial topology. We see that $V_-$ acts as a Lagrange multiplier enforcing the superfield constraints
\be\label{eq:constraint}
\sum_i Q^a_i|\F^i|^2e^{2Q_i\cdot A} = r^a.
\ee
These are supersymmetric versions of the $D$-term constraints we found in components, and in fact $D^a=0$ is just the lowest component of~\C{eq:constraint}\ in a $\th^+$ expansion. These constraints should be regarded as a superfield equations that fix $A^a = A^a(|\F|^2)$, even though they might not be admit closed form solutions.\footnote{Note that solutions for $A^a$ may not always exist for all values of $r^a$. Throughout this work, we will assume $r^a$ are chosen such that solutions exist.}

Having eliminated the vector superfields what we are left with has the form of a $(0,2)$ non-linear sigma model action:
\be
S = -{i\over4}\int\d^2x\d^2\th^+\, \left(K_i(\F,\bar\F)\del_-\F^i - K_\ibar(\F,\bar\F)\del_-\bar\F^\ibar\right)
\ee
with the defining one-form
\be\label{eq:KfromGLSM1}
K_i = \bar\F_i e^{2Q_i\cdot A} +2i{\th^a\over2\pi}\del_i A^a.
\ee
We follow the convention $\F^i = \F_\ibar,\bar\F^\ibar = \bar\F_i$ for raising and lowering complex indices. Before deriving the low energy metric and $B$, it helps to first define the quantity
\be\label{eq:delta}
\left(\D^{ab}\right)^{-1} = 2\sum_i Q^a_i Q^b_i |\f^i|^2 e^{2Q_i \cdot A},
\ee
which is obtained by differentiating~\C{eq:constraint}\ with respect to $A$, while differentiating~\C{eq:constraint}\ with respect to $\f^i$ instead yields the useful relations
\be
\del_i A^a = -\bar{\f}_i\D^{ab}Q_i^a e^{2Q_i\cdot A}.
\ee
Equipped with these relations, we can now take derivatives of $K_i$ to compute the induced metric
\bea\label{eq:metric}
G_{i\jbar} = e^{2Q_i\cdot A}\left(\dd_{i\jbar}-2\bar{\f}_i\f_\jbar Q^a_i \D^{ab} Q_j^b e^{2Q_j\cdot A}\right),
\eea
and $B$-field
\be\label{eq:bfield}
B = 2i{\th^a\over2\pi}\del\bar{\del}A^a.
\ee
We note that the metric is Hermitian, as expected from $(0,2)$ supersymmetry. However the $B$-field is closed, so $H=dB=0$ in this class of examples with constant FI parameters. We will return to this point in the coming chapters.

\subsubsection{Local affine coordinates}
Strictly speaking, the metric~\C{eq:metric}\ has too many components, since $\M$ is only $(N-n)$ dimension. The problem is that $\f^i$ are charged fields, but $\M$ should be parameterized by a set of gauge-invariant coordinates.
To see how this works consider a simple class of models with GLSM gauge group $U(1)$. In a patch $U_{(\a)}$ where $\f^\a\neq0$, we can define gauge-invariant coordinates
\be\label{eq:coords}
Z^i_{(\a)} = \left(\f^i\right)\left(\f^\a\right)^{-Q_i/Q_\a}.
\ee
On the intersection $U_{(\a)}\cap U_{(\b)}$, the coordinates then transform as follows:
\be\label{eq:coordtrans}
Z^i_{(\a)} = Z^i_{(\b)} \left(Z^\a_{(\b)}\right)^{-Q_i/Q_\a}.
\ee
However, note that the right hand side of~\C{constraint}\ is invariant.  It follows that
\be
A_{(\a)} = A_{(\b)} + {1\over Q_\a} \log\left|Z^\a_{(\b)}\right|,
\ee
so $A$ is not globally defined. In a patch $U_{(\a)}$ the local expression for the metric depends only on the $N-1$ gauge-invariant variables $Z^i_\al$ ($i\neq\a$) and it is given by
\be
ds^2 = \sum_{i\neq\a} e^{2Q_i A_\al}|dZ^i_\al|^2  - 2\D \left|\sum_{i\neq\a} Q_ie^{2Q_iA_\al}\bar{Z}^i_\al dZ^i_\al \right|,
\ee
where now
\be
\D = \hlf\left( Q_\a^2e^{2Q_\a A_\al} + \sum_{i\neq\a} Q_i^2|Z^i_\al|^2e^{2Q_i A_\al}\right)^{-1}.
\ee

We have seen that $A^a$ are not globally defined functions. In fact, $A^a$ are non-trivial sections of the set of line bundles $\L^a$ over $\M$ defined by the transition functions~\C{eq:coordtrans}. Note that $\del A^a$ transform like connections on $\L^a$, while $i\del\bar{\del}A^a$ are their (globally defined) curvature two-forms. The set of two-forms $F^a = i\del\bar\del A^a$ form a basis for $H^{2}(\M,\Z)$, and in particular
\be
B \simeq \th^a F^a \in H^{1,1}(\M)
\ee
is globally defined and closed.

\subsubsection{Example: $\PP^N$}
To get our bearings with all of the formalism of this section, it helps to keep a simple example in mind. Take a single $U(1)$ gauge group, with $N+1$ chiral superfields $\F^i$, $i=0,1,\ldots,N$, all of charge $Q=1$. Then, the target manifold
\be
\M = \left\{\f\in\CC^{N+1}\Big| \sum_i|\f^i|^2=r\right\}/U(1) \simeq S^{2N+1}/S^1 \simeq \PP^N
\ee
is $N$-dimensional complex projective space, sometimes also denoted $\CC\PP^N$. Note that $r$ controls the size of the $\PP^N$ target space.\footnote{We assume $r>0$ so $\M$ is non-empty.} An alternate way to think of $\PP^N$ as the identification of points $z^i\in\CC^{N+1}$ under the equivalence relation
\be
(z^0,z^1,\ldots,z^N) \sim (\l z^0,\l z^1,\ldots,\l z^N),\qquad \forall \l\in\CC\backslash\{0\}.
\ee
The equivalence of these two descriptions is apparent if we think of the modulus of $\l$ as fixing the $D$-term constraint, and the phase of $\l$ implementing the $U(1)$ quotient.

In this case, the supersymmetric constraint is easily solved for $A$:
\be
|\Phi^i|^2 e^{2A} = r\quad\Rightarrow\quad A = \hlf\log\left(r\over|\Phi^i|^2\right),
\ee
and so $\D={1\over2r}$. $K_i$ takes a particularly simple form:
\be
K_i = \left(r-i{\th\over2\pi}\right){\bar\F_i\over|\F|^2},
\ee
which can be written as a total derivative,  therefore $\PP^N$ must be a K\"ahler manifold.

Working in a patch where $\phi^0\neq0$, we form gauge-invariant fields $Z^i=\phi^i/\phi^0$. The local metric induced on $\M$ has a very special form,
\be
G_{i\jbar} = r\left({\dd_{i\jbar}\over(1+|Z|^2)} - {\bar Z_i Z_\jbar \over(1+|Z|^2)^2}\right),
\ee
where the bracketed portion is known as the Fubini-Study metric on $\PP^N$. We see that $r$ sets the overall size of the $\PP^N$. The components of $B$ are very similar, and in fact they combine nicely with the fundamental form $J$ into a natural complex two-form:
\be
B+iJ = \left({\th\over2\pi}+ir\right)J_{FS},
\ee
where $J_{FS}=-2\del\bar\del A$ is the fundamental forma associated with the Fubini-Study metric. Clearly, $dB=dJ=0$ which is as we expect for a K\"ahler manifold.

\subsection{Renormalization and conformal invariance}\label{ss:RGandCFT}

So far we have seen that the gauge couplings $e_a$ have positive mass dimension, and so GLSMs are all weakly coupled at high energies but they become strongly coupled at energy scales much below $e_a$.
In the UV, we have a weakly coupled description in terms of charged scalar fields $\f^i$ coupled to gauge fields $A_\m^a$. In the deep IR, the strongly coupled gauge theory description is most conveniently replaced by a non-linear sigma model for gauge-invariant composite fields $Z^i$  on a target space $\M$.

Now we wish to understand the stability of these non-linear models under the renormalization group flow. In particular, we are interested in finding the conditions under which the sigma model will flow to a (non-trivial) conformal fixed point. As discussed in Section~\ref{sec:stringsol}, conformally invariant $(0,2)$ sigma models are required when building supersymmetric solutions of string theory. We will study this question both in the weakly coupled UV gauge theory, and also in terms of the IR sigma model data. Not surprisingly, both approaches will yield the same condition for conformal invariance.

\subsubsection{Conformal invariance in the GLSM}
To understand the renormalization of the UV gauge theory, we can compute the quantum effective action.  At sufficiently high energies, so that the gauge theory is very weakly coupled, it suffices to integrate out the high energy modes of all the fields to only one-loop order. However, because these theories are super-renormalizable ($e_a$ has positive mass dimension), it turns out there is a unique divergent one-loop graph given by the one-point function of $D^a$ \cite{Witten:1993yc}.
\begin{figure}[h]
\centering
\includegraphics[scale=0.5]{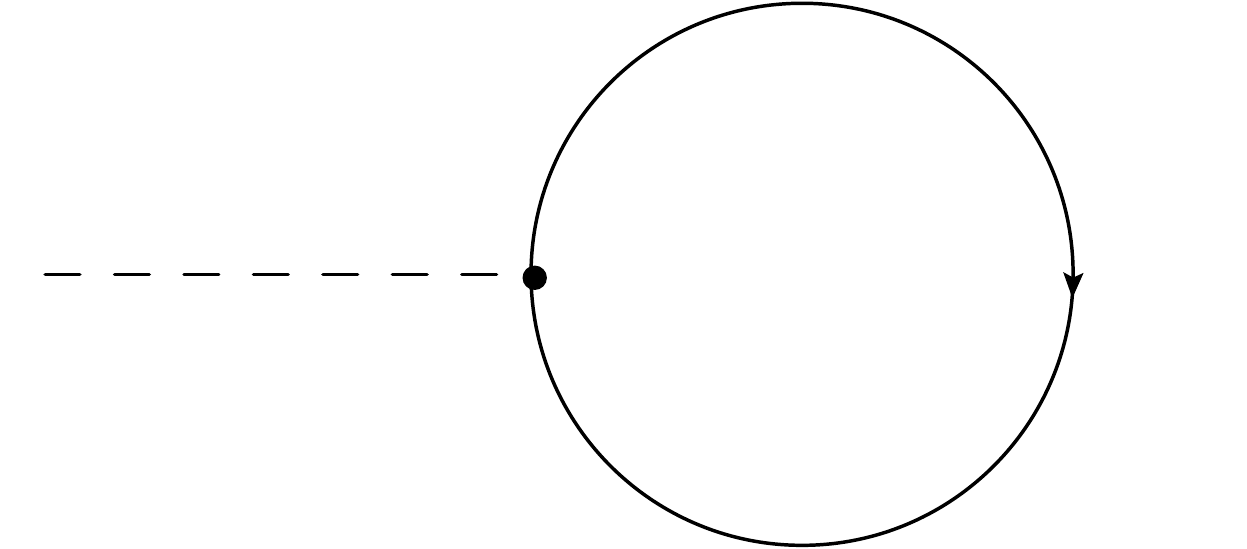}
\vskip 0.2 in \caption{\it The only divergent graph in a GLSM. } \label{fig:D}
\end{figure}
Summing over all internal $\f^i$ fields, and integrating from a UV cutoff scale $\La$ down to some scale $\m$,\footnote{We assume $\m\gg e$ so that the gauge theory remains weakly coupled.} we find the divergent loop integral
\be\label{eq:D1}
\la D^a\ra = \sum_i Q^a_i \underset{\m\leq|k|\leq\La}{\int}{\d^2k\over(2\pi)^2} {1\over k^2} = {1\over2\pi}\left(\sum_i Q^a_i\right) \log\left(\m\over\La\right).
\ee
In the classical FI action~\C{eq:LFI}, $r^a$ is the coefficient of the one-point function for $D^a$, and so we can interpret this term in the effective action as a renormalization of $r^a$:
\be
r^a \rightarrow r^a(\m) = r^a + {1\over2\pi}\left(\sum_i Q^a_i\right) \log\left(\m\over\La\right).
\ee
As we lower the scale $\m$, $r^a(\m)$ runs logarithmically in the direction set by the sum of the charges. In particular, in order to have a conformally invariant theory, which does not change with $\m$, we must impose
\be\label{eq:sumQ}
\sum_i Q_i^a =0 ,\quad \forall a.
\ee
Note that if $\sum_i Q_i^a>0$, as in the $\PP^N$ example, $r^a(\m)$ decreases as we lower $\m$. Since $r^a$ control the size of $\M$, we conclude that when $\sum_i Q^a_i>0$ the target space $\M$ shrinks as we flow to the IR.
On the other hand, when $\sum_i Q^a_i<0$ it follows that $\M$ is expanding as we flow to lower energies.

\subsubsection{Conformal invariance in the NLSM}
Now let us consider this same question from the point of view of the low energy sigma model description. Recall from Section~\ref{ss:classicalgeometry}\ that GLSMs with constant FI parameters always lead to K\"ahler target manifolds $\M$, with $H=0$. In the beginning of Section~\ref{sec:stringsol} we noted that, to lowest order in $\a'$, conformally invariant K\"ahler models satisfy
\be\label{eq:monge}
{\cal R} \equiv \del\bar\del\log\det\left( G\right) = 0,
\ee
where ${\cal R}$ is the curvature two-form associated with the metric $G$. Manifolds with Ricci-flat K\"ahler metrics are called Calabi-Yau, after E.~Calabi who conjectured the criteria for their existence and S.~T.~Yau who completed the proof. What Calabi conjectured~\cite{Calabi:1954}, and Yau later showed~\cite{Yau:1978}, is that if a (compact) K\"ahler manifold $\M$ with metric $G$ has a curvature two-form ${\cal R}$ that represents a trivial class in cohomology:
\be\label{eq:R=0}
[{\cal R}] =0,
\ee
then there is a unique metric $G'$ in the same K\"ahler class as $G$, that is to say $[J']=[J]$, such that $G'$ is Ricci-flat.
The low energy metric~\C{eq:metric}\ induced from the GLSM,
\be\label{eq:G}
G_{i\jbar} = e^{2Q_i\cdot A}\left(\dd_{i\jbar}-2\bar{\f}_i\f_\jbar Q^a_i \D^{ab} Q_j^b e^{2Q_j\cdot A}\right),
\ee
is almost never Calabi-Yau, but we should not expect it to be since this metric will receive quantum corrections as we flow to the IR. In fact, the condition for conformality~\C{eq:monge}\ will also receive quantum corrections, so we are not really interested in exactly Calabi-Yau solutions anyways. However, it was shown in~\cite{Nemeschansky:1986yx}\ that the metric can be corrected, order by order in perturbation theory, to a solution of the full beta-function equations provided~\C{eq:R=0}\ is satisfied.
Therefore, sigma model metrics for which
\be
\del\bar\del\log\det(G)
\ee
is trivial in cohomology should flow to conformal solutions in the IR.

The metric~\C{eq:G}\ is written in terms of projective coordinates, for which $\det G_{i\jbar}$ vanishes. Therefore, we must work in a local patch using gauge-invariant coordinates. First we must generalize the local coordinates~\C{eq:coords}\ suitably for higher rank gauge groups.\footnote{We found a similar discussion in \cite{Adams:2006kb} useful for these definitions.} Each patch must now be labeled by a multi-index $\cA\subset\{1,2,\ldots,n\}$ with $|\cA|=r$. For $\a\in\cA$ we require that the $r\times r$ matrix $Q^a_\a$ be invertible. We then define the patch $U_{(\cA)}=\left\{\f^i\in\CC^n\big|\f^\a\neq0,\forall \a\in\cA\right\}$. Within that patch, we can define the coordinates:
\be
Z^i_{(\cA)} = \f^i \prod_{\a\in\cA} \left(\f^\a\right)^{-(Q^{-1})^\a_a Q^a_i}.
\ee
The transformation properties on intersections $U_{(\cA)}\cap U_{(\B)}$ for the coordinates $Z_{(\cA)}$ and the sections $A_{(\cA)}$ are easy enough to work out,
\bea
Z^i_{(\cA)} &=& Z^i_{(\B)}\prod_{\a\in\cA}\left(Z^\a_{(\B)}\right)^{-(Q^{-1})^\b_a Q^a_i}, \\
A^a_{(\cA)} &=& A^a_{(\B)} + \sum_{\a\in\cA}(Q^{-1})^\a_a \log\left|Z^\b_{(\cA)}\right|.\label{eq:Avar}
\eea
A straightforward but somewhat involved calculation reveals that,
\be
\log\det G_{i\jbar} = 2\left(\sum_{i}Q^a_i\right)A_{(\cA)}^a + 2\log\det Q^a_\a + \log\det\left(2\D^{ab} \right).
\ee
The functions $\D^{ab}$ are globally defined, but as we see from~\C{eq:Avar}, $A^a$ shifts when going between patches.  The only way to ensure that $\del\delbar\log\det G_{i\jbar}$ is trivial in cohomology is to impose $\sum_i Q^a_i=0$ for each $U(1)$ factor. Thus, we recover the same conformal condition,~\C{eq:sumQ}, that we found in the UV gauge theory description.

\chapter{GLSMs with $H$-flux}\label{ch:Hflux}
\section{Field-dependent FI couplings}\label{sec:FIs}
In Section~\ref{ss:classicalgeometry}\ we saw that the low energy limit of a GLSM with constant complexified FI parameters,
\be\label{eq:t2}
t^a=ir^a + {\th^a\over2\pi},
\ee
is a non-linear sigma model with a K\"ahler target manifold $\M$, with a closed $B$-field:
\be
B \simeq \th^a F^a \in H^{1,1}(\M),
\ee
where $F^a=i\del\bar\del A^a$ form a basis for $H^2(\M,\Z)$. In particular, since $B$ is always closed the flux $H=dB$ vanishes.

How then do we engineer a GLSM whose vacuum manifold is a non-K\"ahler with $H$-flux? The simplest thing to do is to include more general superpotential couplings for $\U_-$, such as
\be\label{eq:FI}
S_{FI} = {1\over4}\int\d^2x\d\th^+\, T^a(\F)\U_-^a\Big|_{\thbar^+=0} +c.c.,
\ee
where $T^a$ no longer need to be constant, as in~\C{eq:t2}. If we let $\Th^a = \Re\!T^a$, then we should expect these FI couplings will lead to
\be
B \simeq \Th^a(\f) F^a,
\ee
which is  not closed, and in particular the $H$-flux no longer needs to vanish:
\be
H = d\Th^a\wedge F^a \neq0.
\ee
We will see that this intuitive picture is actually born out in several examples. At this point, we should ask what constraints we must place on $T^a$. It is clear that $T^a$ must be holomorphic in $\F$  in order for~\C{eq:FI}\ to be supersymmetric.
Also, gauge invariance would seem to imply that $T^a(\F)$ should be invariant under the $U(1)$ action.  Models of this type are always non-compact and discussed in Section~\ref{sec:gaugeinvariant}.

Surprisingly, the requirement that $T^a$ must be gauge-invariant is actually too strong! It is possible to add certain non-invariant FI couplings to the $\U_-$ superpotential and still have a sensible quantum gauge theory.
To motivate the appearance of these gauge variant couplings, it helps to draw an analogy with familiar facts from four-dimensional $\N=1$ gauge theory; see, for example,~\cite{Intriligator:1995au}\ a review of this subject.

\subsubsection{Analogy with $\N=1$ super-Yang-Mills}
In four-dimensional supersymmetric gauge theories, the topological $\theta$-angle is paired with the gauge coupling in the combination
\be
\tau = {4\pi i \over g^2} + { \theta \over 2 \pi}.
\ee
The gauge kinetic terms take the form
\be
{1\over 16 \pi}{\rm Im} \left\{  \int  d^2\theta \, \tau \, \Tr \left( W_\a W^\a \right) \right\} = - {1\over 4 g^2}  \Tr (F_{\mu\nu})^2 + {\theta \over 32 \pi^2} \Tr (F\wedge F) + \ldots.
\ee
The quantum renormalization of $\tau$ is highly constrained. Expressed in terms of a complexified strong coupling scale $\Lambda = |\Lambda| e^{i\theta/b}$, $\tau$ takes the schematic form
\be
\tau(\mu) = {b\over 2\pi i} \log\left({\Lambda\over \mu}\right) + f(\Lambda^b, \Phi),
\ee
where $b$ is determined by the one-loop beta-function, and the function $f$ is a single-valued function of chiral fields, collectively denoted $\Phi$. This form for $\tau$ respects holomorphy, and the symmetry $\Lambda^b \rightarrow e^{2\pi i} \Lambda^b$ with $\tau \rightarrow \tau+1$. Note that we must introduce a scale to define the logarithm in four dimensions. Usually the logarithm is generated by integrating out physics at a higher scale.

In two-dimensional $(0,2)$ theories the analog of the superpotential structure $W^\a W_\a$ is the fermionic field strength $\Upsilon_-$, while the analog of the holomorphic coupling $\t(\m)$ is the chiral FI coupling $T(\F)$. For simplicity, let us restrict to a $U(1)$ gauge theory and consider the coupling
\be
{1 \over 4} \int d\th^+\, T(\Phi) \Upsilon_-\Big|_{\thbar^+0} + {\rm c.c.} =   {\rm Re}(T) F_{01} - {\rm Im}(T)D +\ldots.
\ee
The natural periodic $\theta$-angle, given by $ {\rm Re}(T)$, is now paired with the $D$-term which determines the vacuum structure. The analogy with $\t(\m)$ suggests we consider
\be\label{generalf}
T(\Phi) = \sum_i {N_i\over2\pi i} \log(\Phi^i) + T_0(\Phi),
\ee
where $N_i$ are integers and $T_0$ is single-valued. Unlike in four dimensions, we do not need to introduce a scale to define the logarithm since two-dimensional scalar fields are dimensionless. Models with logarithmic FI couplings will be the focus of Section~\ref{sec:compact}, as well as everything that follows it. Related models appeared at the same time as the author's work in~\cite{Blaszczyk:2011ib}.

\subsubsection{Log superpotentials?}
A logarithmic coupling appears problematic in the fundamental theory for two reasons: first, the theory is no longer gauge-invariant if any $\Phi^i$ are charged. However, the violation of gauge invariance involves a shift proportional to $\U_-$, which is precisely of the type that can be canceled by a one-loop gauge anomaly. A similar cancelation between classical and quantum gauge variations appeared in~\cite{Adams:2006kb}. Further details of this cancelation will be presented in Section~\ref{ss:condition}.

The second issue is defining the log at the quantum level. This looks problematic if the moduli space of the theory can access loci where singularities occur.  Fortunately, the $D$-term constraints are now also modified. Consider a model where the fields $\Phi^i$ have charges $Q^i_a$ under each $U(1)$ gauge factor labeled by $a$. The usual symplectic reduction involves solving $D$-term constraints
\be
\sum_i Q_i^a |\phi^i|^2 = r^a,
\ee
and then quotienting by the abelian symmetry group. The log modifies these constraints as follows:
\be\label{manydterms}
\sum_i Q_i^a |\phi^i|^2 - {N_i^a\over2\pi} \log |\phi^i| = r^a.
\ee
For suitable choices of $Q_i^a$ and $N_i^a$, the singular locus of the log can be removed. This is a generalization of  symplectic reduction. After quotienting by the abelian group action, the resulting space is expected to be complex and non-K\"ahler. The modifications to the $D$-term constraints, and the resulting moduli spaces will be explored further in Sections~\ref{ss:firstlook}\ and~\ref{ss:modulispace}.

We expect these spaces to be topologically distinct from Calabi-Yau spaces as was the case for the metrics found in~\cite{Dasgupta:1999ss}.
Our construction also gives a natural class of supersymmetric gauge bundles over non-K\"ahler manifolds which we will not explore in detail here. Clearly, there are many interesting questions to study.
Based on intuition from type II flux vacua, it does seem likely that this class of string vacua will be significantly larger than the currently known heterotic string compactifications.

\section{Gauge-invariant FI couplings}\label{sec:gaugeinvariant}
As we discussed in Section~\ref{sec:FIs}, the simplest way to include torsion in a GLSM is to make the FI terms field-dependent. In this section, let us add the couplings
\be\label{eq:f}
-{i\over4}\int\d^2x\d\th^+\, f^a(\F)\Upsilon_-^a + c.c.
\ee
and restrict our attention to gauge-invariant $f^a$. The case of gauge non-invariant $f^a$ needed for compact models will be considered in Section~\ref{sec:compact}.
Since the $f^a$ are required to be gauge-invariant, this forces us to introduce fields with negative charges.\footnote{One might also consider rational functions which are gauge-invariant. This will generically introduce singularities but it might be possible to excise the singular loci with a suitable superpotential. We will restrict to globally defined $f^a$ in this section. } This means these models will always be non-compact in the absence of superpotential couplings.

Including these generalized FI terms only modifies the analysis of Section~\ref{ss:classicalgeometry}\ by replacing
\bea
r^a &\rightarrow& R^a(\F) = r^a - \Re(f^a), \\
{\th^a\over2\pi} &\rightarrow& \Th^a(\F) = {\th^a\over2\pi} + \Im(f),
\eea
and
\be
t^a \rightarrow T^a(\F) = \Th^a(\F) +iR^a(\F) .
\ee
In particular, at energies small compared to the scale of the gauge couplings $e_a^2$ the vector superfields $V_-^a$ simply enforce the constraints
\be \label{eq:constraint1}
\sum_i Q^a_i|\F^i|^2 e^{2Q_i\cdot A}   = R^a(\F),
\ee
which we use to solve for $A=A(|\F|^2)$. Differentiating~\C{eq:constraint1}\ with respect to $A$ still leads to the quantity
\be
(\D^{ab})^{-1} = {\del R^a\over\del A^b} = 2\sum_i Q_i^a Q^b_i |\F^i|^2 e^{2Q_i\cdot A},
\ee
while differentiating with respect to $\Phi$ now gives the modified relation
\be
\del_i A^a = \D^{ab}\left(\del_i R^b - Q_i^b\bar\F_i e^{2Q_i\cdot A}\right).
\ee
The low-energy action for the gauge theory with field-dependent FI couplings is again a non-linear sigma model for $\Phi^i$, but now specified by
\be
K_i = \bar{\F}^i e^{2Q_i\cdot A} + 2i\Th^a \del_i A^a.
\ee
This leads to the following low energy metric
\bea\label{eq:metric1}
G_{i\jbar} = e^{2Q_i\cdot A}\dd_{i\jbar} - 2\del_i A^a(\D^{ab})^{-1}\del_\jbar A^b -i\del_iT^a\del_\jbar A^a + i\del_i A^a \del_\jbar \bar T^a
\eea
and $B$-field
\be\label{eq:B1}
B= 2i\Th^a \del\bar\del A^a = (f^a-\bar f^a)\,\del\bar\del A^a.
\ee
As a check, note that these expressions reduce to those of Section~\ref{ss:classicalgeometry}\ when $r$ and $\th$ are simply constants. Furthermore, unlike the cases where $\th$ is constant, these spaces contain non-zero $H$-flux:
\be\label{eq:H1}
H = dB = 2id\Th^a\wedge\del\bar\del A^a,
\ee
which because of $(0,2)$ supersymmetry is automatically related to the fundamental form
\be
J = {i\over2}(\delbar K - \del K^*) = -2i R^a \del\bar\del A^a
\ee
in the desired manner
\be
H= i(\bar\del-\del)J.
\ee

$\Th$ is a gauge-invariant quantity, and so it is globally defined on $\M$. So, while $H$ is non-zero it is still trivial in cohomology. This is obvious from the fact that we can continuously vary the FI couplings $f^a(\F)$, and hence $H$ cannot be quantized. This is not terribly surprising, since the quantization condition on $H$ is really only needed for compact target spaces, but the gauge invariance of $f$ requires the presence of negatively charged fields, leading to non-compact solutions to the $D$-term constraint.

\subsection{A special case corresponding to UV $B$-fields}\label{ss:UVBfield}

The case of quadratic $f^a$ is particularly interesting. In this case, we can rewrite the superpotential coupling~\C{eq:f}\ as a $D$-term that preserves linearity of the theory. Since $f^a$ is quadratic, we require pairs of fields with equal and opposite charge, $\F^i$, $\F^j$ where $Q^a_i=-Q^a_j$. Notice that we can now write $f_{ij}^a = Q_j^a b_{ij}$ for some \textit{anti-symmetric} $b_{ij}.$\footnote{While $f_{ij}^a$ is symmetric in $i,j$, $b_{ij}$ must be anti-symmetric because $Q_i^a=-Q_j^a$.} We now see that
\bea
\int\d^2x\d\th^+\, (f^a_{ij} \F^i\F^j)\U_-^a &=& \int\d^2x\d\th^+\, (Q^a_jb_{ij} \F^i\F^j)\U_-^a, \non\\
&=& \int\d^2x\d\th^+\, \Dbar_+ \big(b_{ij}\F^i\nabla_-\F^j\big), \\
&=& \int\d^2x\d^2\th^+\, b_{ij}\F^i\nabla_-\F^j.  \non
\eea
Only for this case of quadratic $f^a$ can we equivalently write these generalized FI couplings as a choice of UV $B$-field coupling,
\be \label{b}
S = {i\over4}\int\d^2x\d^2\th^+\left(b_{ij} \Phi^i\nabla_-\Phi^j - b_{\ibar \jbar} \bar{\Phi}^i\nabla_-\bar{\Phi}^j \right)
\ee
with $b_{ij} = -b_{ji} = b_{\ibar\jbar}^*$. In fact,~\C{b}\ is the most general non-trivial linear deformation of $K_i$ consistent with gauge invariance.\footnote{The other possibility, $K_i = b_{i\jbar}\Phi^\jbar$,  contributes a total derivative.} We should also point out that this coupling cannot appear in a $(2,2)$ theory constructed from chiral superfields, since would have to come from a holomorphic term in the K\"ahler potential, which can always be removed by a K\"ahler transformation. Usually a $B$-field in closed string theory with trivial target space and a flat metric has no effect on the physics. Indeed~\C{b}\ is trivial for neutral fields since a holomorphic deformation of $K_i$ does not alter the physical couplings. Only the presence of the (real) gauge field $V_-$ makes this coupling non-holomorphic and relevant for the low-energy physics.

We suspect this form for the field-dependent FI parameters might be useful for implementing worldsheet duality along the lines of~\cite{Adams:2003zy}.

\subsection{An example: the conifold with torsion}
\label{ss:conifold}

Let us use the conifold as a nice non-compact example. Take a single $U(1)$ gauge group coupled to two chiral fields $X^i$  $(i=1,2)$ with charge $Q_i=+1$, and two fields $Y^m$ $(m=1,2)$ of charge $Q_m=-1$.\footnote{We will ignore the fermionic sector for now, though appropriately charged left-moving fermions should be included to cancel the gauge anomaly.} In the absence of any $f(X,Y)$ coupling, the $D$-term condition is
\be
|X|^2 - |Y|^2  = r.
\ee
The target space of this GLSM is the total space of the vector bundle $\cO(-1)\oplus\cO(-1)$ over $\PP^1$. The size of the $\PP^1$ base is controlled by $r$. In the limit $r\rightarrow0$, the space develops a conifold singularity, while finite $r$ corresponds to a resolved conifold.

Let us restrict to a quadratic $f= f_{im}X^i Y^m$. In this example, the superfield constraint~\C{constraint}\  becomes
\be\label{Aconstraint}
e^{2A}|X|^2 - e^{-2A}|Y|^2 = R(X,Y) = r- Re(f_{im}X^i Y^m),
\ee
whose solution is given by
\bea\label{eq:Asol}
e^{2A} ={R+\sqrt{R^2+4|X|^2|Y|^2}\over2|X|^2} = \left({\sqrt{R^2+4|X|^2|Y|^2}-R\over2|Y|^2}\right)^{-1}.
\eea
Plugging this expression for $A$ into~\C{eq:metric1}\ for the target space metric gives
\bea
G_{i\jbar} &=& e^{2A}\dd_{i\jbar} - { e^{4A} \bar x_i x_{\jbar} - (f_{im}y^m)(\bar f_{\jbar\bar n}\bar y^n)\over \sqrt{R^2+4|x|^2|y|^2}}, \non\\
G_{i\mbar} &=& {\bar x_i y_\mbar - (f_{in}y^n)(\bar f_{\bar m\jbar}\bar x^\jbar)\over \sqrt{R^2+4|x|^2|y|^2}}, \\
G_{m\nbar} &=& e^{-2A}\dd_{m\nbar} - { e^{-4A} \bar y_m y_{\nbar} - (f_{mi}x^i)(\bar f_{\bar n\jbar}\bar x^\jbar)\over \sqrt{R^2+4|x|^2|y|^2}} \non,
\eea
when written in projective coordinates. The flux is most conveniently just written in the form~\C{eq:H1}:
\be
H = 2id\Th\wedge\del\bar\del A,
\ee
with $A$ given by~\C{eq:Asol}. The important point is that $d\Th$ is a trivial one-from, and so $H$ is trivial in cohomology in this example (or any example with gauge-invariant $f$). It would be interesting to see if the renormalization group flow preserves these deformations of the conifold solution, or if these perturbations are irrelevant and the solution relaxes back to the flux-free solution.

\section{Anomalous FI couplings}
\label{sec:compact}

\subsection{The condition for anomaly cancelation}\label{ss:condition}

To construct compact models, we are interested in couplings that we can add to the classical action which are \textit{not}\ gauge-invariant. The classical violation of gauge invariance must be of a form that matches the quantum one-loop gauge anomaly. The sign of the anomaly is rather important for us, so we have presented a detailed derivation of the anomaly in Section~\ref{sec:axanom}. The anomaly shifts the action by
\be\label{eq:minimalgauge}
\dd S = {\cA^{ab} \over {4\pi}}\int\d^2x  \La^aF_{01}^b
\ee
where $\Lambda^a$ is the gauge parameter,  and
\be
\cA^{ab} = \sum_i Q_i^a Q_i^b - \sum_\a Q_\a^a Q_\b^b
\ee
is the anomaly coefficient with  charges $Q_i$ for right-movers and  charges $Q_\a$ for left-movers. In superspace, this reads
\be
\dd S = \left( {\cA^{ab} \over {16\pi}}\int\d^2x \d\th^+\, \La^a \U_-^b + c.c. \right).
\ee
Note that a background NS5-brane can be viewed as a small instanton in the gauge bundle and so would shift the action like a left-mover. An anti-NS5-brane would  induce a shift with opposite sign. The sign of the anomaly  determines whether a positive or negative coefficient of the log corresponds to NS5-brane or anti-NS5-brane flux which is why the sign is of importance for us.

There are basically two classical couplings that we can consider. The first is the log-type FI coupling
\be\label{logfi}
S_1 = {i \over 8\pi}\int\d^2x \d\th^+\, N_i^{a} \log\left( \F^i \right)   \U_-^a + c.c.
\ee
for some choice of  $N_i^{a}$. The simplest assumption is to take $N^i_a\in \Z$. This ensures invariance under the global transformation $\Phi^i \rightarrow e^{2\pi i} \Phi^i$ in any topologically non-trivial instanton sector. However, this appears to be too strong a condition. To cancel the basic the minimal gauge anomaly for a charge one left or right-mover given in~\C{eq:minimalgauge}, we actually need to allow half-integer $N^i_a$.

How this weaker condition is consistent in odd charge instanton sectors is a fascinating question; we will not pursue this question here, beyond commenting that perhaps an odd number of fermion zero modes in those sectors kills the path-integral rendering the theory consistent.   It will also be very interesting to see if the instanton analysis leading to the usual quantization condition on the $N^i_a$ is modified by the dynamical theta angles which, in turn, could relax the half-integrality condition further.
 We will see that the quantization of $N_i^a$ leads to a quantized $H$-flux unlike the models of Section~\ref{sec:gaugeinvariant}. Under a gauge transformation, this term will shift the action by the following amount
\be
\dd S_1 = -\left( {N_{i}^{a} Q_i^b \over {8\pi}}\int\d^2x \d\th^+\, \La^b \U_-^a + {c.c.} \right).
\ee
Notice that only the symmetric part of $N_i^{a} Q_i^b$ can be canceled by the anomaly since $\cA^{ab}$ is manifestly symmetric.

One might imagine replacing the monomial argument of the log in~\C{logfi}\ with a more complicated function with definite charge under the gauge symmetries like a polynomial. The difficulty with such a choice is ensuring invariance of the theory under $\Phi^i \rightarrow e^{2\pi i} \Phi^i$ for each $i$ separately. It would be very interesting if cases generalizing the monomial (or product of monomials) could be made sensible.

To produce an antisymmetric shift, consider the following term
\be\label{s2}
S_2 = {1\over4\pi} \int\d^2x\d^2\th^+\, C^{ab} A^a V_-^b
\ee
where $C^{ab}$ is an antisymmetric tensor to be determined. The $(2,2)$ extension of this coupling interestingly appeared in~\cite{Morrison:1995yh}. Under a gauge transformation,
\bea
\dd S_2 &=& {1\over4\pi} C^{ab}\int\d^2x\d^2\th^+\left({i\over2}(\bar{\La}^a - \La^a)V_-^b - \hlf A^a\del_-(\La^b+\bar{\La}^b) + {i\over4}\left(\La^a-\bar{\La}^a)\del_-(\La^b+\bar{\La}^b) \right)\right) \non\\
&=& -{1\over4\pi} C^{ab}\int\d^2x\d^2\th^+\left(\hlf\La^a(\del_-A^b + iV_-^b) + \hlf \bar{\La}^a (\del_-A^b - iV_-^b) \right) \non \\
&=& \left(- {1\over8\pi} C^{ab}\int\d^2x\d\th^+\, \La^a\U_-^b + c.c. \right).
\eea
Note that the terms quadratic in $\La^a$ either cancel after integration by parts or are purely (anti)-holomorphic and so only contribute a total derivative. Comparing $\dd S_1$ and $\dd S_2$ we see that $T$ must be chosen so that
\be
C^{ab} = \sum_i N_i^{[a} Q_{i}^{b]}.
\ee
Together the classically anomalous terms in the action take the form
\be
S_{anom} = {1\over4\pi}\int\d^2x\left[\d^2\th^+ N_i^{[a} Q_{i}^{b]} A^a V_-^b + \left({i\over2} N_i^{a} \int\d\th^+\, \log (\F^i) \U_-^a +c.c.\right)\right].
\ee
Under a gauge transformation,
\bea
\dd S_{anom} &=& -{Q_i^aN_i^{b}+N_i^{[a} Q_{i}^{b]}\over8\pi}\int\d^2x\d\th^+\, \La^a\U_-^b + c.c. \\
&=&-{Q_i^{(a}N_{i}^{b)}\over8\pi}\int\d^2x\d\th^+\, \La^a\U_-^b + c.c. \non,
\eea
so the requirement of a consistent theory is
\be
\sum_i Q_i^{(a}N_{ i}^{b)} ={1\over 2}\cA^{ab}.
\ee
So far, our discussion is largely focused on the classical physics of these models along with the quantum condition for gauge invariance. Standard $(0,2)$ theories are perturbatively conformal if the $\sum_i Q_i^a=0$ for each $a$. Since we are modifying a superpotential coupling, albeit with a log, we suspect that this condition is unchanged as long as the theory has a moduli space that excludes singularities of the log couplings. We will see later that there are many choices of $N^{a}_i$ for which this is the case.

If one is uncomfortable with the log interaction, it can be replaced by more familiar couplings as follows:\footnote{We would like to thank Allan Adams for suggesting this replacement.} for each $\F^i$, introduce an axially gauged field $Y^i$ transforming in the following way under a gauge transformation
\be
Y^i \rightarrow Y^i + i Q_i^a \La^a.
\ee
Now consider the couplings
\be \label{defY}
S_Y = {i \over 8\pi}\int\d^2x \d\th^+\,  \left( N_i^{a} Y^i   \U_-^a +  \G_{Y^i} \left\{ e^{Y^i} - \Phi^i \right\} \right) +  c.c.
\ee
where $\G_{Y^i}$ are standard chiral Fermi superfields. Solving the superpotential constraint from $\G_{Y^i}$ sets $e^{Y^i} = \Phi^i$. This form again suggests that the renormalization of the theory should not be problematic as long as singular loci are excluded from the moduli space. It is worth noting that the metric expressed in terms of $Y$-fields is not flat. One could also consider a flat metric for the $Y$-fields which leads to models of the type studied in~\cite{Adams:2006kb}.

\subsection{Supersymmetry anomaly}\label{ss:susyanomaly1}

Working in components in WZ gauge, it appears that introducing log interactions that break gauge invariance also leads to a classical breaking of $(0,2)$ supersymmetry. This is surprising since the action expressed in superspace appears supersymmetric. Indeed the theory is supersymmetric if we choose not to fix Wess-Zumino gauge and consider a theory with extra degrees of freedom in the vector multiplet which would usually decouple with this gauge choice. However, choosing Wess-Zumino gauge is not compatible with preserving supersymmetry. Rather a compensating gauge transformation must accompany a supersymmetry transformation in order to preserve this gauge choice. This is the basic source of the supersymmetry anomaly. It is tied directly to the gauge anomaly.

In terms of standard physical fields, we can see this directly from the action as follows: imagine a single charged scalar $\Phi$ with charge $Q$ and the superpotential coupling
\be\label{susyissue}
\int\d\th^+\, \log (\F) \U_-  = 2i (D-iF) \log (\f) - 2\sqrt{2} {\psi_+ \lambda_- \over \phi} .
\ee
The problematic non-cancelation comes from the variation
\be
\delta \psi_+ = \sqrt{2} i  {\bar \e} \, \cD_+\f.
\ee
If $\phi$ were neutral then $\cD_+ \rightarrow \p_+$ and the variation of the second term in~\C{susyissue}\ would cancel against the variation of the first term up to a total derivative.  This is no longer the case when $\phi$ is charged and we pick up a term proportional to $A_+$. In the general case, we find a non-vanishing term
\be\label{anomvariation}
\dd S_1  =  {i\over 2\pi} N^a_i Q^b_i A^b_+ {\bar \e} \lambda^a_- + {c.c.},
\ee
which is exactly the way $S_1$ should transform under a superspace gauge transformation with chiral superfield gauge parameter
\be
\Lambda^a = 2i \th^+ {\bar \e} A_+^a.
\ee
This is exactly the gauge transformation needed to restore Wess-Zumino gauge. Note that $S_2$ given in~\C{s2}\ is also not supersymmetric for the same reason and transforms in a way that precisely cancels the antisymmetric part of~\C{anomvariation}.

To avoid these confusions, we will try to work exclusively in $(0,2)$ superspace so that supersymmetry is manifestly preserved. This comes at the price of having the enlarged super-gauge symmetries, with chiral superfield gauge parameters $\La^a$, being anomalous. We will expand on this point in Section~\ref{ss:susyanomaly}.

\subsection{A first look at the solutions}\label{ss:firstlook}
By including anomalous couplings
\be\label{eq:logfi2}
S = {i \over 8\pi}\int\d^2x \d\th^+\, N_i^{a} \log\left( \F^i \right)   \U_-^a + c.c.,
\ee
we have endowed out GLSMs with field-dependent FI parameters:
\be
R^a(\F) = r^a + {N_i^a\over2\pi}\log|\F^i|,\qquad \Th^a(\F) = {\th^a\over2\pi} - N^a_i\Im\!\log(\F^i)
\ee
The first thing to note is that $\Th^a$ depends on the phase of the chiral fields $\F^i$, and so under a gauge transformation it will shift:
\be
\F^i\rightarrow e^{iQ_i\cdot\La}\F^i \quad\Rightarrow\quad \Th^a\rightarrow \Th^a - \sum_i N^a_i Q_i^b\La^b.
\ee
In particular, $\Th^a$ is no longer globally defined (as it was in Section~\ref{sec:gaugeinvariant}) and so
\be
H\sim d\Th^a\wedge\del\bar\del A^a
\ee
will be a cohomologically non-trivial three-form on $\M$. The full story for $H$ is actually more subtle than this, but we will postpone a complete discussion of this until Chapter~\ref{ch:Charged}. However, the basic picture of logarithmic FI couplings leading to quantized $H$ remains true.

One might worry that the couplings~\C{eq:logfi2}\ are ill-defined when the $\F^i$ for which $N^a_i\neq0$ vanish. However, in many situations the $D$-term constraint excludes these poorly behaved points from the space of vacuum solutions. For simplicity, let us consider the case of a $U(1)$, in which case we have
\be\label{simpleD}
\sum_i Q_i |\phi^i|^2 - {N_i\over 2\pi} \log |\phi^i| = r.
\ee
For the simplest compact model, let us assume all $Q_i>0$ and $r>0$. For large fields $|\phi^i|$, the log terms are irrelevant and we approximate weighted projective space. The dangerous region is when a $\phi^i$ with non-zero $N_i$ becomes small. However, if all $N_i\geq 0$ then this region is excluded.  When $\cA=N_iQ_i$ is positive, this corresponds to an anomaly contribution from NS5-brane flux or a gauge instanton. We see that the flux bounds us away from the sources where one or more $\phi^i$ vanish.

For anti-NS5-brane flux, where at least one $N^i$ is negative, the solution for~\C{simpleD}\ becomes non-compact and develops a throat near the singularity. For this case, we see the brane source and the metric is dominated by the log terms.

\subsection{Moduli spaces and a supersymmetry puzzle}\label{ss:modulispace}

We would like to study in greater detail the moduli spaces that arise when we include the logarithmic FI superpotentials
\be\label{FIparam}
S_{\rm log} =  {i\over 4} \int d^2x d\th^+ \, N^a_i \log (\F^i) \Upsilon_-^a + c.c.,
\ee
of Section~\ref{ss:condition}. To eliminate annoying factors of $4\pi$, we have rescaled the $N$ coefficients so that $4\pi N_i\in { \Z }$. This quantization condition is  consistent with models where the logs are obtained by integrating out massive anomalous multiplets, as we will explain in Section~\ref{sec:axanom}. The integration procedure will be described in Section~\ref{sec:perturbationtheory}; it might be possible to relax this condition for models which are not obtained from this UV completion.

In Section~\ref{ss:firstlook}, we saw that these log couplings modify both the $D$-term constraints and introduce $H$-flux into the resulting geometries. The modified $D$-term constraint for each gauge factor is
\be \label{dterms}
\sum_i Q^a_i |\phi^i|^2 - N_i^a \log |\phi^i| = r^a.
\ee
Most of the results we derive here are for the case of a single $U(1)$ gauge factor. The general case is open and quite fascinating.

\subsubsection{Compactness for a single $U(1)$ factor}

While there are many interesting non-compact toric spaces like the conifold and its torsional generalizations, we are primarily interested in compact spaces here.

If the collection of $U(1)$ charges, $Q$, has at least one positive and one negative component then $\M$ is non-compact. To see this, suppose that $Q_1 > 0$ and $Q_2 < 0$. Restrict to the set where the remaining coordinates are $1$. The remaining equations become
\be
Q_1 |\phi_1|^2 - \log \left(|\phi_1|^{N_1} \right) = r + |Q_2| |\phi_2|^2 + \log \left(|\phi_2|^{N_2} \right) - \sum_{j>2} Q_j.
\ee
Both the left and right hand side are unbounded from above as $|\phi^1|$ or $|\phi^2| \rightarrow \infty$. Equality can therefore be achieved for arbitrarily large values of $|\phi^i|$. Hence the spaces are non-compact.
For compact models, we can therefore choose a convention and require $Q\geq 0$. Let us examine  various cases.

\begin{figure}[ht]
\centering
\subfloat[][$|\f|^2-\log|\f|$]{
\includegraphics[width=0.4\textwidth]{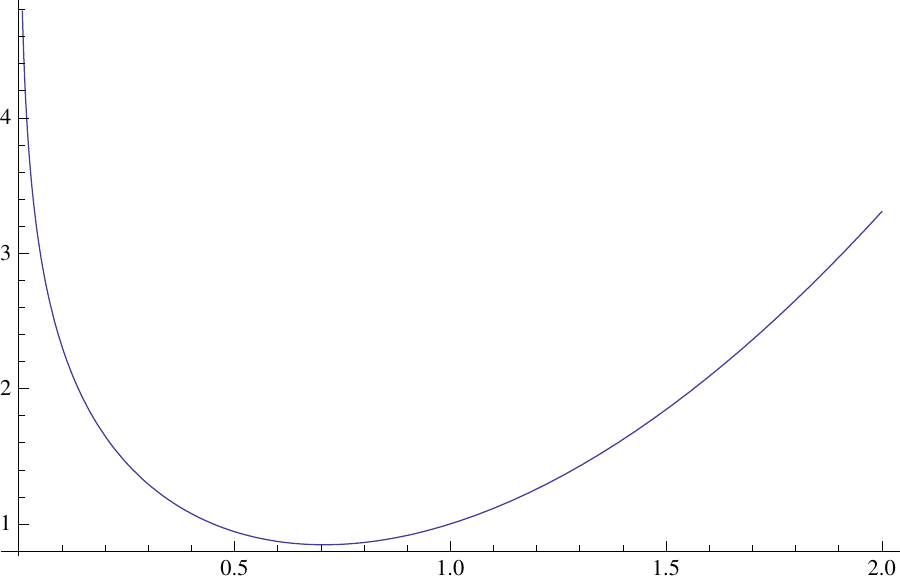}
\label{figure1}}
\qquad\qquad
\subfloat[][$|\f|^2+\log|\f|$]{
\includegraphics[width=0.4\textwidth]{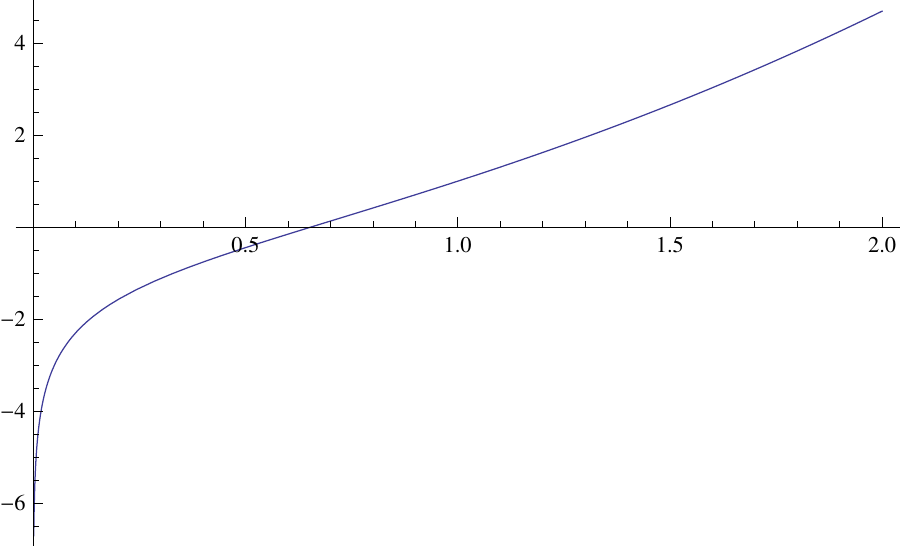}
\label{figure2}}
\caption{\textit{Plots of $|\f|^2 \mp \log |\f|$ against $|\f|$.}}
\label{figures}
\end{figure}

\subsubsection{A single field}
\label{singlefield}
Take the case of a single field, where the $D$-term equation is
\be\label{singlelog}
Q |\phi|^2 - N \log |\phi| = r.
\ee
 If $Q=0$ then $|\phi| = e^{- r/N}.$ Assume $Q\neq 0$ and $N \neq 0$. Rescaling gives
\be
|\phi|^2 - {\hat N} \log |\phi| = {\hat r}, \qquad  {\hat N} = N/Q, \quad {\hat r} = r/Q.
\ee
This equation has a minimum at $|\phi|^2= {\hat N}/2$ if  ${\hat N} >0$.  This defines an ${\hat r}_{min}$ below which there are no solutions:
\be\label{definermin}
{\hat r}_{min} = ({{\hat N} / 2}) \left(1 - \log\left( {{\hat N}/ 2}\right)  \right).
\ee
 Note that ${\hat r}_{min}$ need not be positive! The function is drawn in Figure~\ref{figure1}. The case of ${\hat N} <0$ is drawn in Figure~\ref{figure2}. In this case, there are solutions for all values of ${\hat r}$.

\subsubsection{Two fields}

Now assume $d=2$ scalar fields $\f^i$, and for simplicity choose all charges to be $+1$. There are several possibilities.

\begin{figure}[ht]
\centering
\subfloat[][\textit{No log interactions}]{
\includegraphics[width=0.3\textwidth]{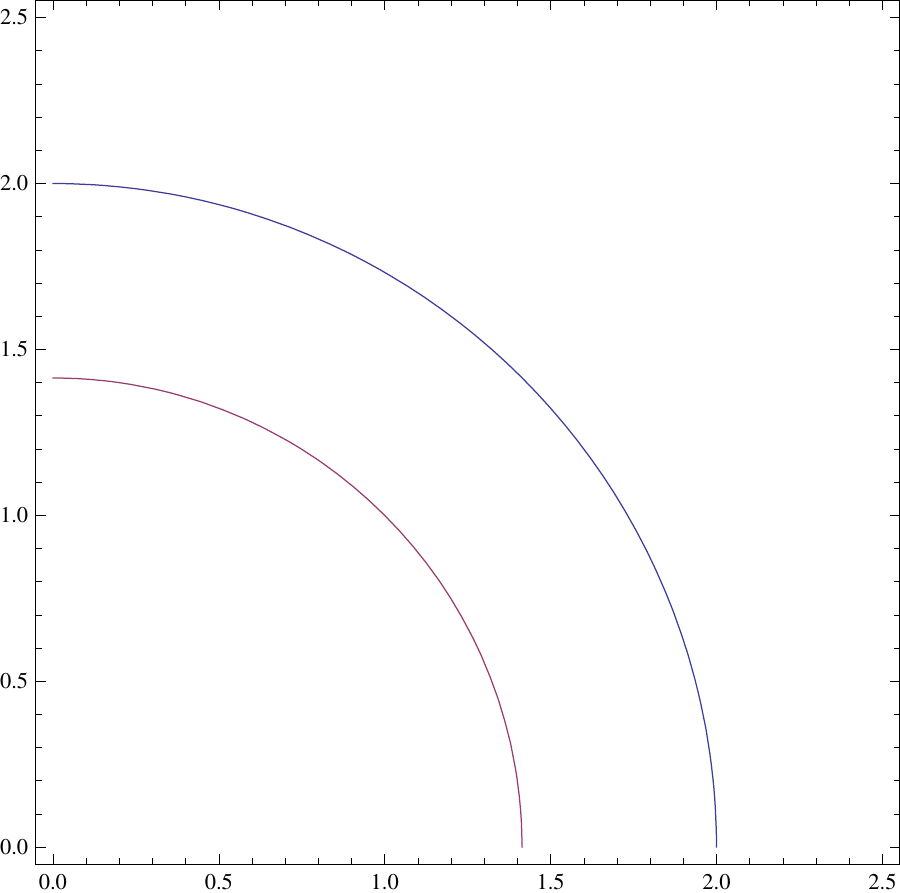}
\label{figure3}}
\
\subfloat[][\textit{A single log interactions}]{
\includegraphics[width=0.3\textwidth]{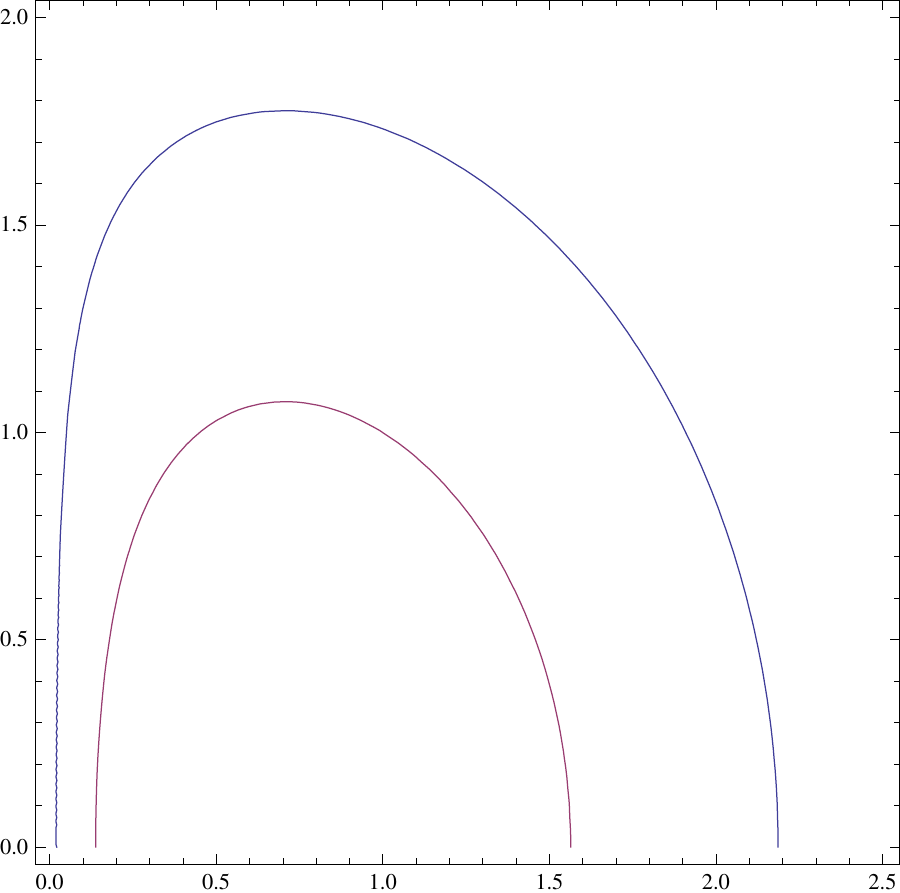}
\label{figure4}}
\
\subfloat[][\textit{Two log interactions}]{
\includegraphics[width=0.3\textwidth]{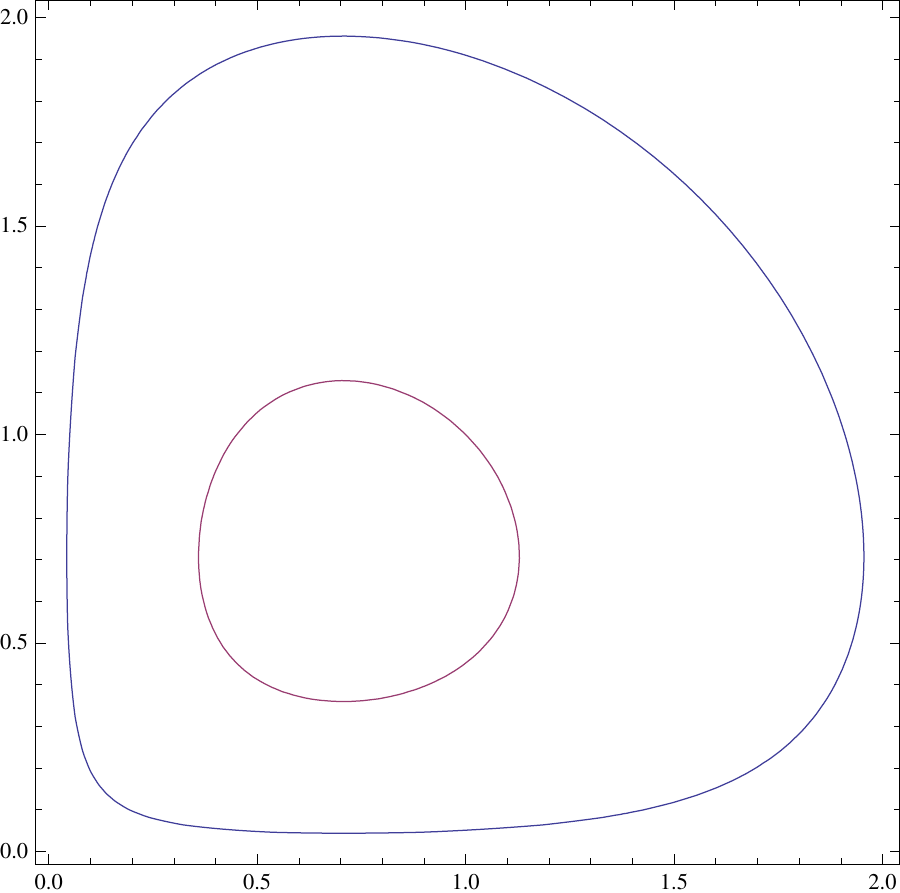}
\label{figure5}}
\caption{\textit{Contour plots of $|\f^1|$ versus $|\f^0|$ for $r=2$ and $r=4$. }}
\label{figures1}
\end{figure}

\vskip0.5cm
\noindent\underline{No log interaction}
\vskip0.5cm
\noindent
First consider the case of two fields with no log interactions. Take
\be  \label{nologdterm}
|\f^0|^2 + |\f^1|^2  = r
\ee
This describes  $S^3$ covered by two patches with either $\f^0\neq 0$ or $\f^1\neq 0$. We can represent this space by the contour in the $(|\f^0|, |\f^1|)$ plane that solves~\C{nologdterm}, which we will call the skeleton of the space. We depict this contour in Figure~\ref{figure3}. Over the skeleton is fibered the phase of $\f^0$ and the phase of $\f^1$. At each axis, one of these two circles degenerates since either $\f^0=0$ or $\f^1=0$. Quotienting by the $U(1)$ gauge group amounts to removing either the phase of $\f^0$ or $\f^1$, depending on the patch. This removed circle is the topologically non-trivial circle of the Hopf fibration of $S^3$. The resulting space is $\PP^1\simeq S^2$.

\vskip0.5cm
\noindent\underline{One log interaction}
\vskip0.5cm
\noindent
Now let us consider a single log interaction,
\be \label{twosphere}
|\f^0|^2 + |\f^1|^2 - N_0 \log |\f^0| = r.
\ee
Note that the case $N_0 <0$ gives a non-compact solution space from the region where $\f^0$ becomes very small and $\f^1$ becomes large. We therefore restrict to $N_0>0$. The log interaction prevents $\f^0$ from vanishing. This means that the phase of $\f^0$ is a globally defined $S^1$. The skeleton for this space is depicted in Figure~\ref{figure4}. Note that the skeleton begins and ends on one axis reflecting the fact that $\f^0$ can now never vanish.

Ignoring the phase of $\f^0$ leaves a space with coordinates $(|\f^0|, \f^1)$. The coordinate $|\f^0|$ takes values in an interval. At the endpoints of the interval, the circle parameterized by the phase of $\f^1$ degenerates. This space is $S^2$. The single log interaction therefore gives $S^2\times S^1$ rather than $S^3$.

We can fix the gauge action by simply setting the phase of $\f^0$ to zero.  The resulting space is an $S^2$, albeit constructed in a way quite different from the preceding $N_0=0$ case. Since $S^2$ is complex, this result is not a priori puzzling.

\vskip0.5cm
\noindent\underline{Two log interactions}
\vskip0.5cm
\noindent

Now let us consider two log interactions,
\be \label{twotorus}
|\f^0|^2 + |\f^1|^2 - N_0 \log |\f^0| - N_1 \log |\f^1| = r,
\ee
with $N_0, N_1>0$.  The log interactions prevent both $\f^0$ and $\f^1$ from vanishing. The phases of $\f^0$ and $\f^1$ define a globally defined $T^2$. The skeleton for this space is depicted in Figure~\ref{figure5}. In this case, the skeleton itself is a circle, and so the total space is $T^3$. Gauge-fixing the $U(1)$ action amounts to removing one circle, leaving $T^2$.

It is really quite surprising that we can construct a torus via a standard Lorentz invariant gauge theory without superpotential couplings.\footnote{By introducing a superpotential, it is easy to build a (2,2) GLSM describing a non-linear sigma model (NLSM) for an elliptic curve in $\PP^2$; however, this cannot be achieved via standard D-term couplings alone as $T^2$ is not toric!} Usually the introduction of circles in the moduli space of gauge theories requires either (Lorentz breaking) impurities introduced in~\cite{Sethi:1997zza, Kapustin:1998pb}, axial gauging~\cite{Gibbons:1996nt}, a special feature of three-dimensional gauge theory (the ability to dualize the photon), or compactification from higher dimensions via Wilson line moduli. Here the log interactions automatically provide globally defined circles.

\subsubsection{Many fields}\label{ss:manyfields}

Let us generalize the preceding discussion to $d>2$. Again choose all charges to be $+1$.
Take
\be  \label{generaldterm}
\sum_{i=0}^{d-1} |\f^i|^2 - \sum_i N_i  \log |\f^i|  = r
\ee
where each $N_i \geq 0$. Suppose $N_i >0$ for $i=0, \ldots, m$, where  $m\leq d-1$. Each log gives a global circle. One circle can be gauged away with the $U(1)$ action. The resulting space is $S^{2d-2-m}\times (S^1)^m$. Including the case with no log interactions gives the following sequence of possible spaces, ranging from $0$ to $d-1$ log interactions:
\be
\PP^{d-1}, \quad S^{2d-2}, \quad S^{2d-3} \times S^1, \quad S^{2d-4} \times (S^1)^2, \quad \cdots \quad , \, S^{d-1} \times (S^1)^{d-1}.
\ee
The appearance of the even dimensional sphere $S^{2d-2}$ in this sequence leads to an immediate worry. For $d=3$, the vacuum manifold appears to be $S^4$ which is known to possess no complex or almost complex structure. How is $(0,2)$ supersymmetry preserved? This strongly suggests that the quantum anomaly must alter the target space topology for any model with a complete UV description that preserves supersymmetry. We will return to this central issue in Chapter~\ref{ch:Charged}, but first we must find a UV completion for theories with logarithmic FI couplings. This will be the one of the main focuses of the next chapter.

\chapter{Invariant logs and NS-branes}\label{ch:Neutral}

In this chapter, we will focus on the logarithmic FI couplings
\be\label{eq:FI2}
S_{\rm log} =  {i\over 4} \int d^2x d\th^+ \, N \log (\F) \Upsilon_- + c.c.,
\ee
of Section~\ref{sec:compact}, except we will restrict ourselves to the case where $\F$ is uncharged. This restriction preserves some of the key features of the log couplings, such as the quantization of $H$, while avoiding the more subtle and confusing ones, like the cancellation between classical and quantum gauge variations.

An immediate benefit of working with this class of models is that they readily admit UV completions as standard GLSMs without field-dependent FI parameters. The FI couplings~\C{eq:FI2}\ emerge at low energies on certain, previously unexplored, branches of the $(0,2)$ GLSM moduli space. This procedure will be explained in Section~\ref{sec:branches}, and it will serve as a guide in uncovering the UV completions for the more intricate cases with charged logs and gauge anomalies.

These theories lead to interesting low energy sigma model descriptions in their own right. These will be explored rather generally in Section~\ref{sec:solutions}, and through specific examples in Section~\ref{sec:exam}. One feature common to all the solutions in this class of models is that they contain explicit magnetic sources of $H$-flux.

\section{Novel $(0,2)$ branches} \label{sec:branches}

To see how the couplings~\C{eq:FI2}\ emerge on certain branches of standard $(0,2)$ GLSMs, start with a conventional $(2,2)$ model but view it from a $(0,2)$ perspective. In terms of $(0,2)$ superfields, the $(2,2)$ abelian vector multiplet contains the $(A, V_-)$ superfields and a neutral chiral superfield $\Sigma$. This vector multiplet can be obtained by dimensionally reducing an ${\cal N}=1$ four-dimensional vector multiplet. The neutral chiral superfield $\Sigma$ captures the two scalars that arise in this reduction.

On the other hand, a $(2,2)$ chiral multiplet decomposes into a pair $(\Phi, \G)$ of $(0,2)$ superfields consisting of a $(0,2)$ chiral multiplet $\F$, and an almost chiral Fermi multiplet $\G$ satisfying~\cite{Witten:1993yc}
\be
\bar D_+ \G = \sqrt{2} E.
\ee
For a Fermi superfield that comes from a $(2,2)$ multiplet, there is a prescribed relation with $E = \sqrt{2}Q\Sigma \Phi$. For general $(0,2)$ models, $E$ can be a more interesting function of all the chiral superfields and we will exploit this freedom. In fact, the $E$ degree of freedom is as rich as a conventional superpotential in terms of physics, but far less well-explored. To see this, note that the total bosonic potential for a $(0,2)$ gauge theory takes the form
\be\label{bosonic}
V_{\rm bos} = {1\over 2 e^2} |D|^2 +|J|^2 +  |E|^2.
\ee
There are two sets of holomorphic data entering~\C{bosonic}, which are on equal footing. The first is the conventional superpotential, $J$, and the second is the choice of $E$.

Let us consider an illustrative case. Take a $U(1)$ gauge theory coupled to $n+1$ $(2,2)$ chiral multiplets of charge $Q_i$. We will distinguish the first $n$ fields $\F^i$ from the $(n+1)$-th chiral field, which we denote $P$. In the absence of any superpotential, the classical target space for this theory is a toric variety; for the case $Q_i=Q_P=1$, the space is $\PP^n$.
From the perspective of a $(0,2)$ model, there are two bosonic potentials. The first is the $D$-term potential,
\be\label{dterm}
D = \sum_i  Q_i |\phi^i|^2 + Q_P |p|^2 -  r,
\ee
where $r$ is the FI parameter. The second is the potential associated to the left-movers $\G^i$ given by
\be
|E|^2 =2 |\s|^2\left(\sum_i Q_i^2|\f^i|^2 + Q_P^2|p|^2 \right).
\ee

Now let us deform this theory away from the $(2,2)$ solution, where $E^i=\sqrt{2}Q_i\S\F^i$, while preserving $(0,2)$ supersymmetry\footnote{Our immediate interest is in geometry rather than the left-moving gauge-bundle, so we will not worry about the question of whether this deformation defines a (semi-)stable or unstable bundle in the Higgs phase. We do, however, require vanishing of the one-loop gauge anomaly for consistency of the theory; this is a quadratic condition on the charges. } by choosing
\be
E^P = \sqrt{2}Q_P\S P, \qquad E^i= 0.
\ee
The classical vacuum structure is found by solving~\C{dterm}\ and the condition
\be
|E^P|^2 = 2Q_P^2|\s|^2 |p|^2.
\ee
Unlike the $(2,2)$ model where solving $|E|^2=0$ implies $\s=0$, there is now a classical branch where $(p=0, \s\neq 0)$. In the $(2,2)$ model, there is a Coulomb branch where $(\phi^i =p=0, \s \neq 0)$, but it is not a classical zero energy branch except for the choice $r=0$. There are, however, vacuum solutions on this branch when one-loop effects are included~\cite{Witten:1993yc}. The central role of the $\s$ field and this Coulomb branch was originally realized in the large $n$ analysis of the $(2,2)$ $\PP^n$ model~\cite{DAdda:1982eh}.  Our new $(0,2)$  branch emanates from the usual Higgs branch at a locus of complex co-dimension one.\footnote{Ilarion Melnikov has amusingly termed these branches ``horns'' sticking out of the conventional Higgs branch.} This is a kind of Higgs branch in which a ``bundle'' direction, $\s$, has become part of the geometry! Such branches are generic in $(0,2)$ models.

For  large $\s$, we can include the leading quantum effect by integrating out the massive field $P$ giving a modified $D$-term,
\be
\sum_{i}  Q_i |\phi^i|^2 - N \log |\s| = {\hat r}.
\ee
The coefficient $N={1 \over 2\pi} Q_P$, while ${\hat r}$ is a renormalized FI parameter. This is a model of the kind described in Section~\ref{sec:compact}, but with a log interaction involving a neutral field. We see that at least a class of those models arise as novel branches of more conventional $(0,2)$ theories. A more detailed derivation of these quantum effects will be provided in Section~\ref{ss:quantum}.

It is very natural to ask whether the theories with charged fields appearing in the log interactions can also be found as branches of conventional $(0,2)$ gauge theories. This indeed appears to be true and comes about as follows: in the example above the scalar field $p$ becomes massive on the new branch along with a gauge non-anomalous combination of left and right-moving fermions. All these fields are integrated out leaving a gauge-invariant model. However, this model is still too closely wedded to its $(2,2)$ origins. There is no reason to consider $E$-couplings which include just a neutral chiral superfield $\S$. One could just as well consider the following $E$-coupling for a Fermi field $\G$,
\be
E = \S_1 \S_2,
\ee
where both $\S_1$ and $\S_2$ are charged. Suppose $\S_1\neq 0$ so $\S_2$ masses up. Necessarily, the associated combination of massed up left and right-moving fermions is now gauge anomalous. We expect a pion-like coupling involving $\log(\S_1)$ which reproduces the anomaly of these massive fermions, together with additional quantum corrections. This should be the right framework to determine the low-energy description of the quantum compactifications described in~\cite{Quigley:2011pv}. This direction, which requires more subtle computations, will be the focus of Chapter~\ref{ch:Charged}.

Our goal in the remainder of this chapter is to study the geometries that arise on the novel branches described above.  The examples considered here only involve neutral log interactions. As we will see, there are both compact and non-compact examples. It is natural to suspect that these models might be conformal for suitable charges, and we will investigate that possibility. Clearly, there are many generalizations.

All of these cases provide natural generalizations of toric geometry, and understanding these geometries is going to be interesting. The example described above takes  the form of a weighted projective space with its K\"ahler class fibered over the $\s$-plane. The point $\s=0$ is special since there is a new massless degrees of freedom; namely, the integrated out $p$-field. This point will correspond to the brane source.

\section{General solutions}
\label{sec:solutions}
As explained in the previous section, there exist branches in the moduli space of $(0,2)$ GLSMs where field-dependent FI parameters of the form
\be\label{f}
{i\over4}\int\d^2x\d\th^+\, N^a\log(\S)\U_-^a + c.c.,
\ee
arise, where $\S$ is a gauge neutral field. These gauge-invariant logarithmic FI couplings are just special cases of those studied in Section~\ref{sec:gaugeinvariant}\ with $f^a=N^a log(\S)$, so we can carry over many of the results from there.

In particular, we have the field-dependent variables\footnote{Note that we have switched the sign $\Th$ in this chapter to make subsequent formulas more convenient.}
\bea
R^a(\s) = r^a + N^a\log|\s|, \qquad \Th^a(\s) = N^a\Im\log\s - {\th^a\over2\pi} ,
\eea
which include possible constant FI parameters $(r^a, \th^a)$. We use $T^a$ to denote the complexified total FI parameter:
\be
T^a = t^a + iN^a\log\s = iR^a - \Th^a; \qquad t^a = ir^a + {\th^a \over 2\pi}.
\ee

The most effective way to determine the induced metric and flux is to first find the induced $K$ in superspace.  The $V_-^a$ superfields only appear as Lagrange multipliers that enforce the superfield constraints,
\be \label{eq:constraint2}
\sum_i Q^a_i|\F^i|^2 e^{2Q^b_iA^b}  = R^a(\S).
\ee
The constraint~\C{eq:constraint2}\ determines the superfields $A^a$ implicitly in terms of $(\F, \bar{\F})$ and $R(\S)$. We will use the notation $A^a$ for both the lowest scalar component of the superfield as well as the superfield itself. Hopefully, the usage is clear from context. Equation~\C{eq:constraint2}\ is a generic polynomial in $e^{2Aa}$ so we can only find explicit solutions for $A^a$ for simple charge assignments. However, we can get surprisingly far just knowing that~\C{eq:constraint2} is satisfied.
Recall from~\ref{ss:classicalgeometry} that $A^a$ are non-trivial sections of some line bundles $\L^a$ over $\M$, and $\del A^a$ transform as connections on $\L^a$. In components,
\be
\del_i A^a = -\bar{\f}_i\D^{ab}Q_i^a e^{2Q_i^c A^c},\qquad \del_\s A^a = \D^{ab}\del_\s R^b,
\ee
which follows from differentiating~\C{eq:constraint2}, and we define
\be\label{eq:delta1}
\D^{ab} = {\del A^b\over \del R_a} = \left(2\sum_i Q^a_i Q^b_i |\f^i|^2 e^{2Q_i^c A^c}\right)^{-1}.
\ee
The invariant two-forms $i\del\bar\del A^a$ give the curvatures of $\L^a$ and provide a basis for (a portion of) $H^2(\M,\Z)$.

We can express $K$ in terms of the $A^a=A^a(|\F|^2,R)$ superfields,
\bea\label{KfromGLSM}
K_i = \bar{\F}^i e^{2Q^a_iA^a} - 2i \Th^a \del_i A^a, \qquad K_\s = \bar{\s} - 2i\Th^a\del_\s A^a.
\eea
Taking appropriate derivatives of $K$ we can easily compute the induced sigma model metric
\bea \label{eq:metric2}
G_{i\jbar} = e^{2Q_i\cdot A}\left(\dd_{i\jbar}-2\bar{\f}_i\f_\jbar Q^a_i \D^{ab} Q_j^b e^{2Q_j\cdot A}\right), \qquad G_{\s\bar{\s}} = 1+ {N^a\D^{ab}N^b \over2|\s|^2},
\eea
and $B$-field
\be\label{eq:bfield2}
B = -2i\Th^a\del\bar{\del}A^a.
\ee
One should not worry too much about the detailed form of these solutions because they will be modified under RG flow; however, we do expect the RG flow to preserve the coarse, topological features.
For example, note that these metrics take the form of warped products over the $\s$-plane, with no off-diagonal mixing between the fiber and base. In addition, the $B$-field roughly takes the form $\Th^a F^a$ where $F^a\sim i\del\bar{\del}A^a$ is the curvature of the line bundle $\L^a$. Even though the $F^a$ are closed, the field-dependence of $\Th^a$ means that $B$ is \textit{not} closed, and there is a non-zero flux $H\sim d\Th^a\wedge F^a$. In particular, the components of $H$ are
\bea\label{H1}
H_{i\s\bar{\s}} = -i \left( \del_\s \Th^a \del_{i\bar\s} A^a - \del_{\bar\s} \Th^a \del_{i\s} A^a \right), \qquad
H_{i\s\jbar} = -i \del_\s \Th^a \del_{i\bar j} A^a,
\eea
which requires use of the relation $\del_{i\jbar}A^a = -\hlf\del_{R_a} G_{i\jbar}$. It is natural to identify
\be J^a_{i\jbar}(R) =i \del_{R_a} G_{i\jbar}, \ee
with the generators of $H^2$ for the toric fiber.


\subsection{Obstructions to conformality}
\label{ss:obs}
The geometric data one obtains directly from a GLSM construction almost never gives $SU(n)$ structure on the nose. It is reasonable to assume that as long as the cohomology class $\left[{\cal R}^{(-)}\right]$ is trivial, the metric will flow to the one with $SU(n)$ structure in the IR. We used similar reasoning for the standard Calabi-Yau case in Section~\ref{ss:RGandCFT}.
As we saw in Section~\ref{ss:alternate}, once we have a metric with ${\cal R}^{(-)}=0$ the associated fundamental form $J$ determines the $H$-flux as well as the Lee form $\xi$ which, if exact, fixes the dilaton.

We therefore expect a $(0,2)$ sigma model to define heterotic string background if the class $\left[{\cal R}^{(-)}\right]$ is trivial.
The GLSM provides a natural choice of coordinates on the target space. From the induced couplings given in~\C{eq:metric2}\ and~\C{eq:bfield2}, we can determine the components of the induced Lee form in these distinguished coordinates:
\be\label{lee}
\xi_i = \del_i\log G_{\s\bar{\s}},\qquad \xi_\s = \del_\s\log\det G_{i\jbar}.
\ee
Now $G_{\s\bar{\s}}$ is a globally defined object, but $G_{i\jbar}$ is not. This observation combined with the form of $\o^{(-)}$ given in~\C{omegaminus},
\be
\o^{(-)} = i(\del-\delbar)\log\det G -2i\xi + 2i\bar{\xi},
\ee
implies a single obstruction; namely, that
\be \del \bar\del \log\det G_{i\jbar} \ee
be a trivial class. This is just the familiar requirement that the toric fibers have vanishing first Chern class, or in terms of GLSM data that $\sum_i Q_i^a=0$. This somewhat surprising result tells us that so long as the fiber metric can flow to a Calabi-Yau solution, then the total space fibered over the $\s$-plane with $H$-flux can flow to an solution with $SU(n)$ structure.

We should be a little careful about the claims of the previous paragraphs. Although it sounds very reasonable, it has not yet been proven that the triviality of $\left[{\cal R}^{(-)}\right]$ implies the existence of an $SU(n)$ structure metric. An analogue of Yau's proof of the Calabi conjecture for KT manifolds is needed to show that vanishing of the cohomological obstruction is sufficient. This kind of result is a little less interesting for $(0,2)$ models compared with $(2,2)$ models because we expect ``most'' compact KT metrics to involve small volumes of order the string scale, like the solutions of~\cite{Dasgupta:1999ss, Becker:2009df}.

 Actually, the metric and flux for a conformal model with a large volume limit will not satisfy just the supergravity equations of motion, but the equations of motion including $\alpha'$ corrections. What is really needed for these theories is a statement about RG flow that generalizes the analysis of~\cite{Nemeschansky:1986yx}\ to $(0,2)$ models. This would involve a classification of the cohomological obstructions that could appear under renormalization. Again for most compact models, an analysis that goes beyond $\alpha'$ perturbation theory is desirable.

The last issue is whether the dilaton equation can be solved. We must ensure that the Lee form is exact with a real potential. This is non-trivial to see starting with GLSM data. The GLSM expression for the induced Lee form given in~\C{lee}\ is not even closed. However, the form is completely determined by the metric. Under RG flow, we expect the metric to flow to one appropriate for a conformal field theory and the IR Lee form should be determined by that metric. The GLSM Lee form is not exact but it is given by gradients of real functions. In the simplest conformal model, we will give evidence that the Lee form actually becomes exact with a real potential by studying the large $Q_P$ limit.

We also note that the GLSM expression~\C{lee}\ has no components proportional to $d\,{\rm Im}\left( \log\s \right)$ which generates $H^1$ for our examples. It seems plausible that RG flow will not produce a component non-trivial in cohomology, but a sharp argument is desirable.
It is worth contrasting this situation with the well-studied $S^3\times S^1$ SCFT, which we reviewed in Section~\ref{ss:S3xS1}. There solutions with $SU(2)$ structure exist, but the Lee form is not exact with a component along the $S^1$ direction. These theories do not define good string backgrounds unless $S^1$ is replaced with ${\mathbb R}$ trivializing the Lee form. The result is the NS5-brane background. Even though all the models described in this chapter admit non-trivial circle factors, the Lee form found in the GLSM never has components along those directions. This is evidence for the assertion that the IR fixed points of these theories can be used to construct heterotic string backgrounds.

\subsection{Quantum corrections} \label{ss:quantum}

In Section~\ref{sec:branches}, we explained how a theory with log couplings can arise from a standard GLSM. Let us now study how this  happens in greater detail. The idea is to use $E$-couplings to generate a mass for a chiral and Fermi superfield pair ($P,\G_P)$, along a branch where $\langle\s\rangle\neq0$. Since we wish to assign canonical dimension 0 to $\S$, we must introduce a mass scale for the $E$-coupling. In a $(2,2)$ theory, $\S$ is part of the vector multiplet so this scale would naturally be set by the two-dimensional gauge coupling $e$. However, in a $(0,2)$ theory we are free to introduce another mass scale, which we call $m_0$. Then the $E$-couplings we want to consider are
\be
E_P = m_0 \S P,\qquad E_i=0.
\ee
When $\s\neq0$, the scalar field $p$ (along with its right-moving fermionic superpartner $\psi_P$ and the left-moving fermion $\gamma_P$) becomes massive with a mass $m=m_0|\s|$. Below the scale $m$, we should integrate out the superfields $P$ and $\G_P$ which generates the field-dependent FI couplings~\C{f}, where $N^a={Q_P^a\over 2\pi}$. In particular, the bare FI parameters get modified as follows,
\be\label{renormr}
r_0^a \rightarrow r^a + N^a \log|\s|,
\ee
where $r^a = r_0^a + N^a \log(m/\La)$ is the renormalized FI parameter and $\La$ is a UV cutoff scale.

In addition to these FI couplings, integrating out $(P,\G_P)$ also modifies the kinetic terms of the vector multiplet:
\be\label{kinetic}
\L_{D,F} = \hlf\left({\dd^{ab}\over e_a^2} + 2\pi {N^a N^b \over m^2} +\ldots\right)\left(D^a D^b + F_{01}^a F_{01}^b \right),
\ee
where the ellipses denote terms that are more suppressed than $O(1/m^2)$. In the IR limit where we send $e_a^2,m_0^2\rightarrow\infty$, we see that the these kinetic terms decouple  provided $\s$ is not too small. In actuality, since we are discussing a quantum mechanical theory in two dimensions, there is no well-defined expectation value for $\s$. Rather there is a branch with $|\s| \neq 0$ which can be studied in a Born-Oppenheimer approximation as long as $|\s|$ is sufficiently large.

In the approximation where we neglect the kinetic terms~\C{kinetic}, $D^a$ and $A^a_\mu$ act as Lagrange multipliers. The constraint of sufficiently large $\s$ should not be too surprising since it just means we are in a regime where we can trust integrating out $(P,\G_P)$ at one-loop. Near $\s=0$, there can be large quantum corrections but we will still be able to study the basic features of our solutions near this point.

There is one more quantum correction induced by integrating out $(P,\G_P)$, which is a correction to the $\S$ kinetic terms. In the large $m_0$ limit,
\be\label{sigmakinetic}
\L_\s = \left(1+{1\over 8\pi |\s|^2} + \ldots\right)|d\s|^2,
\ee
with additional corrections suppressed by ${1/ m_0^2}$. Again, this is only reliable away from $\s=0$. This correction has an important effect since the induced sigma-model metric is significantly modified:
\be\label{renormG}
G_{\s\bar{\s}} = 1 + {1\over2|\s|^2}\left({1\over4\pi} + N^a\D^{ab} N^b\right).
\ee

Finally, since we are considering the theory below the scale $m$, we should also integrate out the high-energy modes of the rest of the fields. As in the usual case, the main effect of this integration is to modify the FI parameters in a way determined by the sum of the charges. Our previous expression for the FI parameters~\C{renormr}\ becomes
\be
r^a + N^a \log|\s| +{1\over2\pi}\left(\sum_i Q_i^a\right)\log\left(\mu\over\La\right),
\ee
where $\mu$ is some IR cutoff scale that we need to introduce since the fields $\f^i$ are massless. In this Wilsonian effective action, no further $\s$-dependent corrections are possible because the FI couplings are controlled by holomorphy. In particular, when $\sum_i Q^a_i=0$ the FI parameters do not run below the scale $m$ and the theory can flow to a non-trivial conformal point. It is reassuring to see the same condition we found in Section~\ref{ss:obs}\ for the low-energy sigma-model also emerges here from RG flow of the GLSM.

To summarize, the picture we find goes as follows: far above the scale $m_0$, we have a standard GLSM with chiral fields $\left(\F^i,P,\S \right)$ with charges $(Q_i^a,Q_P^a,0)$. The FI parameters run according to $Q_P^a +\sum_i Q_i^a$. As we run down to the scale $m_0$, we integrate out $P$ which generates the couplings~\C{f}\ as well as the corrections to the $\s$ kinetic terms given in~\C{sigmakinetic}. Below the scale $m_0$, we have a GLSM with log interactions and the running of the FI parameters is controlled by $\sum_i Q_i^a$. When the sum of these charges vanishes, $r^a$ does not run and the theory can flow to a conformal fixed point.

In the deep IR, the theory flows to a conformal sigma-model whose target is a toric space fibered over the $\s$-plane. The sizes of various two-cycles in the fibers are controlled by the field-dependent quantities $R^a=r^a+N^a\log|\s|$. In general, one combination of the $r^a$ parameters can always be absorbed into the zero-mode of $\s$, along with its corresponding $\th^a$. In this sense, the log interactions remove moduli from the sigma model.

When all $R^a$ are large, the non-linear sigma model geometry should provide a reliable guide to the physics. However, in general there will be regions where some or all of the $R^a$ become small, or even negative. In these regions, one expects another description (like an orbifold SCFT) to be the appropriate description. The correct description can, nevertheless, be determined from the GLSM starting point. This is very much like the phase structure of~\cite{Witten:1993yc}, but with some of the FI parameters promoted to dynamical fields. In some regions, the description is geometric while in others non-geometric. The entire structure glues together to form a single quantum field theory.

\section{Examples}\label{sec:exam}

\subsection{A non-compact massive model}
\label{ss:example1}
The case with multiple $U(1)$ factors can become complicated quickly. Our strategy will be to search for examples of interesting spaces with the number of $U(1)$ factors small. The first interesting case involves just a single $U(1)$ factor. This is a case which should allow us to isolate the essential physics that differentiates these models from conventional branches of $(0,2)$ theories.

Let us begin in the UV with a collection of $n+1$ charge $+1$ chiral fields $\F^i$, along with our distinguished field $P$ with a charge $Q_P$ that can be positive or negative. In the positive case, the conventional $(0,2)$ Higgs branch is a weighted projective space. In the negative case, the Higgs branch is the total space of a line bundle over projective space.

Now imagine moving to the branch in which the $\Sigma$ field becomes massless while $P$ masses up, as described in Section~\ref{ss:quantum}.  The $D$-term constraint on this branch is given by,
\be\label{Dterm1}
\sum_{i=1}^{n+1}  |\phi^i|^2 = R(\s) = r +  N \log |\s|
\ee
where $N={1 \over 2\pi} Q_P$. The space looks like projective space $\PP^n$ (parameterized by the $\f^i$) fibered over the $\s$-plane. The allowed range of $\s$ is fixed by the positivity of $R$, and it depends on the sign of $N$,
\bea
|\s| \in \alt{ (e^{-r/N},\infty), & N>0 \\  (0,\,e^{r/|N|}), & N<0 }
\eea
Interestingly, the effective description is always valid for $N>0$ since $|\s|$ is bounded away from 0, but the $N<0$ models can access that region (corresponding to $R\rightarrow\infty$) where there are large quantum corrections. We will therefore focus on the case $N>0$. It will turn out to be natural to work with the complex variable
\be T=iR -\Th=t +iN\log\s \ee
rather than $\s$.  Note that $T$ has periodicity $T\simeq T+2\pi N$.

Since $\sum_i Q_i=n+1$, we already know that the theory has a mass gap.\footnote{Actually, we should be more careful about whether there is really a mass gap if the $E$-couplings are set to zero. It is possible that the left-moving fermions with $E=0$ flow to a chiral current algebra in the IR. This happens, for example, for the Schwinger model with flavors. We wish to thank Ilarion Melnikov for explaining this possibility. Here we focus on whether the right-moving sector, which characterizes the geometry, flows to a SCFT.  Whether this is possible depends on the sum of the $U(1)$ charges.} Nevertheless, many of the key features that show up in all models with log interactions appear in this  example. This class is particularly nice since $A$, solving~\C{eq:constraint2}, has a very simple form
\be
A= \hlf \log\left(R\over|\f^i|^2\right),
\ee
which allows us to determine $\D= \del_R A = {1\over2 R}$. Plugging these expressions into~\C{metric1}-\C{H1}, and including the quantum correction~\C{renormG}, we find the induced metric and $B$-field
\bea
ds^2 &=& R g_{FS}(\f) + \left({e^{2(R-r)/N} + 1/8\pi\over N^2} +{1\over4R}\right)|dT|^2, \label{metricPn}\\
B &=& \Th \left(J_{FS}(\f) + i{dT\wedge d\bar{T}\over4 R^2}\right),
\eea
where $g_{FS}$ and $J_{FS}$ are the Fubini-Study metric and K\"ahler form on $\PP^n$. Notice that the radius of the $\Th$ circle diverges at the boundaries $R=0$ and $R\rightarrow\infty$, but never vanishes in the interior. In particular, the size of $S^1_\Th$ does not vanish.

The fundamental two-form and flux of the total space are,
\bea
J &=& R J_{FS}(\f) +i \left({e^{2(R-r)/N} + 1/8\pi\over N^2} +{1\over4R}\right)dT\wedge d\bar{T}, \\
H &=& d\Th\wedge J_{FS}(\f),
\eea
and, as a consistency check, it is easy to see that these satisfy the SUSY relation:
\be
H = dB = i(\bar{\del}-\del)J.
\ee

\begin{figure}[ht]
\centering
\includegraphics[scale=1]{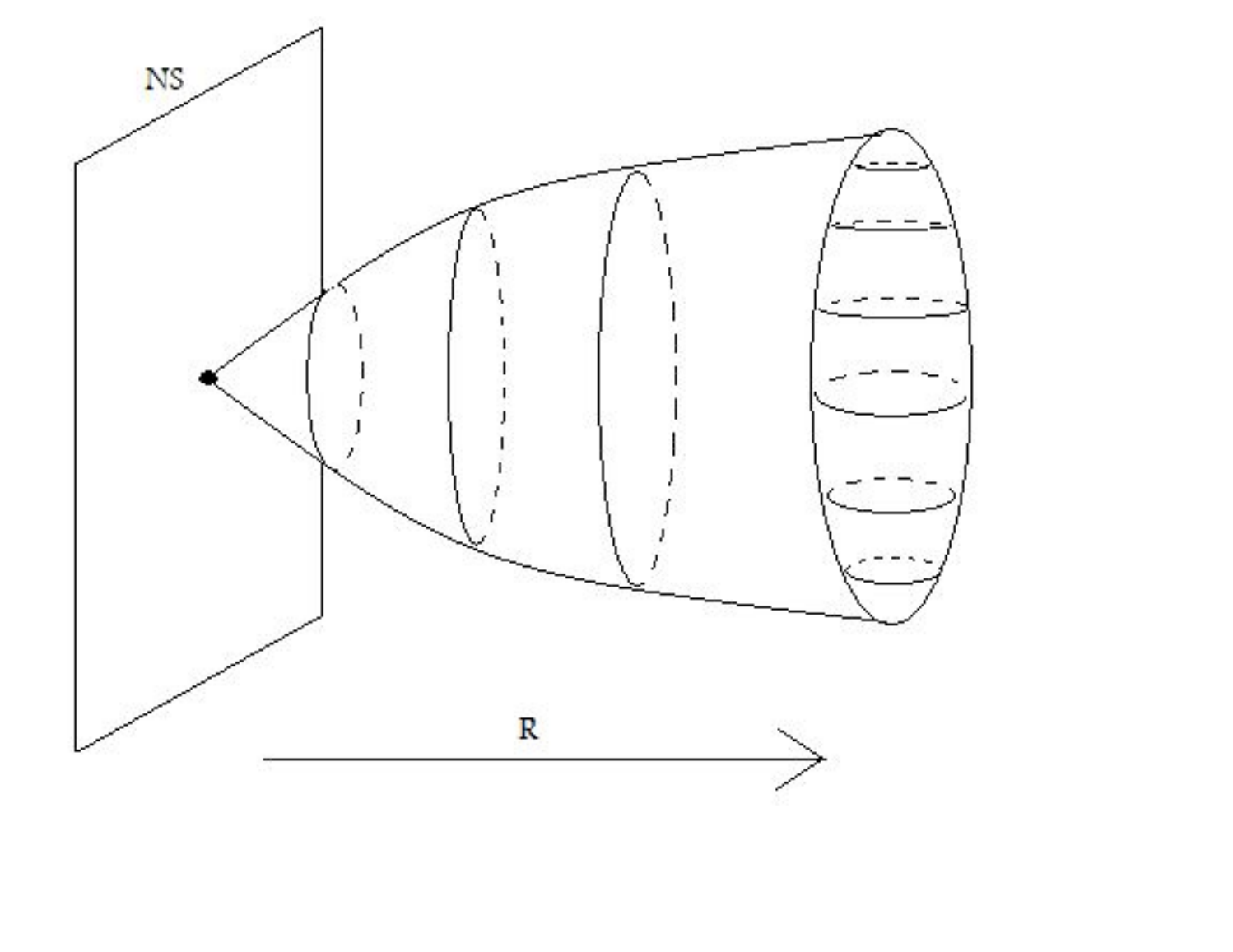}
\vskip 0.2 in \caption{\it A sketch of the brane geometry in the massive $(0,2)$ model. } \label{figure10}
\end{figure}

Notice that near $R=0$, the $\PP^n$ fiber is shrinking to zero size. In particular the two-cycle class $\cC$, dual the K\"ahler form $J_{FS}$, is pinching off. This trivializes $\cC$ in the total space. Even though $\cC$ is trivial in homology, when we integrate $H$ over $\cC\times S^1_\Th$ we get a non-zero result:
\be
\int_{\cC\times S^1_\Th} H = 2\pi N = Q_P,
\ee
at any value of $R$. This indicates that there is a collection of $Q_P$ NS-brane sources located at $R=0$. In fact, we can use this structure as the definition of NS-branes in massive $(0,2)$ theories.  We have depicted these geometries in Figure~\ref{figure10}.

Note that we expect new physics to become important at $R=0$ since, according to the $D$-term constraint~\C{Dterm1}, all the $\f^i=0$. This point is therefore a Coulomb branch since all charged fields vanish, but the physics at this point can still be gapped if there is a non-zero theta-angle, much like the conventional Coulomb branches of $(2,2)$ theories~\cite{Coleman:1976uz, Witten:1993yc}.
To get a better understanding of what is happening near this Coulomb point, we write the metric near $R=0$ as
\be
ds^2 = R g_{FS}(\f) +{| dT|^2\over4R}. \label{metricR0}
\ee
We could also obtain this metric from~\C{metricPn}\  by taking $N\rightarrow\infty$. Notice that the FI parameter $r$ no longer appears in the metric, so the metric and $B$-field have no tunable moduli in this limit. This ``near-horizon" metric has a few interesting equivalent forms. First by writing $R=e^U$, we find
\be
ds^2 = e^U\left(g_{FS}(\f) + {1\over4}ds^2_{PD} \right),
\ee
where $ds_{PD}^2 = dU^2 + e^{-2U}d\Th^2$ is the metric on the Poincar\'e punctured disk. So this space is conformal to a product of two symmetric spaces.

Another equivalent form is as a cone over $\PP^n\times S^1$, though in a peculiar way. Letting $R=\tilde{R}^2$ gives
\be
ds^2 = d\tilde{R}^2 + \tilde{R}^2 g_{FS}(\f) + {d\Th^2\over4\tilde{R}^2}.
\ee
The radius of the $S^1$ goes to zero as $\tilde{R}\rightarrow\infty$ but blows up at the origin, while the size of $\PP^n$ varies in the opposite way. Furthermore, the shrinking $\PP^n$ leads to a conical singularity at the origin.\footnote{The exception is the case $n=1$ where we find a collapsing $\PP^1\cong S^2$ and $\tilde{R}=0$ is a smooth point in $\R^3$. }

To further explore the nature of this singularity, we can perform a T-duality along the $\Th$ direction. It helps to first extract the factor of $N$ from $\Th$ so that it has canonical periodicity $2\pi$. The $T$-dual space then turns out to be the orbifold $\CC^{n+1}/\Z_{Q_P}$, with metric
\be
\widetilde{ds}^2 =d\tilde{R}^2 + \tilde{R}^2 g_{FS}(\f) + {4\tilde{R}^2\over N^2}\left|d\tilde{\Th}^2-N A(\f)\right|^2,
\ee
where $dA=J_{FS}$ and $\tilde{\Th}$ is the coordinate of the dual circle. This orbifold can be viewed as a cone over an $S^1$ bundle over $\PP^n$, where the twist charge of this fibration precisely matches the NS-brane charge $Q_P=2\pi N$ in the original space. The dual space does not contain any $H$-flux.

This picture is in agreement with the basic duality relating NS5-branes with $A$-type $ALE$-spaces. Indeed the precise field theory to which these models flow will depend strongly on the choice of left-moving sector. In particular, whether instantons are localized at the orbifold point.

\subsection{A compact massive model}\label{ss:compactmassive}

It should be clear from the previous example that a model with log interactions and a single $U(1)$ gauge group will always be non-compact, since nothing prevents $R\rightarrow\infty$. An easy way to get compact models of this type is to include a second gauge group, and choose $D$-terms so that $\s$ is bounded. As a nice class of examples, consider two sets of chiral fields: $n+1$ chirals $\F^i$ with charges $(1,0)$ and $m+1$ chirals $\tilde{\F}^k$ with charges $(0,1)$. As before, we integrate out a field $P$ to generate the log interactions, but now with charges $(Q_P,-Q_P)$ under the two gauge groups. For definiteness, let us assume $Q_P$, and hence $N$, is positive. This leads to the following set of $D$-term equations:
\bea\label{twodterms}
\sum_i |\f^i|^2 &=& r^1 + N\log|\s|, \\
\sum_k |\tilde{\f}^k|^2 &=& r^2 -N\log|\s|.
\eea
After quotienting by $U(1)^2$, the solution space takes the form $\PP^n\times \PP^m$ fibered over the $\s$-plane, except now the range of $\s$ is bounded:
\be
e^{-r^1/N}\leq |\s| \leq e^{r^2/N}.
\ee
Notice that $|\s|$ is always bounded away from $0$. In fact, $|\s|$ can be made large by tuning $(r^1, r^2)$ which makes the inclusion of just the leading one-loop quantum corrections quite reliable. One of the two projective spaces collapses at each of the $|\s|$ boundaries resulting in a conical singularity and a Coulomb branch for the relevant $U(1)$. Only for the case $n=m=1$ is the total space smooth. For this special case, the $\PP^1\times\PP^1$ fibration over the $\s$-plane is actually $S^5\times S^1$ with the $S^1$ factor corresponding to the $\Th$ circle. Note that this is a complex space!

\begin{figure}[ht]
\centering
\includegraphics[scale=0.4]{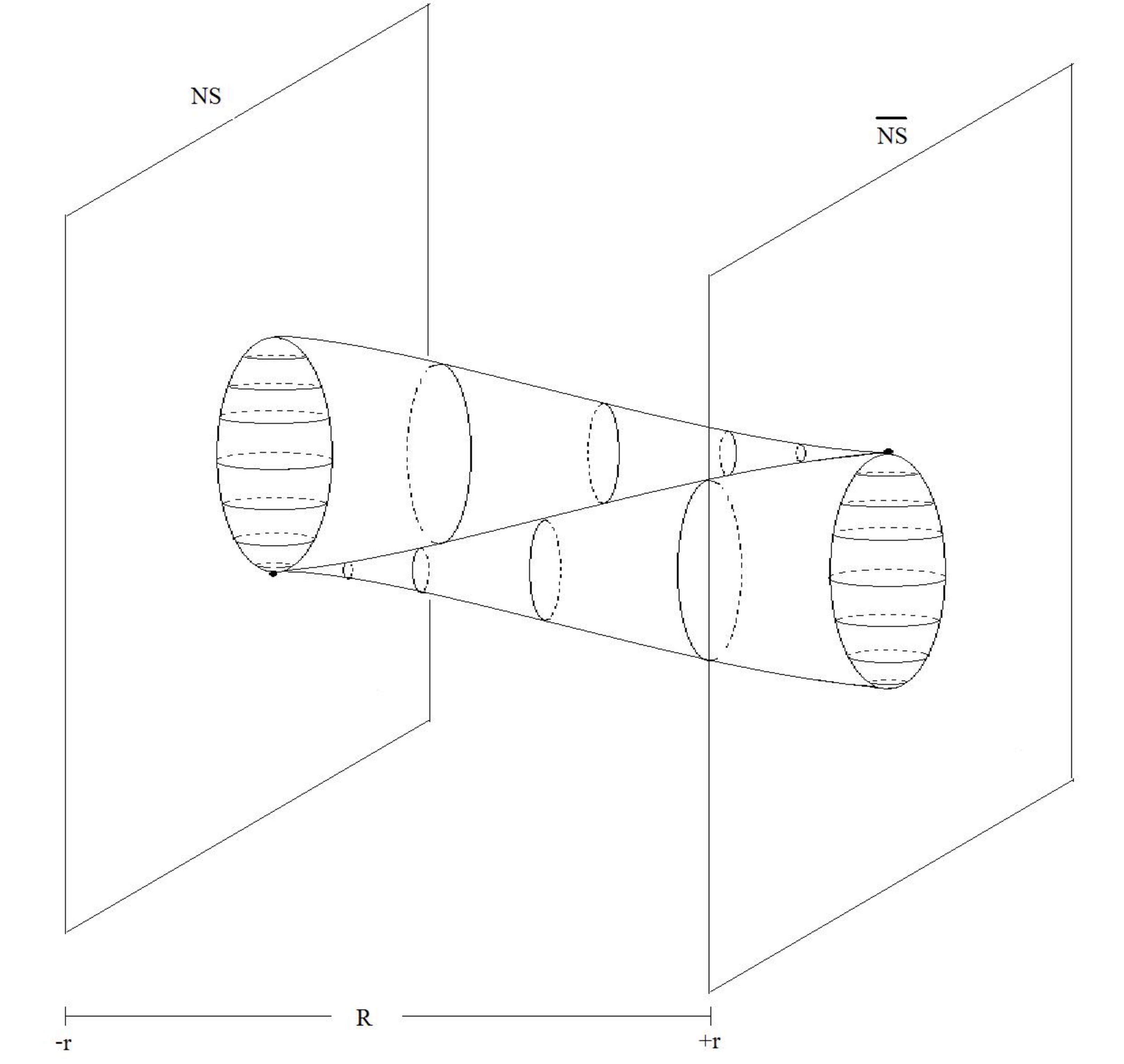}
\vskip 0.2 in \caption{\it A sketch of $S^5 \times S^1$ constructed by gluing together branes and anti-branes. } \label{figure20}
\end{figure}

Another useful way to think about these spaces is to take another combination of gauge groups. Consider the $U(1)$ diagonal in the original $U(1)\times U(1)$ group and its complement. For these combinations, the $D$-term equations become
\bea
&&\sum_i |\f^i|^2 + \sum_k |\tilde{\f}^k|^2 = 2r \equiv  r^1 +r^2,  \\
&&\sum_i |\f^i|^2 - \sum_k |\tilde{\f}^k|^2 = 2R \equiv r^1- r^2 +2N\log|\s|,
\eea
where the factors of 2 have been chosen for later convenience. Now we see that from the original two FI parameters, only the sum $2r=r^1+r^2$ has any physical meaning; it fixes the size of a $\PP^{n+m+1}$ inside the total space. The other combination of FI parameters just gets absorbed into the variable $R$, which now takes values in the range $[-r,r]$. Similarly for the $\theta$-angles where $2\th=\th^1+\th^2$ measures the $B$-field threading the $\PP^{n+m+1}$, while $2\Th = 2N\Im\log\s +\th^2-\th^1$ parameterizes the free circle.

The metric and flux for this class of models is a straightforward generalization of the previous cases:
\bea
ds^2 = \left(r+R\right)g_{FS}(\f) + \left(r-R\right)g_{FS}(\tilde{\f}) + \left({e^{{2R-r^1+r^2\over2N}}  +\coeff{1}{8\pi} \over N^2} + {r/2\over r^2-R^2}\right) |dT|^2  \\
H = d\Th\wedge \left(J_{FS}(\f) - J_{FS}(\tilde{\f})\right).\qquad\qquad\qquad\qquad\qquad\qquad\qquad\qquad\qquad\quad
\eea
Denoting the non-trivial two-cycle classes of $\PP^n$ and $\PP^m$ by $\cC$ and $\tilde{\cC}$, we can again integrate $H$ to get finite results
\be
\int_{\cC\times S^1} H = +Q_P,\qquad \int_{\tilde{\cC}\times S^1} H = - Q_P,
\ee
even though the classes $\cC$ and $\tilde{\cC}$ are trivial in the total space. This indicates that these spaces contain a stack of $Q_P$ branes at $R=-r$ and $Q_P$ anti-brane sources at $R=+ r$. These spaces are basically two copies of the non-compact example of Section~\ref{ss:example1}\ corresponding to brane and anti-brane sources glued together to form a compact geometry. We have sketched this geometry in Figure~\ref{figure20}.

\subsection{Non-compact conformal models}\label{ss:noncompactconformal}

We now turn to the construction of a class of conformal solutions. In this subsection, we only consider cases with $G=U(1)$. As in the preceding non-conformal examples, when we focus on a rank one gauge group nothing bounds the value of $R$, so these models will all be non-compact. While the arguments of Section~\ref{ss:obs}\ give us confidence that we only need to impose $\sum_i Q_i=0$ to guarantee that the IR non-linear sigma model is conformal, we have little hope of following the RG flow to determine the exact geometric data that characterizes the IR solution. We do expect coarse features, like the NS-brane charge and the topological structure of the metric, to remain invariant. We can study this data in specific models.

The analysis also simplifies considerably if we examine a large charge limit. For a single $U(1)$ gauge group, the components of the induced Lee form are
\be \label{lee1}
\xi_\s = \del_\s \log \D,\qquad \xi_i = \del_i \log\left(1+ {1/4\pi + N^2 \D\over2|\s|^2}\right).
\ee
This form is not exact, nor do we expect it to be exact in the UV; however, it has no component along the ${\rm arg}(\s)$ direction so there is no immediate obstruction preventing flow to something exact in the IR. If we consider the $N\rightarrow\infty$ limit, we \textit{do} find an exact Lee form, namely
\be
\xi\underset{N\rightarrow\infty}{\longrightarrow}d\log\D.
\ee
In this limit, we can try to identify the dilaton by setting
\be
e^{2\varphi}=\D.
\ee
The precise form of $\D$ depends on the charge assignments, and $\D$ itself may be subject to renormalization.  However, this does provide an indication that the GLSM solutions simplify in the large $N$ limit, and begin to exhibit features expected in the IR conformal field theories.

The simplest cases to consider are just extensions of those in Section~\ref{ss:example1}, with $n+1$ chiral fields of charge $+1$, together with one chiral field $\Phi^0$ with charge $-n-1$. Note that the addition of a negatively charged field means that $R$ is no longer positive definite. In the UV, for $R>0$ the resulting spaces will be fibrations of the local Calabi-Yau geometry $\cO(-n-1)\rightarrow\PP^{n}$ over the $\s$-plane, with $H$-flux supported on a $2$-cycle in the CY fiber and along the $arg(\s)$ direction (the $\Th$ circle) in the base.

Even for this class of solutions, writing down an explicit form for the induced metric and flux is difficult since this requires knowledge of the function $A(\f,\bar{\f},R)$ defined in~\C{eq:constraint2}. Solving for $A$ requires finding the roots of a degree $n+2$ polynomial, which can only be solved in closed form for $n\leq2$.

To better understand these models, first consider the simplest case possible:  $n=0$. We expect the fibers to look like $\cO(-1)\rightarrow\PP^0$ which is nothing more than a copy of $\CC$. Take $Y=2 \f^0\f^1$ as the gauge-invariant coordinate for the fiber (the factor of 2 has been chosen for later convenience). The total space has real dimension four. In the $N\rightarrow\infty$ limit, the induced target space fields have the simple form
\bea\label{metric2}
&& ds^2 = {|dY|^2 + dR^2 + d\Th^2\over4\left(R^2+|Y|^2\right)^{1/2}}, \\
&& e^{2\varphi} = \D = {1\over 2 \left(R^2+|Y|^2\right)^{1/2}}, \\
&& H = \hlf d\Th\wedge d\Om_2, \label{H2}
\eea
where $d\Om_2$ is the volume form of the $S^2$ embedded in $(Y,\bar{Y},R)$ space. The factor of $\hlf$ appearing in $H$ is reassuring, since this guarantees that $H/2\pi$ is integrally quantized
\be
{1\over2\pi}\int H = {1\over4\pi}\int d\Th\wedge d\Om_2 = 2\pi N = Q_P.
\ee
Up to an overall factor of 2 in the metric, the fields~\C{metric2}-\C{H2}\ are precisely those for a set of $Q_P$ NS5-branes smeared over a transverse circle. It is natural to conjecture that this configuration is the endpoint of the renormalization group flow, and that this result persists even for finite values of $N$ where we expect $1/N^2$ corrections stemming from~\C{renormG}. If we pull out an overall factor of $N$ from all the coordinates, and write $R=Y_3$, then we expect the sigma-model solution to be
\bea
&& ds^2 = e^{2\varphi} \left(d\vec{Y}\cdot d\vec{Y} + d\Th^2\right), \\
&& e^{2\varphi} = e^{2\varphi_0} + {N\over 2 |\vec{Y}|}, \\
&& H = {N\over2} d\Th\wedge d\Om_2.
\eea
Notice once again that the constant FI parameters do not appear in the solutions; they have been absorbed into the fields $R$ and $\Th$. Note that the dilaton blows up at $\vec{Y}=0$ and we should not trust the string loop expansion, as usual for NS5-branes. Although the solutions we have found correspond to smeared NS5-branes, it would be interesting to see if worldsheet instantons  localize the solutions in the $\Th$ direction in a manner similar to~\cite{Tong:2002rq}.

So the simple case $n=0$ corresponds to smeared NS5-branes sitting at a point in $\R^3$, or said differently, they are wrapping a $\PP^0\subset\R^3$. A natural guess for $n>0$ is that the NS5-branes wrap the $\PP^n$ base of $\cO(-n-1)\rightarrow\PP^n$, while the complex line bundle together with the $R$ direction form a transverse $\R^3$. These configurations probably cannot be realized in string theory for $n>3$, but the GLSMs are sensible nonetheless.\footnote{Even $n=3$ is subtle to interpret, since this corresponds to Euclidean NS5-branes wrapping a $\PP^3$.} A similar situation also arose in \cite{Hori:2002cd}.
It would be interesting to study the mirror duals of these models, along the lines of~\cite{Mertens:2011ha}.

Note the important difference between the $n>0$ and the $n=0$ cases (as well as the models studied in \cite{Hori:2002cd}) because the size of the $\PP^n$ varies with $R$; in particular, the $\PP^n$ has zero size at $R=0$. For $R\leq 0$, we should replace the non-linear sigma-model with a $\CC^n/\Z_n$ orbifold CFT. If $N>0$, we expect large quantum corrections in the region $R\rightarrow-\infty$, since that is where $|\s|\rightarrow 0$. In a standard GLSM, these two descriptions would appear as different ``phases" of the same theory, but now they appear within the same geometry just at different values of $R$.


\subsection{Compact conformal models?}\label{ss:compactconformal}

There are many ways to generalize the models of Section~\ref{ss:noncompactconformal}. Take any standard GLSM and include  $\s$-dependent FI terms, while imposing $\sum_i Q^a_i=0$ for each $U(1)$.  The result should be a non-compact, non-K\"ahler $SU(n)$ structure background. Another fascinating direction is to try to build \textit{compact} models which mimic the usual hypersurface or complete intersection construction. This means introducing a superpotential and studying the zero locus. Without a superpotential, it is not possible to find conformal compact solutions because there are no positivity arguments bounding $|\s|$ in models with negatively charged fields.

We will end by describing some difficulties one encounters trying to build compact conformal models. It is useful to revisit the structure of superpotentials in $(2,2)$ models~\cite{Witten:1993yc}. For a hypersurface in a space like $\cO(-n-1)\rightarrow\PP^n$, we consider a $(2,2)$ superpotential of the form
\be
\int d^2\theta \, \phi^0 W(\phi),
\ee
where $W$ is degree $n+1$ in the charge $+1$ fields, while $\phi^0$ is the distinguished field with charge $-n-1$. In $(0,2)$ superspace, this corresponding superpotential has the form
\be\label{twotwo}
S_J = -{1\over \sqrt{2}}\int\d^2x\d\th^+\, \left( \G^0 \cdot W(\phi) + \G^i \phi^0 J_i(\phi) \right)+ c.c.,
\ee
where $J_i = \partial_i W$. Here $\G^0$ is the $(2,2)$ left-moving partner of $\phi^0$. The advantage of starting with the  field content of a $(2,2)$ model is that anomaly cancelation is guaranteed to work. Note that this is a highly non-generic superpotential! Otherwise, there would be no interesting moduli space at all.

Let us try to generalize the compact non-conformal example of Section~\ref{ss:compactmassive}. It is useful to think about this model from the perspective of the two symmetric $D$-term constraints~\C{twodterms}. Without the $\s$-couplings,  we would have made this a conformal model  by introducing a bifundamental field $\phi^0$ with charges $(Q_0, {\tilde Q}_0)$ where:
\be
Q_0 =-  \sum_i Q_i, \qquad {\tilde Q}_0 =-  \sum_i {\tilde Q}_i, \qquad Q_i, {\tilde Q}_i>0.
\ee
The real difficulty in finding an analogue of the complete intersection construction is writing down a superpotential which satisfies transversality. We would like to find a $W$ with charges $(-Q_0, -{\tilde Q}_0)$ under the two $U(1)$ actions such that the only solution to the $D$-term conditions and the constraints,
\be\label{Wcondition}
W= \phi^0 J_i=0,
\ee
is $\phi^0=0$ and $W=0$. This would be a possible compact conformal solution. There are additional desirable conditions  to impose on the choice of charges, described in many places like~\cite{Quigley:2011pv}. For example, we might demand a $U(1)_L$ symmetry, but let us not worry about those additional constraints at the moment.

A $W$ with this charge assignment is necessarily constructed from summing monomials of the form $\phi^n {\tilde \phi}^m$. Any interesting example will have $n>1$ or $m>1$. Taking $J_i = \partial_i W$, we see that there is a flat direction in the potential when either $\phi=0$ or ${\tilde \phi}=0$. This flat direction is usually lifted by the $D$-term constraints which, in the absence of the log interactions, force some $\phi^i$ and some ${\tilde\phi}^i$ to be non-vanishing. In our case, the $\s$-coupling permits a solution to the $D$-term constraints~\C{twodterms}\ with either all $\phi^i=0$ or all $\tilde\phi^i=0$.

We still have a non-compact direction where $\phi^0 \neq 0$ for this attempt which does not stray very far from the structure~\C{twotwo}\ appearing in $(2,2)$ models. For more general $(0,2)$ models, there is a great deal of freedom to play with the structure of the superpotential and the choice of left-moving fermions. Increasing the number of left-moving fermions, subject to the quadratic constraint imposed by anomaly cancelation, increases the number of $J_i$ constraints.

Even with this freedom, it is hard to lift the flat directions in the potential. Indeed, these difficulties suggest that it might not be possible to find compact conformal solutions in this class of classically gauge-invariant models. If so, there should be an argument explaining the existence of flat directions from spacetime physics. At this stage, we hesitate to make a stronger statement because there is a very large space of possible generalizations to explore.

\chapter{Target spaces for anomalous couplings}\label{ch:Charged}

We are now in a position to fully address the solutions of gauge theories with field-dependent FI couplings of the form
\be\label{sketch}
S_{\rm log} =  {i\over16\pi}\int \d^2x\d\th^+  \,  N^a_i\log(\F^i)\U^a_- +c.c.,
\ee
where, unlike in Chapter~\ref{ch:Neutral}, the chiral fields $\F^i$ will now be charged. As explained in Section~\ref{sec:compact}, the gauge variation of~\C{sketch}\ may be cancelled if the gauge symmetry is anomalous. Usually, an anomalous gauge symmetry signals an inconsistency in the theory, but when couplings such as~\C{sketch}\ are present the anomaly is actually required to maintain gauge invariance of the action. In particular, the anomaly coefficient
\be
\cA^{ab} = \sum_i Q^a_i Q^b_i - \sum Q_\a^a Q_\a^b,
\ee
where $Q^a_i$ are the charges of the right-moving fermions and $Q^a_\a$ are the charges of the left-movers, must satisfy
\be
\cA^{ab} = \sum Q^{(a}_i N^{b)}_i.
\ee
This chapter will focus on embedding the couplings~\C{sketch}\ into a UV complete gauge theory, and on the effects they have on the low-energy target manifolds.

\section{Setting the stage}\label{sec:stage}

\subsection{The UV completion}\label{ss:UV}

In Chapter~\ref{ch:Neutral}, we found that we could obtain the logarithmic couplings~\C{sketch}\ on certain branches of the $(0,2)$ GLSM moduli space by integrating out non-chiral pairs of massive fermions. This necessarily led to situations where the fields $\F^i$ appearing the logs were neutral. Since the massive fermions we integrated out are non-chiral, if the high energy theory was free from anomalies then the low energy theory will be as well.

In order to generate FI couplings with charged fields appearing in the logs, we must integrate out fermions with chiral gauge couplings. If the initial theory is anomaly free, so the gauge symmetry is well defined, then after integrating out the chiral fermions we will end up with an anomalous spectrum. The coupling~\C{sketch}\ must therefore be generated in the low energy effective action to preserve the gauge invariance of the theory.

The essential physics we wish to understand is the effective action that describes integrating out an anomalous combination of left and right-moving fermions. To explain the basic setup, let us consider a single $U(1)$ gauge field. Consider the case where $E$ takes the form
\be\label{chargedE}
E = m \S P,
\ee
where both $\S$ and $P$ are charged chiral superfields. This is possible in $(0,2)$ theories but not in $(2,2)$ theories. The mass scale $m$ is needed if we assume canonical dimension $0$ for all scalar fields.

The lowest component of $\S$ is a complex scalar $\s$, while the lowest component of $P$ is $p$. In $(2,2)$ theories, $\S$ usually denotes a neutral field. Since $\S$ is charged here, there is really no reason to distinguish $\S$ from any other chiral multiplet like $\Phi$ or $P$ other than conformity to familiar notation. If $P$ has charge $Q_P$ and $\S$ has charge $Q_\S$ then
\be
Q_\G = Q_\S + Q_P.
\ee
In later discussions, it will be useful for us to note that models with just  $E$-couplings are equivalent to models with just superpotential $J$-couplings. Rather than the $E$-coupling of~\C{chargedE}, we could equally well consider the superpotential coupling
\be
S_J = -{1\over \sqrt{2}}\int\d^2x\d\th^+\, \hat\G  \S P + c.c.,
\ee
where $Q_{\hat\G} = - Q_\G$ and $\bar{\mathfrak{D}}_+ \hat\G = 0$. With this equivalence in mind, let us start by considering a model with just the $E$-coupling given in~\C{chargedE}.

We can now consider the effect of the $E$-coupling. If $\S \neq 0$, the coupling~\C{chargedE}\ masses up the anomalous combination of left and right-moving fermions contained in $\G$ and $P$. The net anomaly from the massive $\G$ and $P$ fields,
\bea\label{expanomaly}
  {1\over 4\pi}Q_P^2 - {1\over 4\pi} Q_\G^2 = - {1\over 4\pi} Q_\S ( Q_P + Q_\G),
\eea
must be reflected in any low-energy effective action.

 Let us take the mass scale $m$ to be much larger than any other scale in the problem. 
 The other natural dimensionful parameter is the gauge coupling, $e$, with mass dimension one.
 In general, the physics depends on the dimensionless combination $em^{-1}$. We will usually work in the limit where $m\gg e$ so we can treat the gauge dynamics perturbatively.
We can then integrate out the anomalous combination of massive fields at one-loop.

It is worth noting that by scaling the charges, the anomaly can be made arbitrarily large with either a positive or negative sign. Setting the FI parameter $r\gg1$, the deep infrared theory will be in the same universality class as a non-linear sigma model. For conventional branches, the corresponding geometry is K\"ahler; for conformal models, the geometry is Calabi-Yau up to small corrections. In our case, which is really the generic situation, the sigma model geometry must reflect the UV gauge anomaly in an essential way.  This is the key issue we want to understand.

\subsection{The UV moduli space}\label{ss:UVmod}

Let us examine the classical moduli space of the basic UV model of Section~\ref{ss:UV}, which contains a single $U(1)$ gauge multiplet and charged chiral matter $(\S,P)$ along with a charged Fermi superfield $\G$. Consistency requires an anomaly free theory and this combination of fields satisfying~\C{expanomaly}\ is anomalous. Imagine adding a collection of superfields, $\F^i$ and $\G^\a$, which supplement the basic fields $(\G, \S, P)$. The only characteristic of these additional fields is that they do not couple directly to $(\G, \S, P)$. They do, however, contribute to the gauge anomaly which must vanish:
\be\label{uvanomaly}
\left( Q_P^2 + Q_\S^2 - Q_\G^2 \right)+  \left( Q_{\F^i}^2  -  Q_{\G^\a}^2 \right)=0.
\ee
Here $Q_{\F^i}^2$ and $Q_{\G_\a}^2$ denote the contributions of potentially many fields.

The UV theory has no log interactions so the moduli space is obtained by minimizing the bosonic potential
\be
V = {1\over 2 e^2} D^2 + |E|^2.
\ee
The condition $D=0$ requires
\be
Q_P |P|^2 + Q_\S |\S|^2 + \sum Q_\F |\F|^2 = r.
\ee
After quotienting by the $U(1)$ action, this constraint gives a weighted projective space if all charges are positive. If some charges are negative, the space is a non-compact toric variety. Vanishing of the $E$-term carves out the hypersurface
\be
\S P = 0
\ee
in this projective space. This is the classical moduli space with two branches, where either $\S\neq 0$ or $P\neq 0$, and a singular locus where $\S=P=0$ and the two branches touch. We expect this classical picture to be drastically modified by quantum effects. On the branch with $\S\neq 0$, integrating out the anomalous massive  pair $(\G, P)$ at one loop generates a log interaction of the form~\C{sketch}. The two branches of the classical moduli space with either $\S\neq 0$ or $P\neq 0$ are already disconnected by this log interaction, which prevents either $\S$ or $P$ from vanishing. A study of the resulting vacuum equations with log interactions leads, however,  to a puzzle about how supersymmetry is preserved; for example, the branch with $\S\neq 0$ can be a non-complex sphere.
The resolution of this puzzle requires a careful examination of the quantum corrections, and one of our main results is that the sphere is in fact replaced by a ball with a finite distance boundary.  The appearance of such boundaries should be a very generic feature in (0,2) target geometries.

\subsection{Outline}

The picture that emerges from our analysis is a target space constructed by a procedure that generalizes a holomorphic quotient. In conventional branches of abelian gauge theory, the moduli space is realized via a symplectic quotient: solve the $D$-term equations and quotient by the gauge group action, which is equivalent to a holomorphic quotient by the complexified gauge group. This promotion of a $U(1)$ compact gauge quotient to a $\CC^\ast$ quotient is natural in supersymmetric gauge theory. The structure of superspace automatically admits the action of the complexified gauge group as a symmetry group if we do not choose a particular gauge like Wess-Zumino gauge.

There is a tension between the supersymmetry requirement that we implement a holomorphic quotient and the inclusion of charged log couplings of the form~\C{sketch}\ in a low-energy effective action. This comes about because the solution of the $D$-term equations is no longer unique when there are charged log interactions. A unique solution is needed to complexify the gauge group action. In Section~\ref{ss:modulispace} we saw that this led to a puzzle: among the target spaces is $S^4$ which does not admit any complex structure. Worldsheet supersymmetry, however, requires a complex manifold. If this is the target manifold, worldsheet supersymmetry would break spontaneously, which is unexpected.  

Before finally resolving this puzzle in this chapter, we revisit the chiral anomaly in Section~\ref{sec:axanom}. What is of particular importance to us is the normalization of couplings in the effective action obtained by integrating out anomalous multiplets. Specifically, a subtle factor of two in~\C{expanomaly}\ when compared with the global chiral anomaly. On very general grounds, we determine the dependence of the low-energy effective action on the phase of the Higgs field which masses up the anomalous multiplet, whether or not the model has any supersymmetry. This is part of the data determining the effective action.

In Section~\ref{sec:perturbationtheory}, we start with a UV complete non-anomalous gauge theory and integrate out a single  massive $(\G, P)$ pair at one-loop. This is a fairly subtle calculation since we are dealing with a chiral supersymmetric gauge theory and there is no regulator that can preserve all the symmetries of the theory. To perform this integration and determine the Wilsonian effective action, we develop a set of Feynman rules for $(0,2)$ supergraphs. Those rules are likely to be useful in wider contexts. The effective action is computed both in a general gauge choice and in the specific case of unitary gauge.

The effective action includes a coupling like~\C{sketch}\ that reproduces the gauge anomaly of the massed up multiplet, but it also includes two additional critical contributions: the first is a correction to the metric of the $\S$-field. This modification is  similar to what one finds when integrating out non-anomalous multiplets. The second contribution comes from path-integrating over the high energy modes of the remaining light superfields, whose fermion content is anomalous. This last contribution vanishes when integrating out massive multiplets with no net gauge anomaly.

In Section~\ref{sec:NLSM}, we study the non-linear sigma model (NLSM)  obtained by classically integrating out the gauge fields of the  gauged linear sigma model (GLSM). This is a manifestly $(0,2)$ sigma model, but one defined in terms of a metric, $G$, and $B$-field which transform unconventionally under holomorphic changes of coordinate. At first sight, this is very peculiar. For example, there is a natural patch where we use the $\CC^\ast$ gauge symmetry to set the Higgs field $\S=1$. This gauge choice is always possible because the log interactions prevent $\S$ from vanishing. In this patch, we find that the metric $G$ is K\"ahler. However, in other patches the metric is not K\"ahler. This is only possible if the metric does not transform as a tensor with an invariant line element. This phenomenon does not happen for GLSMs in which anomaly-free massive multiplets are integrated out, but it does happen here.

To find a metric with conventional transformation properties, we are forced to leave the off-shell $(0,2)$ formalism and work only with manifest $(0,1)$ supersymmetry. With some hindsight explained in Section~\ref{ss:NLSManomalies}, it is clear that this had to be the case for models with tree-level torsion. In Section~\ref{ss:invariant}, we define a natural metric $\widehat G$ invariant under holomorphic coordinate reparameterizations. It is this metric which provides a conventional notion of geometry to our target spaces. Using this metric, we see that these spaces are non-K\"ahler, and we compute the associated $H$-flux in Section~\ref{ss:Hflux}. We provide a detailed study of a class of examples, of which the supposed $S^4$ target space is a member, in Section~\ref{ss:examples}, and find that the gauge anomaly leads to singular boundaries in the target space metrics. The resolution to the puzzle with $S^4$ is that the space is basically cut in half, leaving a 4-ball. We close this chapter in Section~\ref{ss:gencase}\ with some generalizations of this construction.

\section{The abelian gauge anomaly}
\label{sec:axanom}

In this section we review the familiar  problem of the abelian gauge anomaly in two dimensions; our aim is to give a careful treatment of normalizations of terms in the effective action that are involved in the anomaly cancelation central to the rest of this chapter.

\subsection{Conventions} \label{ss:anconv}
We work with a flat infinite volume Euclidean worldsheet in conventions
of~\cite{Polchinski:1998rr}.\footnote{That is $z \equiv y^1+iy^2$; $\p_z \equiv \ff{1}{2}(\p_1-i\p_2)$;  $d^2z \equiv i dz\wedge d\zb = 2 d^2 y$; $\delta^2(z,\zb) \equiv \ff{1}{2} \delta^2(y)$, and $\p_z \zb^{-1} = 2\pi \delta^2(z,\zb)$.}  Our starting point is a free action for $r$ left-moving fermions $\gamma^\alpha$ and $n$ right-moving fermions $\psi^i$:
\be
S_0 = \int \frac{d^2z}{2\pi}~ \left[ \gammab^\alpha \pb_{\zb} \gamma^\alpha + \psib^i \p_z \psi^i \right].
\ee
The non-zero two-point functions are
\be
\la \gammab^\alpha_1 \gamma^\beta_2 \ra_0 = \delta^{\alpha\beta} z_{12}^{-1}, \qquad
\la \psib^i_1 \psi^j_2 \ra_0 = \delta^{ij} \zb_{12}^{-1},
\ee
where the subscript on a field indicates the insertion point, e.g. $\gamma^\alpha_1 \equiv \gamma^\alpha(z_1,\zb_1)$.  This theory has a large global symmetry group $\SO(r)\times\SO(n)$, but we will concentrate on its $\GU(1)^r \times\GU(1)^n$ subgroup with chiral currents $\cJ_L^\alpha = J^\alpha dz$ and $\cJ_R^i = \Jb^i d\zb$, where the operators $J^\alpha$ and $\Jb^i$ are defined by free-field normal ordering:
\begin{align}
J^\alpha = i~ \normd{\gammab^\alpha\gamma^\alpha}~, \qquad  \Jb^j = i~\normd{\psib^j\psi^j}~.
\end{align}
We will be interested in coupling this theory to a background $\GU(1)$ gauge field,
\be {\bf A} = A dz + \Ab d\zb. \ee
Note that until Section~\ref{ss:susyanomaly}, we use $A$ and $\Ab$ to refer to standard bosonic gauge-fields rather than superfields. In order to examine chiral currents in this background, we will introduce a slightly more general interaction term:
\begin{align}
S_\tint & = \int  \frac{d^2z}{2\pi}~\left[ A^i \Jb^i + \Ab^\alpha J^\alpha \right].
\end{align}
The $\GU(1)$ gauging sets $A^i = q_i A$ and $\Ab^\alpha = Q_\alpha \Ab$.

\subsection{The partition function and gauge invariance}\label{ss:partf}
There is no difficulty in evaluating the partition function $Z[{\bf A}] \equiv \la e^{-S_\tint}\ra_0$.  It is given by $Z[{\bf A}] = e^{W[{\bf A}]}$  with\footnote{There are no connected $n$-current correlation functions for $n>2$; with a more general non-abelian gauging, there will be a finite number of additional terms of higher order in the gauge field.  For instance, for $\SU(2)$ $W$ has just an additional $O({\bf A}^3)$ term.}
\begin{align}
W = -\frac{1}{2} \int \frac{d^2z_1 d^2z_2}{(2\pi)^2} \left[ \frac{\Ab^{\alpha}_{1} \Ab^{\alpha}_{2}}{z_{12}^2}+ \frac{A^{i}_{1} A^{i}_{2}}{\zb_{12}^2}\right]~.
\end{align}
While easily computed, $W$ is not gauge-invariant.  Under $\delta_\ep {\bf A} = -d \ep$, we find the local variation (we now set $A_i = q_i A$ and $\Ab_\alpha = Q_\alpha \Ab$)
\bea
\delta_\ep W &=& \int \frac{d^2z}{2\pi} \ep (k_L \p_z  \Ab + k_R \pb_z A), \cr &=& \frac{k_L+k_R}{4\pi} \int d^2z \delta_\ep( \Ab A) +\frac{(k_L -k_R)}{4\pi} \int d^2 z \ep (\p_z\Ab-\pb_z A),
\eea
where $k_L = \sum_\alpha Q_\alpha^2$ and $k_R = \sum_i q_i^2$.  The form of the gauge variation can be brought into a canonical topological form by a choice of counter-terms~(see, e.g., \cite{AlvarezGaume:1984dr} for a thorough discussion), and, indeed, the first term in $\delta_\ep W$ can be canceled by setting
\begin{align}
S_{\tct} = \int \frac{d^2z}{4\pi} A^i N_{i\alpha} \Ab^\alpha,\label{counterterm}
\end{align}
where $N_{i\alpha}$ satisfies $q_i N_{i\alpha} Q_\alpha = (k_L+k_R)$.  We parameterize $N$ by
\begin{align}
N_{i\alpha} = \frac{q_i Q_\alpha(k_L+k_R)}{k_L k_R} + M_{i\alpha},
\end{align}
where $M_{i\alpha}$ is annihilated by $q_i$ and $Q_\alpha$.
Including this counter-term, we obtain an improved partition function ${\widetilde Z}[{\bf A}] = e^{ {\widetilde W}[{\bf A}]}$, with
\begin{align}
\label{eq:gaugetransfW}
\delta_\ep {\widetilde W} = \frac{i(k_L-k_R)}{4\pi} \int \ep \cF,
\end{align}
where $\cF = \cF_{12} dy^1\wedge dy^2$, and $\cF_{12} = -2i (\p_z \Ab-\pb_z A)$. This is the reason for the factor of ${1\over 4\pi}$ appearing in~\C{expanomaly}.

We have reached the familiar conclusion that the partition function will be gauge-variant unless $k_L = k_R$.  Although it has been noted that the chiral Schwinger model remains unitary despite the anomaly~\cite{Jackiw:1984zi,Boyanovsky:1987ad}, for our applications we will insist that the total gauge anomaly of the GLSM cancels.

\subsection{Chiral currents}\label{ss:chicur}
Even when $k_L = k_R$, so that the gauge symmetry is non-anomalous, there will be an anomaly in the global chiral symmetries.  To study these, we define improved currents
\begin{align}
\JJ^\alpha \equiv 2\pi \left. \frac{\delta S}{\delta \Ab^\alpha}\right|_{\cA} = J^\alpha + Q_\alpha A, \qquad
\JJb^i \equiv 2\pi \left. \frac{\delta S}{\delta A^i} \right|_{\cA} = \Jb^i + q_i \Ab,
\end{align}
where we have included contributions from $S_{\tct}$.  To leading order in the gauge field,
\be
\la J^\alpha(z)\ra_{\cA}  = \int \frac{d^2 w}{2\pi} \frac{Q_\alpha \p_{w} \Ab (w)}{w-z} + O({\bf A}^2), \qquad
\la \Jb^i (z) \ra_{\cA} =  \int \frac{d^2 w}{2\pi} \frac{q_i  \pb_{\wb} A(w)}{\wb-\zb} + O({\bf A}^2),
\ee
so that
\begin{align}
\pb_z \la \JJ^\alpha(z) \ra = -\frac{i Q_\alpha}{2} \cF_{12},\qquad
\p_z \la \JJb^i (z) \ra = \frac{iq_i}{2} \cF_{12}.
\end{align}
It follows that the theory retains  $\GU(1)^{r-1} \times\GU(1)^{n-1}$ chiral currents; the non-chiral gauge current is conserved, and one chiral current is anomalous.\footnote{The reader may worry that our expressions for the currents do not appear to be gauge-invariant.  The resolution is simple: the normal ordering prescription $\normd{\gammab\gamma}(w) = \lim_{z\to w} (\gammab(z)\gamma(w) - (z-w)^{-1})$ is not a priori gauge-invariant, and the improvement terms compensate for that.}  We can give this result an interpretation \`a la Fujikawa~\cite{Fujikawa:1979ay}:  the properly regulated gauge-invariant fermion measure has non-trivial transformations under chiral rotations $\delta_{\zeta} \gamma^\alpha = i \zeta^\alpha\gamma^\alpha$ and $\delta_{\xi} \psi^i = i \xi^i \psi^i$, which are interpreted as shifts of the effective action by
\begin{align}
\label{eq:Fuji}
\delta S = \frac{i}{2\pi} \int (\xi_i q_i - \zeta_\alpha Q_\alpha) \cF.
\end{align}
Note that this is larger by a factor of two than the gauge variation in~(\ref{eq:gaugetransfW}).

\subsection{A background Higgs field}\label{ss:higgsb}
For our applications we are interested in coupling the chiral fermions to an additional gauge-charged bosonic scalar field $\vphi$ via a Yukawa interaction.  In this section we will make some observations on the effect of integrating out massive fermions on the Higgs branch. For simplicity, we concentrate on the case where just two fermions, a left-moving $\lambda$ and a right-moving $\chi$, have a non-trivial Yukawa coupling:
\begin{align}
S_{\tYuk} = m \int d^2 y ~ \left[ \vphi \chi \lambda + \vphib \lambdab \chib\right].
\end{align}
We take $|m \vphi|$ to be much larger than any other scale in the theory and integrate out the massive fermions.\footnote{This is the $d=2$ abelian analogue of the well-known work~\cite{DHoker:1984ph} in $d=4$ non-abelian chiral gauge theory.}  As a final simplification, we assume that $|\vphi|$ is frozen at some value, so that only its phase $\vtheta$ plays a role.
Gauge invariance of the Yukawa coupling requires $Q_\vphi = -Q_\chi-Q_\lambda$, and the phase $\vtheta$ transforms by $\delta_\ep \vtheta = Q_\vphi \ep$ under gauge transformations.

As pointed out in~\cite{DHoker:1984ph}, a naive decoupling argument as $|m\vphi| \to \infty$ fails because both the mass of the fermions and the actual Yukawa coupling of the $\theta,\lambda,\psi$ system diverge in this limit; of course the result of integrating out the massive fields must be a set of couplings for $\vtheta$ and the gauge field ${\bf A}$.  One can actually see the terms emerge explicitly by bosonizing the $\lambda,\chi$ fermions, but we will not need that level of detail.  Instead, we observe that low energy  $\vtheta$--${\bf A}$ couplings must reproduce the contribution to the anomaly from the massive fermions, given by~(\ref{eq:gaugetransfW}) as
\begin{align}
\delta_\ep W'_{\lambda,\chi} = \frac{i(Q_\lambda^2-q_\chi^2)}{4\pi} \int \ep \cF = \frac{i(Q_\lambda^2-q_\chi^2)}{4\pi} \int d^2y~\ep \cF_{12}.
\end{align}
To leading order in derivatives and ${\bf A}$, the effective action for $\vtheta$ is fixed up to two undetermined constants, $\kappa$ and $\kappa'$,
\begin{align}
S_{\teff,\vtheta} = \frac{1}{4\pi} \int d^2y~\left[ \kappa D_z\vtheta D_{\zb} \vtheta + i\kappa' \vtheta \cF_{12} \right].
\end{align}
Here $D \vtheta = d\vtheta + Q_{\vphi} {\bf A}$ is the gauge-invariant one-form.  To match $\delta_\ep W'_{\lambda,\chi}$ we see that
\begin{align}
\kappa' =(Q_\lambda-q_\chi).
\end{align}
We can also fix $\kappa$ by matching the chiral symmetries in the UV to those in the IR. The UV theory has a non-anomalous $\GU(1)^{n-1}$ symmetry with\footnote{The argument can be repeated with $\GU(1)^{r-1}$; if $q_\chi=Q_\lambda = 0$, then $\kappa =\kappa'=0$.}
\begin{align}
\delta_\xi \psi^j = i \xi^j \psi^j,\qquad
\delta_\xi \chi =  -i q_j q_\chi^{-1} \xi^j \chi, \qquad
\delta_\xi \vtheta = q_j q_\chi^{-1} \xi^j.
\end{align}
Since we have not introduced a kinetic term for the background field $\vphi$, these are chiral symmetries of the UV theory, and we should be able to recover them in the IR theory in the presence of the quantum-generated $\vtheta$ kinetic term.  As we will now show, this is the case if and only if $\kappa =1 $.

The variation of the effective action receives two contributions.  First, there is the contribution from the light fermions; this has a term from the classical action and a term from the measure as in~(\ref{eq:Fuji}):
\begin{align}
\Delta_1 S_{\teff} = \int \frac{d^2 y}{2\pi}~\xi^j \left[ -\p_z \Jb^j + i q_j \cF_{12} \right].
\end{align}
Second, there are terms from the variation of $S_{\teff,\vtheta}$, which yield
\begin{align}
\Delta_2 S_{\teff} = \int \frac{d^2y}{4\pi} ~\xi^j \left[ \kappa q_j q_{\chi}^{-1} (-2 \p D_{\zb}\vtheta + \ff{i}{2} Q_{\vphi} \cF_{12}) + i (Q_\lambda-q_\chi) q_j q_{\chi}^{-1}\cF_{12}
\right].
\end{align}
All together, we obtain
\begin{align}
\Delta_1 S_{\teff} + \Delta_2 S_{\teff} &= -\int \frac{d^2y}{2\pi}  \xi^j \p_z \left[ \Jb^j + q_j q_{\chi}^{-1} D_{\zb}\vtheta\right] \nonumber\\
&\qquad + \frac{i}{4\pi} \left[\frac{1}{2} + \frac{\kappa Q_{\vphi} + Q_\lambda-q_\chi}{q_\chi} \right] \int d^2y ~\xi^j q_j \cF_{12}.
\end{align}
The second line vanishes if and only if $\kappa = 1$, and the remaining term corresponds to the improved conserved chiral currents for $\GU(1)^{n-1}$.

To summarize: integrating out the massive fermions $\lambda$ and $\chi$ induces a correction to the kinetic term and axial coupling of the phase of the Higgs field $\vtheta$.  As we just argued, the exact result for these terms is
\begin{align}
S_{\vtheta,\teff} = \frac{1}{4\pi} \int d^2y \left[ D_z\vtheta D_{\zb} \vtheta + i(Q_\lambda-q_\chi) \vtheta \cF_{12} \right].
\end{align}
In the non-supersymmetric setting this of course does not determine the corrections to the kinetic term or potential of the modulus $|\vphi| = \rho$, but as we will discuss in Section~\ref{sec:perturbationtheory}, they do play an important role in determining $S_{\vphi,\teff}$ in the supersymmetric theory.

\subsection{The $(0,2)$ gauge anomaly}\label{ss:susyanomaly}

We close our discussion by turning to the supersymmetrization of the gauge anomaly.
The $(0,2)$ supersymmetric version of the gauge anomalous variation~\C{eq:gaugetransfW}\ is
\be
\dd_\La W = {\cA\over16\pi}\int d^2xd\th^+\ \La \Upsilon + c.c.,\label{(0,2)anom}
\ee
where $\La$ is a chiral superfield gauge parameter and $\cA = \sum_i Q_i^2 -\sum_\a Q_\a^2$. This variation can be produced from the non-local effective action
\be
W[A,V_-] = {\cA\over16\pi}\int d^2xd^2\th^+\ A\, {1\over\del_+}\left(D_+\U - \Dbar_+\bar\U\right).\label{GammaSusy}
\ee
In Appendix~\ref{ap:quad} we will compute this expression directly from a loop diagram. For now, let us note that~\C{GammaSusy}\ possesses all the characteristics we desire for a representative of the two-dimensional gauge anomaly: it is expressed entirely in terms of the gauge fields, it is quadratic in the gauge fields, it is inherently non-local and so cannot be canceled by any local counter-term; most importantly, its gauge variation agrees with~\C{(0,2)anom}. Similar non-local representations of the anomaly have appeared in non-supersymmetric and $(0,1)$ gauge theories~\cite{Hwang:1985tm}.

To expand $W[A,V_-]$ in components we do not have the luxury of working in WZ gauge because the action is not gauge-invariant. Instead, we must work with the full non-gauge-fixed form of the gauge fields:
\bea
A &=& C + i\th^+\chi + i\thbar^+\bar\chi +\th^+\thbar^+ A_+, \\
V_- &=& A_- - \th^+\left( 2i\bar\l -\del_-\chi\right) -\thbar^+\left(2i\l + \del_-\bar\chi \right) +\th^+\thbar^+\left(2D +\del^2C\right),
\eea
which transform as follows:
\bea
\dd_\La A &=& {1\over2i}\left(\La-\bar\La\right) = \Im\La -{i\over\sqrt{2}}\th^+\psi_\La -{i\over\sqrt{2}}\thbar^+\bar\psi_\La -\th^+\thbar^+\del_+\Re\La \label{Avar}, \\
\dd_\La V_- &=& -{1\over2}\del_-\left(\La+\bar\La\right) \cr &=& -\del_-\Re\La - {1\over\sqrt{2}}\th^+\del_- \psi_\La +{1\over\sqrt{2}}\thbar^+\del_-\bar\psi_\La +\th^+\thbar^+\del^2\Im\La.
\eea
After performing the superspace integral in~\C{GammaSusy}\ we obtain the component action
\bea\label{GammaComp}
W &=& {\cA\over4\pi} \int d^2x \left(A_+\,{1\over\del_+}F_{01} - \bar\chi\l + \chi\bar\l -CD\right).
\eea
Using~\C{Avar} and integration by parts, we find the local gauge variation
\be
\dd_\La W = {\cA\over4\pi}\int d^2x \left(\Re(\La)\,F_{01} - {1\over\sqrt{2}}\psi_\La\l + {1\over\sqrt{2}}\bar{\psi}_{\La}\bar\l - \Im(\La) D\right),
\ee
as expected. Expanding the field strength $2F_{01}=F_{+-}=\del_+A_- - \del_-A_+$ shows that the non-locality of~\C{GammaComp}\ can be confined to a single term: $\hlf A_+\,{\del_-\over\del_+} A_+ $, with the rest of the effective action comprised of purely local terms. In superspace, we can find an analogous split into local and non-local pieces by inserting $\U = \Dbar_+(\del_-A +iV_-)$ into~\C{GammaSusy}. After some straightforward manipulations, we arrive at
\be
W[A,V_-] = {\cA\over8\pi}\int d^2xd^2\th^+\left(\Dbar_+A {\del_-\over\del_+}D_+A - AV_-\right). \label{anomaly}
\ee
This is the form of the anomaly we will use throughout this work.

\section{Computing the effective action}\label{sec:perturbationtheory}

We now turn to the computation of the one-loop effective action. Rather than study a mass term for an anomalous multiplet generated by an $E$-coupling, it will be more convenient to use a $J$ superpotential coupling. The two formulations are equivalent as explained in Section~\ref{ss:UV}.

\subsection{The setup}\label{ss:setup}

Consider a theory with charged chiral superfields $P$, $\S$ and a Fermi superfield $\hG$, coupled together by the superpotential
\be \label{superpot}
\L_J = -{m\over\sqrt{2}}\int d\th^+ \hG \S P +c.c.,
\ee
where $\bar{\mathfrak{D}}_+ \hat\G = 0$ and
\be Q_\hG + Q_\S + Q_P=0\ee
to ensure gauge invariance of the superpotential.
This set of fields is generally anomalous, so we will include additional charged fields $\left(\Phi^i,\G^\a\right)$ ensuring that the net gauge anomaly vanishes:
\be
Q_P^2 + Q_\S^2 - Q_\hG^2 +\cA = 0, \qquad {\rm with} \qquad \cA = \sum_i Q_i^2 - \sum_\a Q_\a^2.
\ee
Note that  $\cA = 2Q_\S Q_P$. In this section, the fields $\Phi^i,\G^\a$ will only act as spectators ensuring the cancelation of the gauge anomaly; for clarity, we will suppress these fields.

\subsection{The form of the effective action}\label{ss:effact}

When $\S$ develops an expectation value, the gauge theory is Higgsed and the $(P,\G)$ multiplets become massive. When the mass of the $(P,\G)$ fields is large compared to the scale of the gauge coupling, $e$, they should be integrated out leaving an effective theory of the Higgs field, $\S$, and the vector multiplets $A$ and $V_-$. We therefore would like to compute the effective action
\be
e^{iW[\S,\bar\S, A,V_-]} = \int\left[\cD P\, \cD\G\right] e^{iS_0[P,\G,\S,A,V_-]}.
\ee
We know that $W$ must be a local integral over both fermionic coordinates for $(0,2)$ superspace; see Appendix~\ref{ap:superFeynman}. Furthermore, by expanding in powers of ${1\over m^2}$, $W$ must also be expressible as a local integral in position space. Dimensional analysis and Lorentz invariance imply that
\bea \label{Gammaeff}
W &=& \int d^2x d^2\th^+  \left[ f_V(A,\S,\bar\S) V_- + f_{A}(A,\S,\bar\S) \del_- A \right. \cr
&& \left. \qquad\qquad \quad+ \left(f_\S(A,\S,\bar\S)\del_-\S +c.c.\right)  +\ldots \right],
\eea
where the ellipses denote terms that are suppressed by ${1\over m^2}$. Such terms do not contribute to $W$ in the low-energy limit. Note that the $f_A$ and $f_\S$ terms are not  uniquely defined. Rather they should be identified under the equivalence relation
\be
f_A \sim f_A + \del_A f,\qquad f_\S \sim f_\S + \del_\S f,
\ee
for any function $f=f(A,\S,\bar\S)$. This identification shifts the effective action by a total derivative.

\subsection{Unitary gauge}\label{ss:unitary}

Integrating out the massive charged fields requires care because they contain an anomalous set of fermions. This situation has been considered in the past by D'Hoker and Farhi in the context of integrating out the top quark from the Standard Model~\cite{DHoker:1984pc}, and more generally in~\cite{DHoker:1984ph}. One approach is to combine the phase of the Higgs field with the charged fermions to give gauge-invariant fermions, which can then be integrated out without worry. This is a valid procedure, as long as the Higgs field does not vanish and so its phase is well-defined.

In a supersymmetric Higgs theory, we can go one step further. Using the enlarged gauge symmetry present in superspace, we can gauge fix the full Higgs chiral superfield $\S$ to unity while simultaneously rewriting the remaining charged fields in terms of gauge neutral fields.
We do this by effectively fixing unitary gauge. We transform all the fields by a super-gauge transformation with parameter
\be
\La = {i\over Q_\S} \log\S \label{unitary}.
\ee
Since we are transforming \textit{all} the charged fields, including $\Phi^i$ and $\G^\a$, there is no anomalous shift of the action. We end up with a set of gauge-invariant fields:
\bea
\tilde P = P\, \S^{-{Q_P/ Q_\S}}, \ && \quad \tilde A = A + {1\over Q_\S}\log|\S|, \\
\tilde \G = \hG\,\S^{-{Q_\G/ Q_\S}}, \ && \quad \tilde V_- = V_- + {1\over Q_\S}\del_-\Im\log\S \non.
\eea
Note that $\S$ has been gauged away with $\tilde\S \equiv \S/\S = 1$, so only the physical degrees of freedom remain - namely, a massive vector multiplet coupled to chiral superfields. In this gauge, the effective action simplifies tremendously:
\be
W = \int d^2xd^2\th^+ \left[f_V(\tilde A)\tilde V_- + f_A(\tilde A)\del_-\tilde A +\ldots\right].
\ee
The second term is a total derivative which can be ignored in perturbation theory. In unitary gauge, the low energy effective action is therefore completely determined by $f_V(\tilde A)$. A second useful feature of unitary gauge is that the superpotential coupling~\C{superpot}\ reduces to a standard mass term that combines $(\tilde P,\tilde\G)$ into a single massive multiplet with mass $m$. There are no higher order $F$-term interactions.

Unitary gauge is often problematic for carrying out loop computations because the massive vector propagator does not decay sufficiently rapidly at large values of momentum. However, this will not be an issue for us since we will be treating the vector multiplets as background fields and only integrating out the massive chiral fields $(\tilde P,\tilde\G)$. This approach is justified since the mass of the vector multiplets is set by the gauge coupling $e$. As described in Section~\ref{ss:UV}, we are considering the ratio ${e\over m} \ll1$.

\subsection{Computing the effective action in unitary gauge}\label{ss:computing}

\begin{figure}[h]
\centering
\includegraphics[scale=0.5]{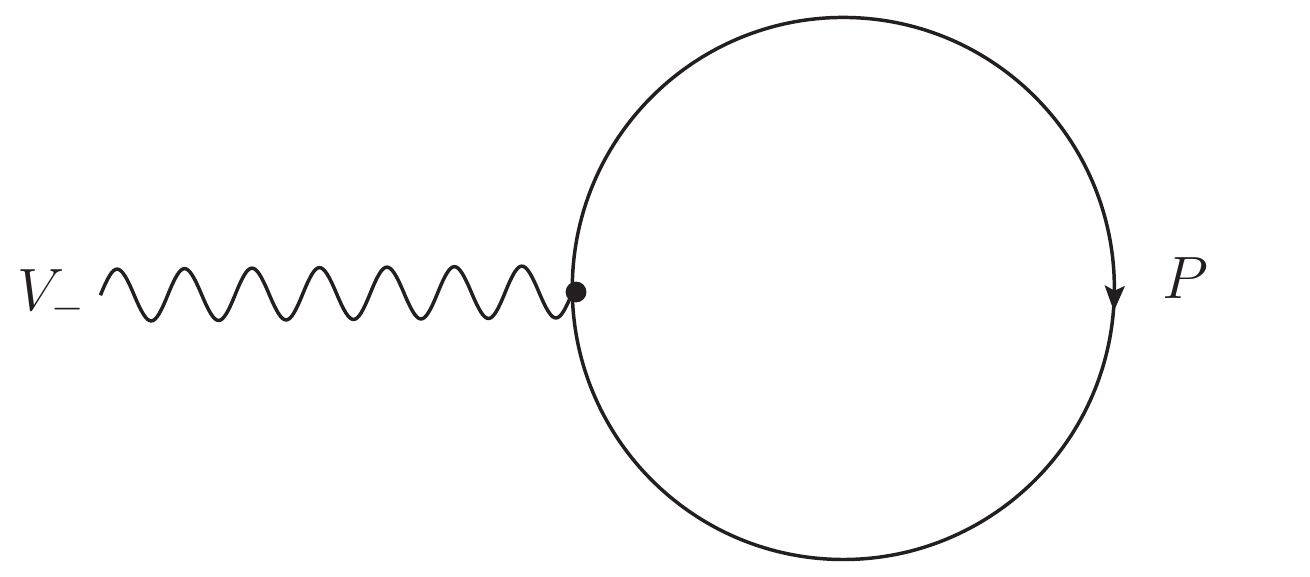}
\vskip 0.2 in \caption{\it The only contribution to the effective action in unitary gauge. } \label{V1pt}
\end{figure}

We can make our lives easier by noting that $f_V(\tilde A)$ is completely determined by its zero-mode dependence: if we expand $\tilde A$ about some constant value $\tilde A_0$ then
\be f_V(\tilde A_0 + \tilde A) = f_V(\tilde A_0) + \tilde A \, f'_V(\tilde A_0) +\ldots.\ee
So we really only need to determine $f_V(\tilde A_0)$, which means we only need the $\tilde A_0$-dependence of the $1$-point function $\langle \tilde V_-\rangle$. The Feynman rules for supergraphs in the presence of a constant background $\tilde A_0$ are derived in Appendix~\ref{ap:superFeynman}. There is a single diagram to compute, shown in Figure~\ref{V1pt}, which involves a loop of $P$ connected to $\tilde V_-$.
 This leads to the result,
\be
iW[\tilde A_0,\tilde V_-] = \int d^2x d^2\th^+\, \tilde V_-(x)\left(Q_P\over2\right)\I_{0,1}\left(m^2 e^{2Q_\S \tilde A_0}\right),
\ee
where the integrals
\be
\I_{p,q}(M^2) = \int_{\ell_E^2\geq\m^2} {d^2\ell\over(2\pi)^2}\, {\left(\ell^2\right)^p \over \left(\ell^2+M^2\right)^q}
\ee
are evaluated in Appendix~\ref{sec:Loops}. The integral $\I_{0,1}$ has a logarithmic divergence. After renormalization at a scale $\mu_r$, we find
\bea
W[\tilde A_0,\tilde V_-] &=& -{Q_P\over8\pi}\int d^2x d^2\th^+\, \tilde V_-(x) \log\left(\m^2 + m^2 e^{2Q_\S \tilde A_0}\over \m_r^2\right) \\
&=& -{Q_P\over8\pi}\int d^2x d^2\th^+\, \tilde V_-(x) \left(2Q_\S \tilde A_0 +\log\left(m^2 \over \m_r^2\right) +\ldots\right), \non
\eea
where we have dropped terms that are suppressed by $\left({ \m \over m}\right)$. Restoring the full $\tilde A$-dependence and recalling that $\cA = 2Q_\S Q_P$, we find the low-energy effective action:
\be
W[\tilde A,\tilde V_-] = -{\cA\over8\pi}\int d^2x d^2\th^+\, \tilde A(x) \tilde V_-(x) \label{Gammafinal}.
\ee
We have dropped a field-independent correction to the FI parameter. We give an alternate computation of this term by directly computing the $\langle A V_-\rangle$ correlation function in Appendix~\ref{ap:quad}.

\subsection{Computing the effective action without unitary gauge}\label{ss:nogauge}

While the result~\C{Gammafinal}\ is all that is needed to determine the data of the low energy sigma-model in the patch where $\S=1$, we would also like to know how this the effective action looks in other patches; for example, a patch where we set a chosen $\Phi^i=1$ instead of $\S$. Simply undoing unitary gauge, by using the inverse of the gauge transformation~\C{unitary}, turns out to be rather subtle. Instead it will prove easier to recompute $W$ without fixing unitary gauge. We will recover~\C{Gammafinal}\ by gauge fixing this more general result. As a bonus, this gauge-unfixed result will generalize straightforwardly to the case of multiple $\S$ fields giving large masses to multiple $(P,\G)$ pairs. This gives us a picture of the sigma model geometry on the cover of the $\CC^\ast$-action; i.e., a picture in terms of homogeneous rather than inhomogeneous (or gauge-fixed) coordinates.

\begin{figure}[h]
\centering
\subfloat[][]{
\includegraphics[width=0.475\textwidth]{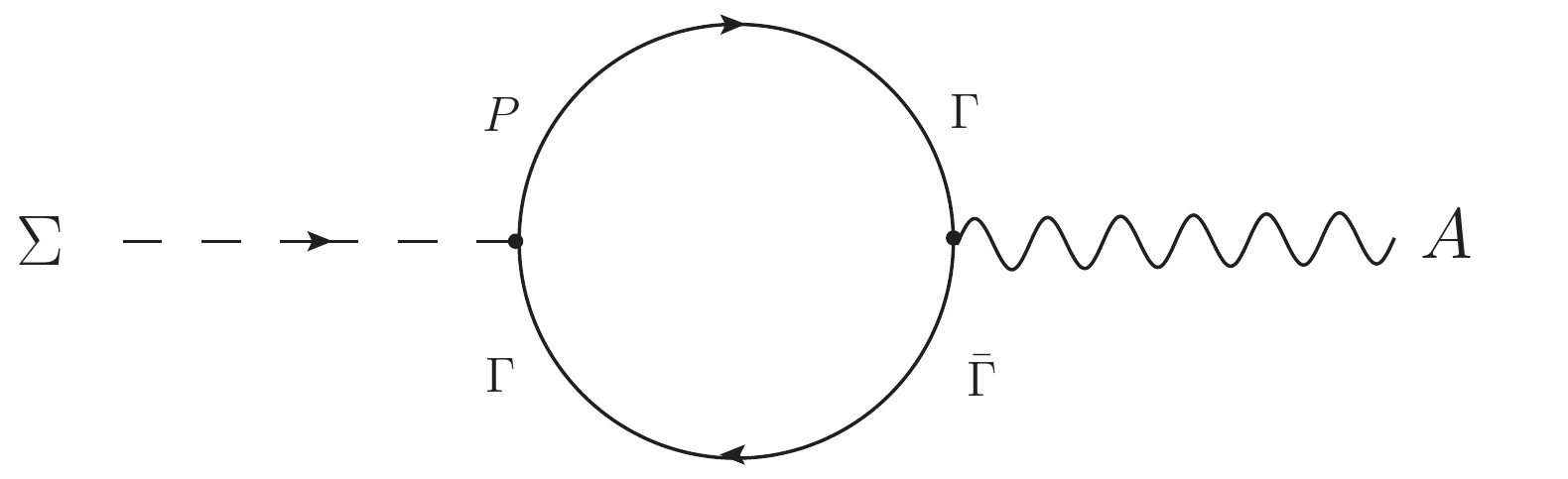}
\label{ASigma1}}
\
\subfloat[][]{
\includegraphics[width=0.475\textwidth]{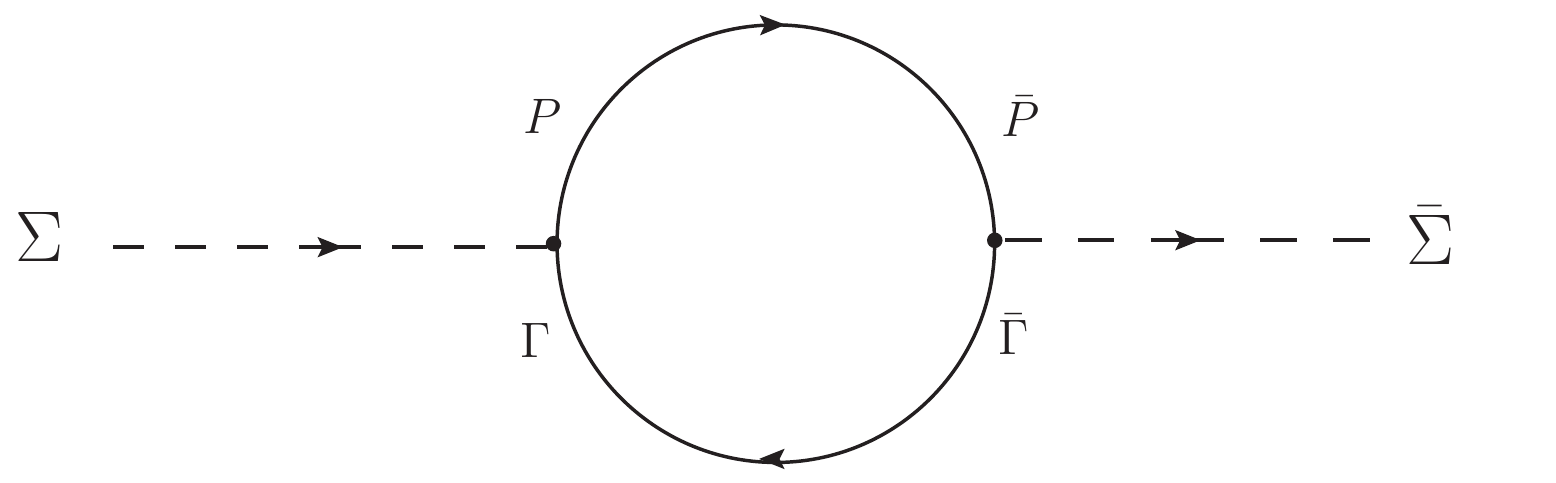}
\label{SSbar1}}
\caption{\textit{The remaining contributions to the effective action without gauge-fixing.}}
\label{otherloops}
\end{figure}

In this situation we must compute all three functions, $(f_V,f_A,f_\S)$, appearing in~\C{Gammaeff}. Once again, we will perform the computation around some constant background fields, but now we expand about a point $(A_0,\S_0)$ in moduli space, rather than just $A_0$.  The computation of $f_V$ goes through exactly as before except for the replacement $m\rightarrow m\S_0$:
\bea
f_V(A_0,\S_0,\bar\S_0) &=& -{Q_{P}\over8\pi} \log\left(\m^2+m^2|\S_0|^2e^{2Q_{\S}A_0}\over \m_r^2\right) \label{fV}\\
&=& -{1\over8\pi}\left(2Q_{P}Q_{\S}A_0 + Q_{P}\log|\S_0|^2 +\ldots \right). \non
\eea
We see that the $AV$ term is unchanged, but we now have an additional $\log|\S|$ term. This term vanishes in the gauge $\S=1$. The fact that $f_A$ and $f_\S$ appear at linear order in derivatives means we cannot get them from a $1$-point function and we have to go to $2$-point correlators. The relevant diagrams are shown in Figure~\ref{otherloops}.

From figure 4\subref{ASigma1}\ we find the result
\be
Q_{\G}m^2\bar\S_0e^{2Q_\S A_0}\int {d^2q\over(2\pi)^2}d^2\th^+\ q_-A(q)\S(-q) \int dx\ (1-x) \I_{0,2}(\D),
\ee
where $\D=m^2|\S_0|^2e^{2Q_\S A_0} + x(1-x)q^2$. The same diagram with $AP\bar P$ replacing the $A\G\bar\G$ vertex happens to vanish. Expanding in $\left({\m\over m}\right)$, we can see that this correlator requires the following term in the effective action:
\be
f_A(\S_0, \bar\S_0) = -i{Q_\G\over8\pi}\log\left(\S_0\over\bar\S_0\right).
\ee
Finally, we compute the loop shown in~\ref{SSbar1}, which corrects the $\langle \S\bar\S\rangle$ propagator. The result is
\be
\hlf m^2 e^{2Q_\S A_0}\int {d^2q\over(2\pi)^2}d^2\th^+\ q_-\S(q)\bar\S(-q) \int dx\ x\I_{0,2}(\D).
\ee
Again expanding in the limit $m\gg\m$, we find that this term originates from
\be
f_\S(\S_0,\bar\S_0) = -{i\over32\pi}{\log(\bar\S_0)\over\S_0}.
\ee
These results combine very nicely into a sensible effective action. The function $f_\S$ is clearly a renormalization of the $\S$ kinetic term. To ensure gauge invariance of this term, we must extract the appropriate couplings to the vector multiplets. Writing
\bea 2Q_P &=& (Q_P-Q_\hG)-Q_\S \\
2Q_\hG &=& -(Q_P-Q_\hG)-Q_\S \non
\eea
for the remaining $\log(\S)$ terms, we can write the one-loop effective action at a scale $\mu\ll m$ in the form
\bea
W^{\rm 1-loop} &=& -{i\over16\pi}\int d^2xd^2\th^+\left[\left(\log|e^{Q_\S A}\S|\over\S\right)\nabla_-\S +c.c.\right] \non\\
&&-\left(Q_\S^2 +\cA\over8\pi\right) \int d^2xd^2\th^+ AV_- \label{Gammaunfixed}\\
&&+i{(Q_P-Q_\G)\over16\pi}\int d\th^+\ \log(\S)\U_-\ +c.c. ,\non
\eea
where $\nabla_-=\del_-+Q_\S \left(\del_-A+i V_- \right)$.

\subsection{The structure of the effective action}\label{ss:structureeffective}

Before examining the structure of the effective action, we should comment on the validity of the one-loop approximation. We are integrating out a massive multiplet with mass $m |\S|$. As $|\S|$ becomes sufficiently small, there can be large corrections to a one-loop effective action. With this caveat in mind, we note that
the first line of~\C{Gammaunfixed}\ gives a gauge-invariant correction to the $\S$ kinetic terms.

These kinetic terms, together with the remaining quantum corrections of~\C{Gammaunfixed}, are crucial in resolving the supersymmetry puzzle of Section~\ref{ss:modulispace}. We can sketch how this comes about: supersymmetry requires a $\CC^\ast$-action on the space of fields, but our $D$-term equations typically admit multiple solutions in the orbit of this action, as shown in Figure~\ref{figure1}\ for one $D$-term. A horizontal slice of that graph typically has two solutions. This is what led to the sphere topologies of the target manifold and the apparent  supersymmetry breaking. Because of the quantum corrections of~\C{Gammaunfixed}, we find that Figure~\ref{figure1}\ is basically cut in half because the low-energy metric becomes singular before one can access both solutions to the $D$-term equation. Since our one-loop effective action is reliable at large $|\S|$, the small $|\S|$ branch of solutions is not accessible via our analysis.

Turning to the remaining terms of~\C{Gammaunfixed}, we see that the third line is precisely the pion-like $F$-term coupling needed to reproduce the gauge anomaly of the pair $(P, \G)$. Under a gauge transformation, this $F$-term transforms anomalously like~\C{(0,2)anom}\ but with  coefficient
\be
-(Q_P-Q_{\hat\G})Q_\S = Q_P^2-Q_{\hat\G}^2,
\ee
which reproduces the anomaly of the fields we  integrated out. Finally, the term appearing in the second line has a nice interpretation as a local contribution coming from the anomalous measure of the remaining light charged fields $(\S,\Phi^i,\G^\a)$. Recall that we can represent the gauge anomaly by the non-local effective action
\be\label{repeatanomaly}
W= {\tilde\cA\over8\pi}\int d^2xd^2\th^+\left(\Dbar_+A {\del_-\over\del_+}D_+A - AV_-\right),
\ee
described in Section~\ref{ss:susyanomaly}, where $\tilde\cA=Q_\S^2+\cA$.

This 1PI effective action is non-local because we have integrated out massless degrees of freedom. We are actually studying the local Wilsonian effective action, where we only integrate down to a fixed scale $\m$. This IR cut-off has the effect of smoothing out the non-local term:
\be
\Dbar_+\del_-A {1\over\del^2}D_+\del_-A \,\, \rightarrow \,\, \int_0^1 dx\ \Dbar_+\del_-A \left({-x(1-x)\over \m^2-x(1-x)\del^2}\right)D_+\del_-A.\label{smoothing}
\ee
Indeed, in Appendix~\ref{ap:quad} we find the local expression appearing on the right hand side of~\C{smoothing}\ when we only integrate down to a scale $\m$, rather than all the way to zero. If we go to momentum space replacing $\p^2$ by $q^2$, we see that the non-local expression~\C{smoothing}\  vanishes in the limit $\m^2\gg q^2$.

What remains is the local term of~\C{repeatanomaly}. By including the local $AV_-$ term in our effective action, we are essentially changing the anomalous variation of the measure for the light chiral fermions from the usual
\be {\tilde\cA\over16\pi}\int d\th^+\ \La\U_- +c.c. \ee
to
\be
\dd S = {\tilde\cA\over8\pi} \int d^2xd^2\th^+\ \dd\left(\Dbar_+A {\del_-\over\del_+}D_+A\right) = {\tilde\cA\over8\pi} \int d^2xd^2\th^+\ (\La+\bar\La)\del_-A +O(\La^2).
\ee
Generalizing the result~\C{Gammaunfixed}\ to multiple sets of $(\S,P,\hG)$ is now straightforward. Furthermore, setting $\S=1$ does indeed recover the unitary gauge result~\C{Gammafinal}\ with the correct coefficient $-\cA/8\pi$. Finally, we note that setting $Q_\S=0$ forces $\cA=0$ and $Q_\hG =-Q_P$, and then~\C{Gammaunfixed}\ reduces to the gauge-invariant cases studied in~\cite{Quigley:2012gq}\ with NS-brane sources. The coefficient of the log interaction in the neutral $\S$ case is a factor of $2$ larger than the charged case considered here for reasons explained in Section~\ref{sec:axanom}.

\subsection{Effects from other fields}\label{ss:others}

Since we are computing the Wilsonian effective action at a scale $\m$, we should in principle also integrate over the high-energy modes of the light fields $\Phi^i,\G^\a,\S,A,$ and $V_-$. Fortunately, up to a field-independent shift of the FI parameter, the path integration over these light fields do not affect our results. More details can be found in Appendix~\ref{ap:others}.

\section{The non-linear sigma model}\label{sec:NLSM}

Now that we have evaluated the one-loop corrected effective action, including the effects from integrating out anomalous pairs of massive multiplets, we are in a position to study the non-linear sigma models that emerge at low energies. We will extract a low-energy non-linear sigma model in a semi-classical fashion by sending the gauge coupling $e^2\rightarrow\infty$. In this limit, the gauge fields are effectively non-dynamical and we can integrate them out classically. We will separately consider the case with a single $\S$ field and the case of multiple $\S$ fields.

\subsection{Effective NLSM actions, supersymmetry, and anomalies}\label{ss:NLSManomalies}

Before we turn to explicit computations of background geometries, we should discuss a few interesting subtleties in extracting a geometric interpretation from effective actions.\footnote{This material is discussed in a number of classic papers~\cite{Hull:1986xn,Sen:1986mg,Howe:1987nw}; the last paper is particularly relevant to the (0,2) discussion.} The essential point is relatively simple:  to extract a geometric interpretation from a NLSM effective action we make a split between the local and non-local contributions, and such a split is inherently ambiguous up to choosing various local finite counter-terms.  These terms are constrained by demanding manifest (super)symmetries and other desirable properties.

To make these comments concrete, consider a (0,1) NLSM.  The defining geometric data for such a theory consist of a metric $G$ and $B$-field $B$, as well as a choice of connection on the left-moving gauge bundle.
When we expand the classical action in components, we find that the right-moving fermions couple to a connection with torsion given by $dB$.   In the quantum theory $B$ acquires non-trivial spacetime gauge and Lorentz transformations, and $H = dB +\ff{\alpha'}{4}\textrm{CS}$ is the physical gauge-invariant field strength.  This seems to lead to a small paradox:  either the right-moving fermion kinetic term is gauge-variant, or it is not supersymmetric~\cite{Hull:1986xn,Sen:1986mg}.

The resolution follows by computing the effective action in a manifestly (0,1) supersymmetric fashion.  This is comparatively easy because of the unconstrained nature of (0,1) superfields, and the result is an explicitly (0,1) SUSY form for the spacetime Lorentz- and gauge-variant terms in the one-loop effective action~\cite{Hull:1986xn}.  It is then easy to see that this local anomaly can be cancelled by assigning transformation properties to \emph{both} $G$ and $B$.  The latter transformation is familiar, but the former is unusual and perhaps undesirable if one wishes to use conventional intuition from Riemannian geometry.

Fortunately, there is a simple alternative:  we can add a finite (0,1) SUSY counter-term whose variation exactly matches the $G$-variation.
Now the metric in the two-derivative action can be kept invariant under the gauge transformations; furthermore, expanding the resulting effective action in components, we find that the modification of the local terms is exactly to shift $dB \to H$ in the right-moving fermion kinetic terms.

A similar analysis has also been carried out for (0,2) NLSMs~\cite{Howe:1987nw}.  The classical (0,2) NLSM is determined in terms of a (0,2) potential $K_I$ and a Hermitian metric for the left-moving fermions.  The latter determines a holomorphic connection for the gauge bundle, while $K_I$ is the (0,2) potential that fixes the Hermitian metric and $B$-field via
\begin{align}
\label{eq:GBfromK}
G_{I\bar J}(x,\bar x) = \del_{(I}K_{\bar J)},\qquad B_{I\bar J}(x,\bar x) =  \del_{[I} K_{\bar J]},
\end{align}
where $X^I$ denote the complete set of (0,2) bosonic chiral fields, and $x^I$ are their scalar components.  The gauge-variant part of the effective action can be evaluated in a manifestly supersymmetric fashion (though there are complications because of the use of constrained chiral superfields), and the resulting variation can be cancelled by assigning gauge transformations to the (0,2) potential $K_I$.  However, there is no manifestly (0,2) SUSY finite local counter-term that can be used to reproduce the variation due to the shift of the Hermitian metric.  Thus, to keep (0,2) SUSY manifest, we must work with spacetime Lorentz and gauge-variant (0,2) potential $K_I$; in particular, both the metric and $B$-field shift under the transformations.

This is significant:  in a manifestly (0,2)-SUSY regularization, $G_{I\Jbar}$ is in general a gauge-variant object. If we want to consider a more conventional geometry, where the metric is gauge-invariant, we will need to leave the realm of manifestly off-shell $(0,2)$ supersymmetry and construct an invariant metric $\widehat G_{I\Jbar}$. Such a metric has fundamental form, $\widehat J$, that is related to $H$ via
\be
H = i \left( \p -\bar\p \right) \widehat J,
\ee
and satisfies the Bianchi identity:
\be \label{Jbianchi}
dH = 2 i \p \bar\p \widehat J = {\alpha'\over 4} \left[ \tr R\wedge R - \tr F\wedge F \right].
\ee
Alternatively, we can work with the gauge variant ``metric" $G_{I\Jbar}$, but this can lead to confusion: for instance, an apparently K\"ahler background can be gauge equivalent to a Hermitian background with torsion. We will see explicit examples of this.

Usually, we are not interested in a metric that precisely satisfies~\C{Jbianchi}, anymore than we are interested in the precise $\alpha'$-corrected metric that defines a conformal $(2,2)$ model. A metric solving~\C{Jbianchi}\ on the nose will be very complicated, since the curvatures appearing on the right hand side are evaluated with quantum corrected connections. Rather, we are usually interested in how~\C{Jbianchi}\ is solved at the level of cohomology. Renormalization group flow will take care of generating the precise set of $\alpha'$ corrections. This is a subtle question because the right hand side must be globally trivial, and yet integrate to something non-vanishing on a space with torsion. How this works for the original compact torsional solutions of~\cite{Dasgupta:1999ss}\ has been explored in detail~\cite{Fu:2006vj, Becker:2008rc, Becker:2009df, Becker:2009zx, Melnikov:2012cv, Melnikov:2010pq}.

\subsection{The $(0,2)$ metric and $B$-field}\label{ss:singlesigma}
We begin with the complete low-energy effective action in the limit $e^2\rightarrow\infty$:
\bea\label{effectiveaction}
S &=& \hlf \int d^2xd^2\th^+\left[-{i\over2}\sum_i\left( \bar\Phi^i e^{2Q_i A}\del_-\Phi^i -c.c.\right) - \sum_\a\bar\G^\a e^{2Q_\a A}\G^\a \right. \non\\
&&\qquad -{i\over2} \left(\left(\bar\S e^{2Q_\S A} + {\log\bar\S\over8\pi\S}\right)\del_-\S -c.c.\right) +\Th(\S)\del_-A \\
&&\qquad \left. +\left(\sum_i Q_i|\Phi^i|^2 e^{2Q_i A} + Q_\S|\S|^2 e^{2Q_\S A} - {\cA\over4\pi} A -R(\S) \right)V_- \right] \non,
\eea
where we have introduced the natural field-dependent quantities
\be\label{defineR}
R(\S) = r + {Q_P\over2\pi}\log|\S|,\qquad \textrm{and}\qquad \Th(\S) = {\th\over2\pi}+{Q_{\hat\G}\over2\pi}\Im\left(\log\S\right).
\ee
These combine naturally into the complex quantity
\be\label{defineT}
T \equiv \Th+iR =  t +i\left(Q_P-Q_{\hat\G}\over4\pi\right)\log\S - i{Q_\S\over4\pi}\log\bar\S, \qquad t = ir + {\th\over 2\pi},
\ee
which is only holomorphic when $Q_\S=0$; examples with $Q_\S=0$ were studied in~\cite{Quigley:2012gq}.
Let us recall that all the effects of integrating out the massive anomalous pair $(\G, P)$ are encoded in the $\S$ couplings of~\C{effectiveaction}. The fields $\Phi^i$ and $\G^\a$ were spectators in that computation, described in Section~\ref{sec:perturbationtheory}. They appear with standard couplings in~\C{effectiveaction}.

This action is not gauge-invariant. This is critical: we are studying a quantum consistent low-energy theory but not a classically consistent theory. Under an infinitesimal gauge transformation, the action changes by
\be
\Th(\S) \, \rightarrow \, \Th(\S) - {\tilde\cA\over4\pi}\, \Re(\La),\label{Thshift}
\ee
where
\be
\tilde\cA=Q_\S^2 +\cA=Q_{\hat\G}^2-Q_P^2=Q_\S(Q_P-Q_{\hat\G})
\ee
is the anomaly coefficient of the low-energy degrees of freedom. For convenience, we recall that $\cA = 2Q_\S Q_P$. As noted in Section~\ref{ss:structureeffective}, this shift in the action is compensated by an anomalous transformation of the path-integral measure.

To simplify notation, we will denote the complete set of chiral fields by
\be
X^I = (\Phi^i,\S) = X_{\bar I}.
\ee
Note that our convention for raising and lowering indices conjugates the index. $V_-$ appears as a Lagrange multiplier in~\C{effectiveaction}, enforcing the constraint
\be
\sum_I Q_I|X^I|^2 e^{2Q_I A}  - {\cA\over4\pi} A = R(\S). \label{constraint}
\ee
This constraint should be viewed as an equation that determines $A$ in terms of $X^I$ and the complex conjugate field $\bar X^{\bar I}$. Note that this equation is actually gauge-invariant under the full $\CC^\ast$-action.

Implicitly solving the constraint~\C{constraint}\ gives a $(0,2)$ non-linear sigma model  action
\be
S = -\hlf\int d^2xd^2\th^+\left[{i\over2} \left(K_I\del_-X^I -c.c.\right) + h_{\a\bar\beta} \bar\G^{\bar\beta}\G^\a \right].
\ee
The metric on the gauge bundle over the target space is
\be
h_{\a\bar\beta} = e^{2Q_\a A}\dd_{a\bar\beta},
\ee
but we will ignore $h$ in the rest of this discussion because our primary concern is with the target space metric itself and its associated $B$-field. These objects are derived from
\bea \label{defineK}
K_I = X_I e^{2Q_I A} + 2i\Th \del_I A + \dd_{I\S}{\log\bar\S\over8\pi \S},
\eea
using~\C{eq:GBfromK}. Their evaluation is greatly facilitated by the relations,
\be
\del_I A = \D\left(\del_I R - Q_I\bar X_I e^{2Q_I A}\right), \label{delA}
\ee
where we have introduced the quantity
\be
\D = {\del A\over \del R} = \left( 2\sum_I Q_I^2 |X^I|^2 e^{2Q_I A} - {\cA\over4\pi}\right)^{-1}. \label{Delta}
\ee
These relations follow from differentiating the constraint~\C{constraint}. Away from $|\sigma|=0$,  the induced target space metric is
\be\label{inducedmetric}
G_{I\bar J} = e^{2Q_I A}\dd_{I\bar J} -2\del_I A \D^{-1} \del_{\bar J}A + i\del_I A\del_{\bar J}\bar T -i\del_I T\del_{\bar J}A +{\dd_{I\s}\dd_{\bar J\bar\s}\over8\pi|\s|^2},
\ee
and the induced $B$-field is
\be
B = 2i\Th\,\del\bar\del A +i\del A\bar\del T -i\bar\del A\del \bar T.\label{B}
\ee
Note that curvature of $B$ has the rather simple form,
\be\label{Bcurvature}
dB = i\left[\del T+\bar\del \bar T\right] \del\bar\del A = i\left[d\Th +i(\del-\bar\del)R\right]\del\bar\del A,
\ee
which satisfies
\be
dB = i(\bar\del-\del) J.\label{dBandJ}
\ee
This relation follows automatically because both $B$ and $J$ are derived from the same $(0,2)$ potential $K_I$.

Both~\C{inducedmetric}\ and~\C{Bcurvature}\ are gauge-variant quantities with respect to the superspace $\CC^\ast$-action. We can see this in a very striking way: there is a very natural choice of gauge in which we set $\S=1$ using the superspace $\CC^\ast$-action. In this gauge $dB=0$ and therefore~\C{dBandJ}\ implies that the corresponding metric should be K\"ahler. This is sufficiently surprising that we will verify K\"ahlerity directly in Section~\ref{ss:patch}. In other gauge choices $dB \neq 0$, and so the metric no longer appears K\"ahler. Clearly, we are missing some important ingredient.

From the target space perspective, the chiral gauge parameter $\Lambda$ can be regarded as a holomorphic function of $X^I$, so that gauge transformations correspond to target space diffeomorphisms. What have found is that neither $G$ nor $B$ transform as tensors under this diffeomorphism. Based on the discussion above this had to be the case, because K\"ahlerity is a coordinate independent property. In hindsight, this might have been expected for reasons explained in Section~\ref{ss:NLSManomalies}. The manifestly $(0,2)$ GLSM is naturally giving us a NLSM with anomalous transformation properties for both the metric and $B$-field.

It is intriguing that this phenomenon does not appear for conventional GLSMs, where only non-anomalous multiplets mass up, but it does appear here. In the conventional case, the $G$ that results from the procedure we have followed defines a genuine metric. The redefinition of $G$ described in Section~\ref{ss:NLSManomalies}\ is still required, but the need for such a redefinition only shows up at one $\alpha'$-loop in the NLSM. In our case, the initial metric given to us by the GLSM is already unconventional. Presumably, this reflects the torsion present at tree-level in the background. To find a conventional metric with a $B$-field that transforms in the usual way under target space gauge and Lorentz transformations, we will need to leave our manifestly $(0,2)$ framework.

\subsection{A preferred patch}\label{ss:patch}

We argued in Section~\ref{ss:singlesigma}\ that it is possible to find a local coordinate patch where the ``metric"~\C{inducedmetric},
\be
G_{I\bar J} = e^{2Q_I A}\dd_{I\bar J} -2\del_I A \D^{-1} \del_{\bar J}A + i\del_I A\del_{\bar J}\bar T -i\del_I T\del_{\bar J}A +{\dd_{I\s}\dd_{\bar J\bar\s}\over8\pi|\s|^2},
\ee
restricts to a K\"ahler metric. The argument relied on the relation
\be
dB = i(\bar\del-\del)J,\label{dBandJ3}
\ee
which is a consequence of working in a manifestly $(0,2)$ supersymmetric formalism. Here we would like to check K\"ahlerity in this patch directly without relying on~\C{dBandJ3}.

We use the $\CC^*$-action to set $\s=1$ and work with affine coordinates:
\be
z^i \equiv { \phi^i/\left(\s\right)^{Q_i/ Q_\S}}.
\ee
This choice is equivalent to fixing unitary gauge for the UV theory, as described in Section~\ref{unitary}. If we ignore any potential singular behavior of the low-energy metric, this is the only patch needed to globally fix the gauge action since $\s$ is always non-vanishing. In this patch, the metric is given by
\be\label{metric1}
G_{i\jbar} = e^{2Q_i A}\left(\dd_{i\jbar} - 2\bar z_i z_\jbar Q_i \D Q_j e^{2Q_j A} \right).
\ee
Notice that with $\s=1$, $(R,\Th)$ of~\C{defineR}\ just reduce to $(r,\th)$, and  $B$ is now exact:
\be\label{kahlerB}
B= i{\th\over\pi}\del\bar\del A
\ee
since $\th$ is constant. This metric and $B$ have precisely the same form expected in a conventional $(0,2)$ sigma-model with one important exception: the function $\D$, which we recall here for convenience
\be
\D =  \left( 2\sum_I Q_I^2 |X^I|^2 e^{2Q_I A} - {\cA\over4\pi}\right)^{-1},
\ee
contains the quantum correction $-{\cA\over4\pi}$. This correction term can be seen by fixing unitary gauge in~\C{Gammafinal}, and all the anomalous behaviour of $G$ can be traced back to this term.

In this patch the $B$-field~\C{kahlerB} is exact, which suggests that the metric is K\"ahler. We can check this explicitly by computing:
\bea
\del_k G_{i\jbar} &&= 2Q_i\del_k A \dd_{i\jbar} e^{2Q_i A} - 2\bar z_i Q_iQ_j\left(\dd_{\jbar k}\D + z_\jbar\del_k\D+ 2(Q_i+Q_j)z_\jbar \D\del_k A\right)e^{2(Q_i+Q_j)A}, \non\\
&&= -2Q_j\D e^{2Q_jA}\left(Q_k\bar z_k\dd_{i\jbar} e^{2Q_k A} + Q_i \bar z_i \dd_{\jbar k} e^{2Q_i A}\right) \\
&&+ 4Q_iQ_jQ_k \bar z_i z_\jbar \bar z_k \D^2 e^{2(Q_i+Q_j+Q_k)A} \left[1-2\D\left(Q_\S^3 e^{2Q_\S A} + \sum_\ell Q_\ell^3|z^\ell|^2e^{2Q_\ell A}\right)\right], \non\cr
\eea
where have used the fact that $Q_i\dd_{i\jbar}e^{2Q_i A} = Q_j\dd_{i\jbar}e^{2Q_j A}$ along with the relations~\C{delA}\ and~\C{Delta}. Each line is separately symmetric under the exchange $i\leftrightarrow k$ and therefore
\be
(\del J)_{ij\bar k} = i\del_{[i}G_{j]\bar k} =0.
\ee
Similarly, $\bar\del J=0$ confirming that the metric~\C{metric1}\ is in fact K\"ahler, as claimed.

Suppose we choose a different gauge-fixing, $\phi^0=1$, and work with affine local coordinates:
\be
\tilde z^i = \phi^i/\left(\phi^0\right)^{Q_i/Q_0},\qquad \tilde\s = \s/\left(\phi^0\right)^{Q_\S/Q_0}.
\ee
In this set of coordinates, $H\neq0$ and the space is non-K\"ahler. This can happen precisely because of the unusual metric transformation properties found in~\C{GLSMtransform}. We are perfectly free to work with this metric and $B$-field as long as we keep track of the unusual patching conditions. Indeed the GLSM naturally gives us this form for $G$ and $B$ in a manifestly $(0,2)$ supersymmetric way. However, if we want to assign a conventional geometry to this NLSM, we need to understand how to define a conventional metric.

\subsection{Defining an invariant metric}\label{ss:invariant}

The projective coordinates naturally parameterize a $\CC^*$-bundle over the target, but we are really only interested in the  quotient of this total space by the $\CC^*$-action. Let us work with sections of this bundle that define local patches. We cover the target space with open neighbourhoods $U_\al = \{x^\a\neq0\}$, and within each such set define local coordinates:
\be
Z^I_{(\a)} = x^I /\left(x^\a\right)^{Q_I/Q_\a}.
\ee
On the intersections $U_\ab =U_\al\cap U_\bt$, we relate the local coordinate systems by
\be
Z^I_\al = \left(\exp{iQ_I\La_\ab}\right) Z^I_\bt,\label{gluing}
\ee
where the (holomorphic) gluing functions $\La_\ab$ are naturally identified with the $\CC^*$ gauge transformations:
\be
\La_{\ab} = {i\over Q_\a}\log Z^\a_\bt.
\ee
These transition functions define the bundle over the target space. Note that $A$, defined implicitly by~\C{constraint}, is not globally defined. On $U_\ab$, $A$ transforms according to
\be
A_\al = A_\bt + A_\ab,\qquad \textrm{with}\qquad A_\ab = {\La_\ab -\bar\La_\ab\over2i}.\label{Aab}
\ee
The one-form $\del A_\al$ acts as a (holomorphic) connection on our line bundle with $\del\bar\del A_\al=\del\bar\del A_\bt$ its invariant curvature two-form. Finally, it will be useful to note that the quantity $T$, defined in~\C{defineT}, shifts in a way similar to $A_\al$ except
\be
T_{(\a\b)} \equiv T_{(\a)} - T_{(\b)} = -{\tilde\cA\La_\ab + Q_\s^2 \bar\La_\ab \over4\pi}.\label{Tab}
\ee
Note that
\be
\del T_\ab = -i{\tilde\cA\over2\pi}\del A_{\ab} ,\qquad \textrm{and}\qquad \bar\del T_\ab = i{Q_\s^2\over2\pi}\bar\del A_\ab.\label{delT}
\ee

From our earlier discussion, we expect $G_\al$ and $B_\al$ to have anomalous transformations on the overlaps $U_\ab$. Indeed, by examining the line element
\be
ds^2 = G^\al_{I\Jbar}dZ^I_\al dZ^\Jbar_\al,
\ee
and applying the transformations~\C{gluing}-\C{Tab}, we find that the metric $G^\al$ has an anomalous transformation law:
\be \label{GLSMtransform}
G^{(\a)}_{I\Jbar} = G^{(\b)}_{I\Jbar} -{\tilde\cA\over2\pi} \left(\del_I\left( A_{(\b)} +A_{(\a\b)} \right)\del_\Jbar\left(A_{(\b)}+A_{(\a\b)}\right) -\del_I A_{(\b)}\del_\Jbar A_{(\b)}\right).
\ee
This is problematic if we wish to interpret $G$ as a metric since the line element $ds^2$ would not be an invariant. However,~\C{GLSMtransform}\ suggests a natural resolution to this puzzle because the quantity
\be\label{Ghat}
\widehat G_{I\Jbar} = G_{I\Jbar} +{\tcA\over2\pi}\del_I A \del_\Jbar A
\ee
does define an invariant line element. In particular,
\be\label{intersections}
\widehat G^\al_{I\Jbar} = G^{(\a)}_{I\Jbar} +{\tilde\cA\over2\pi} \del_I A_{(\a)}\del_\Jbar A_{(\a)} = G^{(\b)}_{I\Jbar} +{\tilde\cA\over2\pi} \del_I A_{(\b)}\del_\Jbar A_{(\b)} =\widehat G^\bt_{I\Jbar}
\ee
is a candidate metric for our target spaces.

\subsubsection{An alternate derivation of $\widehat{G}$}\label{ss:altder}

Let us derive the result~\C{Ghat}\ for the metric from another, more systematic, approach. The idea is to consider the holomorphic vector field
\be
L = \sum_I Q_I X^I \del_I
\ee
that generates the $\CC^*$-action. Next, consider the contraction of this vector field with the fundamental form $J$ associated to $G$. If the $(0,1)$-form
\be
\bar V \equiv i_L\left(-iJ\right)
\ee
is non-zero, then the metric $G$ will not naturally descend to the quotient space. However, the improved fundamental form
\be
{\widehat J} = J - V \left(i_L V\right)^{-1}\bar V
\ee
will be invariant by construction, and we associate the metric ${\widehat G}$ with this improved fundamental form.

In order to compute $\bar V$, it will help to recall that $T=\Th+iR$, defined in~\C{defineT}, is a function only of $\S$ and $\bar\S$; in particular,
\be
\del_I T = i\left({Q_P-Q_{\hat\G}\over4\pi}\right){1\over\S} \dd_{I\S}.
\ee
Furthermore, we recall that
\be
\del_I A = \D\left(\del_I R -Q_I \bar X_I e^{2Q_I A}\right) = \D\left({Q_P\over4\pi\S}\dd_{I\S} -Q_I \bar X_I e^{2Q_I A}\right),
\ee
which leads to
\bea
L^I \del_I A &=& \sum_I Q_I X^I \D \left(\del_I R -Q_I \bar X_I e^{2Q_I A}\right) \non\\
&=& Q_\S \D \S \del_\S R - \hlf\D\left(\D^{-1} + {\cA\over4\pi} \right)\\
&=& \D{Q_\S Q_P \over4\pi} -\hlf -\D {\cA\over8\pi} \non\\
&=& -\hlf \non,
\eea
because $\cA=2Q_PQ_\S$. Now we can evaluate the components of the connection $\bar V$:
\bea
V_{\Jbar} &=& L^I G_{I\Jbar} \non\\
&=& \sum_I L^I \left(e^{2Q_I A}\dd_{I\bar J} -2\del_I A \D^{-1} \del_{\bar J}A + i\del_I A\del_{\bar J}\bar T -i\del_I T\del_{\bar J}A +{\dd_{I\S}\dd_{\bar J\bar\S}\over8\pi|\S|^2}\right)\\
&=& Q_J X_\Jbar e^{2Q_J A} +\left(-2\left(L^I \del_I A\right)\D^{-1} -iL^I\del_I T\right)\del_\Jbar A +\left(i(L^I\del_I A)\del_{\bar\S}\bar T +{Q_\S\over8\pi\bar\S}\right) \dd_{\Jbar\bar\S} \non\\
&=& Q_J X_\Jbar e^{2Q_JA} +\left(\D^{-1} +Q_\S {Q_P-Q_{\hat\G}\over4\pi}\right) \del_\Jbar A +\left(-Q_P+Q_{\hat\G} +Q_\S\over8\pi\bar\S\right)\dd_{\Jbar\bar\S} \non\\
&=& Q_J X_\Jbar e^{2Q_JA} -{Q_P\over4\pi \bar\S}\dd_{\Jbar\bar\S} +\D^{-1}\del_\Jbar A + {\tilde\cA\over4\pi}\del_\Jbar A \non\\
&=& {\tilde\cA\over4\pi}\del_\Jbar A \non,
\eea
where we used the relations $Q_P+Q_{\hat\G}+Q_\S=0$, and $\tcA=Q_\S(Q_P-Q_{\hat\G})$. Finally, we compute
\be
L^I V_I = {\tilde\cA\over4\pi}\left(L^I\del_I A\right) = -{\tilde\cA\over8\pi}.
\ee
Therefore, the correct invariant metric is
\bea
{\widehat G}_{I\Jbar} = G_{I\Jbar} - V_I \left(L^K V_K\right)^{-1} V_\Jbar = G_{I\Jbar} + {\tilde\cA\over2\pi}\del_I A\del_\Jbar A,
\eea
as claimed in~\C{Ghat}.

\subsection{The associated $H$-flux}\label{ss:Hflux}

Now let us consider the $B$-field, and look for an invariant $H$ associated to it. Recall the relations~\C{Bcurvature}\ and~\C{dBandJ},
\be
dB_\al = i\left(\del T_\al +\bar\del\bar T_\al\right)\del\bar\del A_\al = i(\bar\del-\del)J_\al,   \label{dBandJ2}
\ee
where $J$ is the natural $(1,1)$-form associated to $G$. This relation implies
\be
2i\del\bar\del J_\al = d^2B_\al =0.
\ee
However if instead we consider $\widehat J$, the fundamental form associated to $\widehat G$, then we can define
\bea
H_\al &\equiv& i\left(\bar\del-\del\right)\widehat J_\al \non\\
&=& i\left(\bar\del -\del\right)\left(J_\al +i{\tilde\cA\over2\pi}\del A_\al \bar\del A_\al\right) \label{H} \\
&=& dB_\al +{\tilde\cA\over2\pi} \left(\bar\del-\del\right)A_\al \del\bar\del A_\al\non.
\eea
On the overlaps $U_\ab$, $dB_\al$ shifts by
\bea
dB_{\ab} &=& i\left(\del T_{(\a\b)} +\bar\del \bar T_{(\a\b)}\right)\del\bar\del A_{(\b)} \non\\
&=& -i{\tilde\cA\over4\pi}\left(\del\La_{\ab} +\bar\del\bar\La_\ab\right)\del\bar\del A_\bt \label{dBab}\\
&=& -i{\tilde\cA\over2\pi}\,d\left(\Re\La_{\ab}\right) \del\bar\del A_\bt. \non
\eea
The Chern-Simons-like form appearing in~\C{H}\ shifts in exactly the opposite way, which follows from~\C{Aab}, so that $H_\al=H_\bt$. $H$ is therefore an invariant three-form, as desired, which satisfies the Bianchi identity:
\be\label{definedH}
dH = {\tilde\cA\over\pi}\ \del\bar\del A \wedge \del\bar\del A.
\ee
It would be very interesting to understand the relation between the right hand side of~\C{definedH}\ and $\tr R\wedge R$ of the metric $\widehat G$ computed with an appropriate connection.

\subsubsection{An improved $B$-field}\label{ss:imprvdB}

Given the transformation~\C{dBab}\ for $dB$, it is clear that $B$ must transform by
\be
B_\ab = -i{\tilde\cA\over2\pi}\,\left(\Re\La_{\ab}\right) \del\bar\del A_\bt + \textrm{exact}.\label{Bvar}
\ee
Our intuition from the Green-Schwartz mechanism tells us that the exact term in the above equation should vanish identically. Sadly, this is not the case for the $B$-field~\C{B}\ that follows directly from the GLSM construction. However, just as we were able to construct an improved metric, $\widehat G$, so too can we construct a $\widehat B$ that transforms like~\C{Bvar}\ but without the additional exact term.

For this discussion, it will help to introduce the complex quantity
\be
M_\al \equiv T_\al - i{Q_\S^2\over2\pi}A_\al,
\ee
which shifts by
\be
M_\ab \equiv M_\al - M_\bt = T_\ab -i{Q_\S^2\over2\pi}A_\ab = -{\tilde\cA+Q_\S^2\over4\pi}\La_\ab\label{M}
\ee
on overlaps $U_\ab$. Note that this transformation property implies $\bar\del M$ is an invariant:
\be
\bar\del M_\al = \bar\del M_\bt + \bar\del M_\ab = \bar\del M_\bt,
\ee
and furthermore, since $\del\bar\del T=0$,
\be
\del\bar\del M_\al = -i{Q_\S^2\over2\pi}\del\bar\del A_\al.
\ee
With generous use of the relation~\C{delT}, we can now compute the variation of $B$:
\bea
B_\ab &=& 2i\Th_{\ab} \del\bar\del A_\bt +i\del A_\ab\bar\del M_\bt +i\del\bar M_\bt \bar\del A_\ab \non\\
&=& 2i{Q_\S Q_\G\over2\pi}(\Re\La_\ab)\del\bar\del A_\bt +{1\over2}\del \La_\ab \bar\del M_\bt +{1\over2}\bar\del\bar\La_\ab \del\bar M_\bt \\
&=&-i{\tilde \cA\over2\pi}(\Re\La_\ab)\del\bar\del A_\bt + \hlf\del\left(\La_{\ab}\bar\del M_\bt\right) +\hlf\bar\del\left(\bar\La_\ab\del \bar M_\bt\right), \non
\eea
where we have used $\tilde \cA = Q_\S(Q_P-Q_\G) = -Q_\S^2 -2Q_\S Q_\G$. As expected, $B$ shifts according to~\C{Bvar}, but thanks to~\C{M}\ we may write this in the form
\be
B_\ab = -i{\tilde \cA\over2\pi}(\Re\La_\ab)\del\bar\del A_\bt -{2\pi\over\tilde\cA +Q_\S^2}\left(\del\left(M_\ab\bar\del M_\bt\right) +\bar\del\left(\bar M_\ab\del \bar M_\bt\right)\right).
\ee
Now we see that
\be
\widehat B_\al = B_\al + {2\pi\over\tilde\cA +Q_\S^2}\left(\del\left(M_\al\bar\del M_\al\right) +\bar\del\left(\bar M_\al\del \bar M_\al\right)\right)
\ee
has the desired transformation property, namely:
\be
\widehat B_\ab = -i{\tilde \cA\over2\pi}(\Re\La_\ab)\del\bar\del A_\bt.
\ee

\subsection{A class of examples}\label{ss:examples}

Rather than clutter the resulting formulae with $Q$ factors, let us consider the class of models where
\be Q_i=Q_\S=1,\ee
and $\cA>0$. An appropriate set of left-moving charges can always be found which satisfy the gauge anomaly cancelation condition~\C{uvanomaly}\ in the UV theory if $Q_P>0$. As a nice specific case, consider the model described in Section~\ref{ss:manyfields}\  that  gives $S^4$ without accounting for the metric corrections. That model is of this form with $Q_1=Q_2=Q_\S=1$ and no left-moving Fermi superfields; the anomalous set of massive fields integrated out have charges $Q_P=1$ and $Q_\hG=-2$. For that example, $\cA=2$ and $\tilde\cA=3$.

\begin{figure}[ht]
\begin{center}
\[
\mbox{\begin{picture}(230,180)(40,20)
\includegraphics[scale=1]{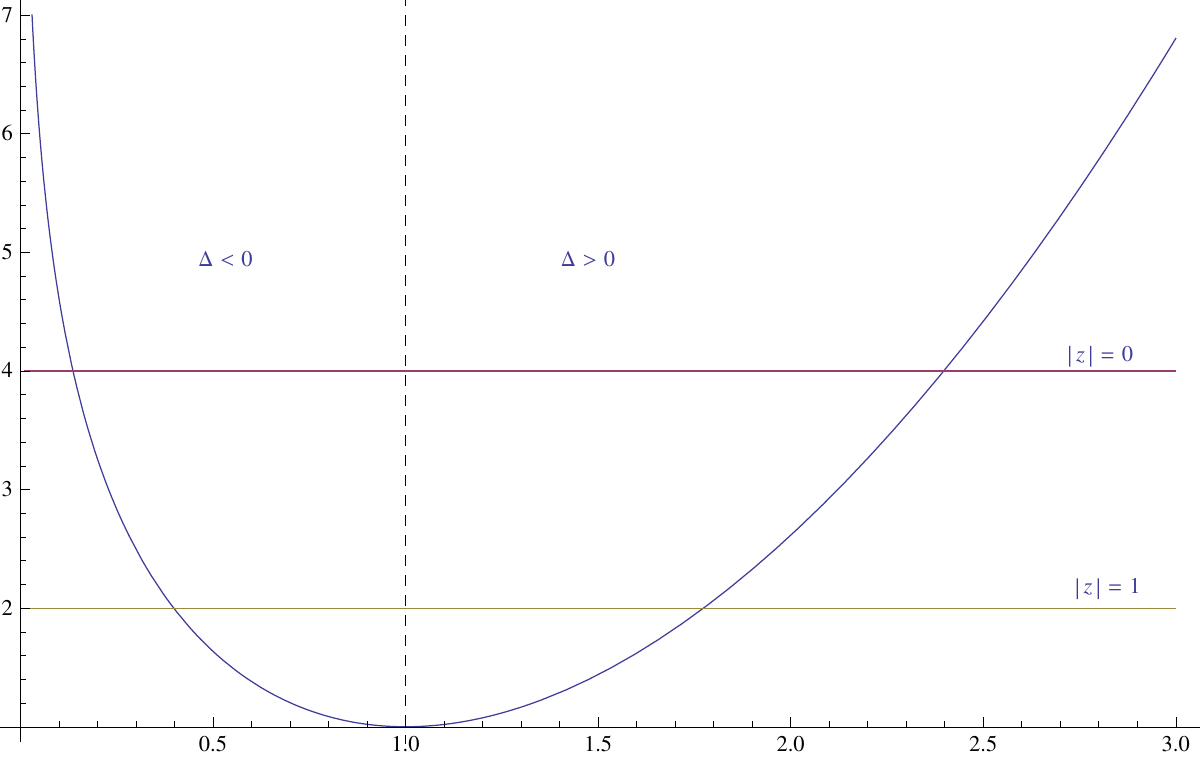}
\end{picture}}
\]
\vskip 0.2 in \caption[{\it A plot depicting solutions of~\C{defineA}\ we increase $|z|$ from $0$ to $|z|_{max}$.}]{\it A plot depicting the solutions of~\C{defineA}\ as we increase $|z|$ from $0$ to $|z|_{max}$, with the latter value corresponding to the unique minimum. } \label{rangeA}
\end{center}
\end{figure}

It is convenient to work in the preferred patch where we set $\s=1$. The resulting metric takes the form
\be\label{samplemetric}
{\widehat G}_{i\jbar} = e^{2 A}\dd_{i\jbar} - 2 \D\left(1-{\tilde\cA\over4\pi}\D\right)  \bar z_i z_\jbar e^{4A},
\ee
where $A$ satisfies
\be\label{defineA}
e^{2A}(1+|z|^2)-{\cA\over4\pi}A =r,
\ee
with
\be
\D = \left(2e^{2A} \left(1+ |z|^2\right) -{\cA\over4\pi} \right)^{-1} = \left(2r+{\cA\over4\pi}(2A-1)\right)^{-1}.
\ee
To orient ourselves, note that~\C{defineA}\ is equivalent to~\C{singlelog}\ after setting $A=\log|\phi|$. We can view the solutions to the equation in the following way: take $|z|$ as an input. For any given $|z|$, there will generically be two solutions for $A$ as long as $r> r_{min}$. This is pictured in Figure~\ref{rangeA}. The value for $r_{min}$ depends on $|z|$ via the relation~\C{definermin}. As we increase $|z|$, $r_{min}$ increases until $r=r_{min}$ at $|z|=|z|_{max}$:
\be
1+|z|_{max}^2 = {\cA \over 8\pi} e^{{8\pi r\over \cA}-1}.
\ee
We note that taking $\cA\rightarrow 0$ sends $|z|_{max} \rightarrow \infty$, which is appropriate for the projective space limit. At $|z|_{max}$, there is a unique solution for $A$ corresponding to the minimum of Figure~\ref{rangeA}. At this critical point, $A$ is determined in a very nice way:
\be
A_{crit}= \left({1\over 2} -{4\pi r\over \cA}\right) .
\ee
It is very useful to note that $\Delta^{-1}$ is just the derivative of~\C{defineA}\ with respect to $A$, and so corresponds to the slope of the graph in Figure~\ref{rangeA}. To the right of the minimum, $\Delta^{-1}>0$, while to the left $\Delta^{-1}<0$ with a zero at the minimum. This is very good news! The two branches of solutions for $A$ were the cause of the supersymmetry puzzles described in Section~\ref{ss:modulispace}. Whatever happens, the conformal factor of the metric~\C{samplemetric}\ will diverge at the minimum of Figure~\ref{rangeA}\ disconnecting these two regions. This is what  singles out a unique solution of the $D$-term equation under the $\CC^\ast$ gauge action.

It is curious that if we approach the critical point from the left branch for $A$ with $\Delta<0$ then the metric~\C{samplemetric}\ is manifestly positive. For very large $r$, $A \sim -{ 4\pi r \over \cA}$ plus small corrections. This means the classical leading term in the metric~\C{samplemetric}\ is very small. This is the region where we do not trust our one-loop effective action, though it is intriguing that the metric is positive with a divergence when one hits the critical point at $|z|=|z|_{max}$. It would be very interesting to find an interpretation of this branch.

\begin{figure}[ht]
\begin{center}
\[
\mbox{\begin{picture}(230,180)(40,20)
\includegraphics[scale=1]{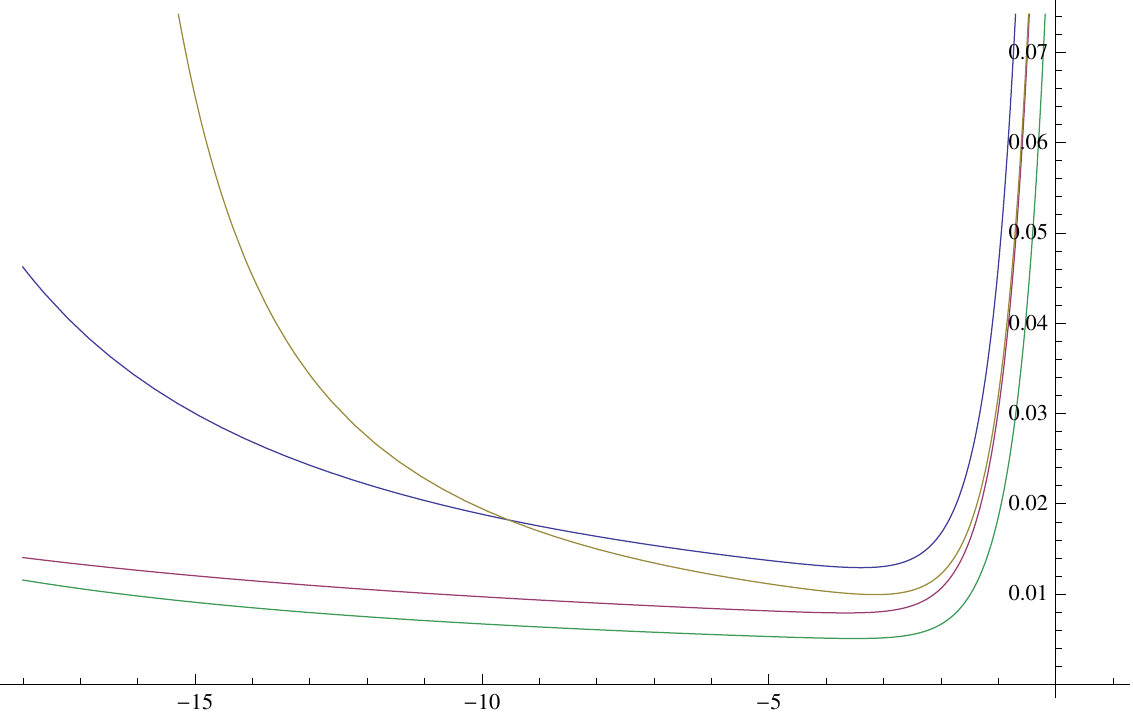}
\end{picture}}
\]
\vskip 0.2 in \caption[{\it A plot of the conformal factor appearing on the left hand side of~\C{vanishingcond}\ vs. $A$}]{\it A plot of the conformal factor appearing on the left hand side of~\C{vanishingcond}\ versus $A$ for the examples $(r=4, \cA=2)$ (blue in the graph), $(r=6, \cA=2)$ (red), $(r=6, \cA=4)$ (yellow) and $(r=10, \cA=4)$ (green). The $x$-axis corresponds to vanishing conformal factor. If one extends the graph sufficiently far to the left, all the conformal factors diverge. Each conformal factor reaches some finite value to the right. } \label{conformalfactor}
\end{center}
\end{figure}

However, we want to approach from the far right where the classical metric is large and we trust our one-loop effective action. In the region to the far right, $\Delta>0$ and small if $r$ is very large. What we need to check is whether the metric~\C{samplemetric}\ encounters a singularity before we hit  $|z|_{max}$.  Because of the sign of $\Delta$, this is possible. To make our life easier, we will rotate coordinates so that
\be
(z_1, z_2, z_3, \ldots) = (z, 0, 0, \ldots).
\ee
We then need to check whether the conformal factor for the metric~\C{samplemetric}\ can vanish for some $|z|<|z|_{max}$. This requires
\be\label{vanishingcond}
1 - 2 \D\left(1-{\tilde\cA\over4\pi}\D\right) |z|^2 e^{2A} =0.
\ee
This is a transcendental equation. To see if the conformal factor vanishes, it is easiest to plot some examples. Figure~\ref{conformalfactor}\ contains a plot of the left hand side of~\C{vanishingcond}; in the examples plotted, we see that the conformal factor becomes small but does not vanish. As $|z|$ approaches $|z|_{max}$, it diverges to the far left of the plot. This behavior is quite remarkable  because generating a large variation of the conformal factor is a hint that string solutions built from these spaces might exhibit hierarchies. It is worth noting that without the correction proportional to $\tcA$ in~\C{samplemetric}, the metric would have vanished with a collapsed circle before we reach the critical point of Figure~\ref{rangeA}.

We can prove positivity of the left hand side of~\C{vanishingcond}\ as follows: using the $D$-term constraint~\C{defineA}, we see that
\be
|z|^2 e^{2A} = \left( {|z|^2 \over 1+|z|^2}\right)\left( {1 \over 2 \Delta}+ {\cA \over 8\pi}\right) < \left( {1 \over 2 \Delta}+ {\cA \over 8\pi}\right)
\ee
for $\Delta>0$. Plugging this into the left hand side of~\C{vanishingcond}\ gives
\be
1 - 2 \D\left(1-{\tilde\cA\over4\pi}\D\right) |z|^2 e^{2A} = {\tcA \cA \Delta^2 \over 16\pi^2} + {\Delta \over 8\pi} \left(2\tcA - \cA\right) >0,
\ee
demonstrating that the metric is strictly positive.

It is going to be very interesting to understand the properties of the metric~\C{samplemetric}\ in more detail, including the behavior of the conformal factor. Here, however, it will suffice to note that there is a boundary at $|z|_{max}$ at which the metric diverges. We would like to see whether this boundary is at finite distance. Near the boundary, the metric is dominated by
\be
{\widehat G}\sim {\tilde\cA\over2\pi}\D^2  |z|^2 e^{4A} d\bar z dz + \ldots,
\ee
with omitted terms non-singular at $|z|=|z|_{max}$. We note that
\be
d|z|^2 = - dA  e^{-2A} \left\{ {\cA\over  2\pi} \left(A - A_{crit} \right)\right\},
\ee
and that
\be
\Delta^{-1} = {\cA \over 2\pi} \left( A -A_{crit} \right).
\ee
These relations permit us to express the metric in terms of $A$ near the boundary at $A_{crit}$,
\bea
{\widehat G} &\sim &  {\pi \tilde\cA\over 2\cA^2 \left( A -A_{crit} \right)^2}  \left\{ {\cA\over  2\pi} \left(A - A_{crit} \right)\right\}^2 \left(dA\right)^2   + \ldots, \non\\
&\sim &  {\tcA \over 8\pi} {\left(dA\right)^2 } +\ldots.
\eea
The point $A=A_{crit}$ is therefore  at finite distance. Our target manifold has developed a finite distance boundary at which the scale factor diverges.

Specifically, the metric for the angular direction for $z$ (rather than the radial direction $|z|$) diverges at the boundary in a way highly reminiscent of the metric for the $SU(2)/U(1)$ WZW model~\cite{Bardakci:1990ad}. It seems quite possible that the metric near the boundary is regular after T-dualizing this circle direction, leading to the fascinating possibility that this space is a kind of non-geometric T-fold; that possibility will be explored further elsewhere.

\subsection{The general case}\label{ss:gencase}

Now that we have a basic understanding of how integrating out an anomalous multiplet affects the low energy geometry, let us take a brief look at a more general class of examples. We will let the gauge group have rank $n$, and use $a,b,\ldots$ to label the different $U(1)$ factors. We will also include multiple $\S^m$ fields, with charges $Q^a_m$. Each $\S^m$ give a large mass to a pair of $(P^m,\G^m)$ fields that we integrate out. The constraint among the charges is $Q_m^a + Q_{P^m}^a + Q_{\G^m}^a=0$. Many of the formulae from Section~\ref{ss:singlesigma} generalize straightforwardly. We will still use the collective notation
\be
X^I = (\Phi^i,\S^m)
\ee
to denote the complete set of chiral fields. We define
\be
R^a(\S) = r^a + \sum_m{Q^a_{P^m}\over2\pi}\log|\S^m|,\qquad \textrm{and}\qquad \Th^a(\S) = {\th^a\over2\pi}+\sum_m{Q^a_{\G^m}\over2\pi}\Im\left(\log\S^m\right).
\ee
Under a gauge transformation, the low energy action changes by
\be
\Th^a(\S) \rightarrow \Th^a(\S) - {\tilde\cA^{ab}\over4\pi} \Re(\La^b),
\ee
where
\be
\tilde\cA^{ab}=\sum_m Q^a_m Q^b_m +\cA^{ab} = \sum_I Q^a_I Q^b_I - \sum_\a Q_\a^a Q_\a^b.
\ee
Integrating out $V^a_-$ enforces the constraints
\be
\sum_I Q_I^a|X^I|^2 e^{2Q^b_I A^b} - {\cA^{ab}\over4\pi} A^b = R^a(\S), \label{constraint2}
\ee
and we are left with a $(0,2)$ non-linear sigma model, characterized by
\bea
K_I = X_I e^{2Q_I^a A^a} + 2i\Th^a(\S)\del_I A^a + \dd_{Im}{\log\bar\S_m\over8\pi\S^m} .
\eea
The constraints~\C{constraint2}\ imply
\bea
&&\del_I A^a =  \D^{ab}\left(\del_I R^b - Q^b_I \bar X_I e^{2Q_K^c A^c}\right), \\
&&\D^{ab} = {\del A^a\over \del R^b} = \left(2\sum_I Q_I^a Q_I^b|X^I|^2 e^{2Q_I^c A^c}  - {\cA^{ab}\over4\pi}\right)^{-1},
\eea
and these allow us to compute the $(0,2)$ metric
\be
G_{I\bar J} = e^{2Q_I^a A^a}\dd_{I\bar J} -2\del_I A^a \left(\D^{-1}\right)^{ab} \del_{\bar J}A^b + i\del_I A^a\del_{\bar J}\bar T^a -i\del_I T^a\del_{\bar J}A^a +{\dd_{Im}\dd_{\bar J\bar m}\over8\pi|\s^m|^2},
\ee
and $B$-field,
\be
B = 2i\Th^a\,\del\bar\del A^a +i\del A^a\bar\del T^a -i\bar\del A^a\del \bar T^a.
\ee
These are the objects that transform anomalously. In this case, the natural expression for an invariant metric takes the form
\be
\widehat G_{I\Jbar} = G_{I\Jbar} +{\tcA^{ab}\over2\pi} \del_I A^a \del_\Jbar A^b,
\ee
with associated $H$-flux,
\be
H = i(\bar\del-\del)\widehat J = dB + {\tcA^{ab}\over2\pi} \left(\bar\del-\del\right) A^a \del\bar\del A^b,
\ee
that satisfies the Bianchi identity:
\be
dH = {\tcA^{ab}\over\pi} \del\bar\del A^a \wedge \del\bar\del A^b.
\ee
In general, these spaces will contain boundaries and flux. As shown previously in~\cite{Quigley:2012gq}, if any $Q^a_m=0$ there will be NS-branes as well. We look forward to further studying these spaces which are simply given to us from chiral gauge theory.
Clearly much remains to be explored.

\chapter{Conclusions}\label{ch:conc}

In this thesis, we have generalized the standard GLSM construction to include field-dependent FI parameters:
\be
S_{FI} = {1\over4}\int\d^2x\d\th^+\, T^a(\F)\U^a_-\Big|_{\thbar^+=0} +c.c.\,.
\ee
As explained in Section~\ref{sec:FIs}, the main effect of these new couplings is to deform the low energy target spaces, $\M$, away from being K\"ahler by inducing non-trivial $H$-flux threading space. The form of $H$ is given by
\be
H = d\Th^a\wedge F^a,
\ee
where $\Th^a = {\rm Re}\,T^a$ and $F^a \in H^2(\M,\Z)$. There is also a corresponding deformation of the $D$-term constraint:
\be\label{eq:Dterm}
\sum_i Q_i^a|\f^i|^2 = R^a(\f),
\ee
where $R^a = {\rm Im}\,T^a$.

\subsubsection{Gauge-invariant $T^a$}
The most straightforward option for $T^a(\F)$ is to choose a gauge-invariant polynomial in the charged chiral fields. This is only possible when there are both positive and negative charged fields in the model. The $D$-term constraint~\C{eq:Dterm}\ then leads to non-compact target manifolds. This was explored in Section~\ref{sec:gaugeinvariant}. The fact that $T^a$ are continuously deformable in these cases means that $H$, while non-zero, will be trivial in cohomology

A more interesting possibility for gauge-invariant $T^a$ was explored in Chapter~\ref{ch:Neutral}. Instead of neutral combinations of charged fields, we considered
\be\label{eq:Ta}
T^a = {N^a\over2\pi i}\log\S,
\ee
where $\S$ is a neutral chiral field. Single valuedness of the path integral requires that $N^a\in\Z$, and so the flux is quantized. Models of this type also contained explicit magnetic sources for $H$ (NS-branes). Among the examples we studied in Section~\ref{sec:exam}, one was a conformal model that yielded the NS5-brane solution of string theory (smeared over a transverse circle). The metric of this example has a ${1\over r}$ singularity at the origin, and this can be traced to the divergence of the dilaton at the location of the 5-brane source. It will be interesting to see if non-perturbative effects localize these solutions in the circle direction, along the lines of~\cite{Tong:2002rq}.

Another key observation of Chapter~\ref{ch:Neutral}\ is that the FI parameters~\C{eq:Ta}\ can be generated along branches of standard GLSMs (with constant FI parameters), by integrating out non-chiral massive fermions. A logarithmic coupling in a fundamental theory is problematic, because it is difficult to define the theory quantum mechanically at the points where the log becomes singular. These are the locations of the NS-branes.  However, when the logarithms appear in a low energy effective description of some UV complete theory, then the singular loci have a nice interpretation as points in moduli space where massive degrees of freedom that have been integrated out are becoming light.

\subsubsection{Non-invariant $T^a$}

We considered a much more exotic set of FI couplings in Section~\ref{sec:compact}, with
\be
T^a = {1\over4\pi i} \sum_i N^a_i \log\F^i,
\ee
but now the chiral fields $\F^i$ charge non-zero charges $Q^a_i$. These FI parameters shift under gauge transformations, destroying the classical gauge invariance of the action. However, the gauge symmetry may be restored if the theory also has an anomalous spectrum, with an anomaly coefficient
\be
\cA^{ab} = \sum_i Q^a_i Q^b_i - \sum Q_\a^a Q_\a^b,
\ee
that satisfies
\be
\cA^{ab} = \sum Q^{(a}_i N^{b)}_i.
\ee
These models appear even more problematic to properly define quantum mechanically than those of Chapter~\ref{ch:Neutral}, since we are now playing two very sick features, singular superpotentials and quantum gauge anomalies, off of each other to produce a sensible theory.

Fortunately, we saw in Chapter~\ref{ch:Charged}\ that these interactions can also be generated from a UV complete theory, where now massive fermions with chiral charges are integrated out. The two sick features mentioned earlier are generated in tandem to ensure that the low energy description of the theory remains sensible.
We also saw that the new couplings in the effective action, reflecting the gauge anomalous nature of the massive multiplet, lead to rather interesting behavior for the low energy metric $\widehat G$. The scale factor for a circle in the space shrinks down to a fairly small but non-vanishing value, determined by a transcendental equation; it then begins to grow until it  diverges at a boundary located at finite distance in the target space. This large variation in the scale factor suggests that string solutions built from these spaces might exhibit hierarchies.
For the example that would have given $S^4$ without including these corrections, we find that
the sphere is roughly cut in half giving a $4$-ball. Near the boundary, the form of the metric  suggests that we might want to study the theory in T-dual variables to find a weakly coupled description. That possibility will be explored elsewhere.

In addition to producing a metric on the target space, the linear model also yields an $H$-flux that should satisfy the heterotic Bianchi identity:
\be\label{bianchi}
dH  = {\alpha'\over 4}\left[ \tr R_+\wedge R_+ - \tr F\wedge F \right],
\ee
where $R_+$ is curvature of the spin connection twisted by $H$-flux.\footnote{See~\cite{Becker:2009df}\ for an explanation about why there is a preferred gravitational connection, $\Omega_+$, used to evaluate the Chern-Simons forms and curvatures.  See~\cite{Hull:1986xn}\ for the $(0,1)$ superspace counter-terms associated with these Chern-Simons corrections; a recent discussion is given in~\cite{Melnikov:2012cv}. }
The corresponding gauge transformations of the $B$-field lead to subtleties in defining the NLSM quantities, but it is clear that the non-anomalous GLSM produces a solution to the Bianchi identity.   To see this directly at the NLSM level will require a better understanding of the boundary.  It might be possible to find similar ``quantum quotient'' constructions in type II string backgrounds with orientifold planes and D-branes, which can also modify Bianchi identities.

Boundaries appear in several settings when studying string compactifications. For example, a strong coupling limit of the $E_8\times E_8$ heterotic string compactified on $\M$ develops a boundary. In that limit, the appropriate description is heterotic M-theory on $\M\times S^1/\Z_2$~\cite{Horava:1996ma}. A closer analogue for the boundary we see is found in the geometry of gauged WZW models. The simplest case is the $SU(2)/U(1)$ WZW model~\cite{Bardakci:1990ad}. The geometry of the covering space is $S^3$ so a straight geometric quotient would give
\be
{S^3 \over S^1} \sim S^2.
\ee
Because of the presence of $H$-flux threading the $S^3$, the metric on the quotient space actually degenerates at the equator of what should be $S^2$ producing a curvature singularity. This degeneration changes the topological type of the target manifold from $S^2$ to a disk. There is simply no room for three-form flux on $S^2$ so this topology change is the only residue of the flux present on the covering space. This has some similarities to what we see in models with a single massive anomalous pair, although our cases are typically not conformal.

\subsubsection{Outlook}
The larger picture that emerges for heterotic compactifications involves three basic building blocks. The first are brane sources obtained by integrating out non-anomalous multiplets. The second are boundaries and fluxes from anomalous massive multiplets. The final ingredient is the gauge bundle specified by the choice of left-moving fermions. If we consider combinations of anomalous and non-anomalous massive multiplets, 
we will generally find target manifolds that are non-K\"ahler spaces with boundaries, branes and $H$-flux. General combinations of these ingredients should produce a large landscape of heterotic quantum field theories.

We expect compact conformal models to appear via complete intersections obtained by turning on additional superpotential or $E$-couplings. These are the models that can potentially be used as string vacua. There are many directions to pursue. A sample of questions include: what are the precise conditions for conformal invariance? This could be investigated perhaps along the lines of~\cite{Koroteev:2010ct, Cui:2011rz, Cui:2011uw, Adams:2012sh}. What are the spacetime spectra and moduli spaces for these models? Are these consistent with the results of~\cite{Melnikov:2011ez}\ computed at large radius? How many vacua exist for massive models? What is the structure of the ground ring? For a discussion of heterotic ground rings and quantum sheaf cohomology,  see~\cite{Adams:2003zy, Katz:2004nn,Guffin:2007mp, Adams:2005tc, Sharpe:2006qd, Tan:2006qt,  Tan:2007bh, Donagi:2011va, Donagi:2011uz}.
Does a weakly coupled description of the high curvature boundary exist? Such a description might follow from a mirror description which generalizes~\cite{Hori:2000kt, Adams:2003zy}. For a review of $(0,2)$ mirror symmetry, see~\cite{Melnikov:2012hk}. What is the right way to describe these target manifolds? Can threshold corrections be computed in these models, perhaps along the lines of~\cite{Carlevaro:2012rz, Carlevaro:2009jx}? Can elliptic genera be computed for these generically non-K\"ahler spaces, perhaps along the lines of~\cite{ Adams:2009zg}\ or \cite{Benini:2013nda}? Finally, a connection between $(0,2)$ theories and 4-manifolds has emerged recently~\cite{Gadde:2013sca}, and it would be interesting to see exactly how this class of models fits into that framework. Clearly, many questions remain and we have only scratched the surface of this fascinating set of models.

\appendix
\chapter{Supergraph computations}
\section{Feynman rules for the $P$ and $\G$ superfields}\label{ap:superFeynman}

Here we derive the Feynman rules needed to compute the one-loop effective action in Chapter~\ref{ch:Charged}. Before we present the derivation, let us establish some conventions.

\subsection{Conventions}\label{sec:conv}

A point in $(0,2)$ superspace will be denoted $z=(x^+,x^-;\th^+,\thbar^+)$; we will denote the difference between two points by $z_{12}\equiv z_1-z_2$. A delta-function on all of superspace is given by
\be
\dd^4(z_{12}) \equiv \dd^2(x_{12})\dd^2(\th_{12}),
\ee
where $\dd^2(x_{12})$ is the usual delta-function in two-dimensions, and the Grassmann delta-function takes the usual form:
\be \dd^2(\th_{12}) = \th_{12}^2 = \th^+_{12}\bar\th^+_{12}. \ee
Because $x^{\pm}\equiv \hlf(x^0\pm x^1)$, the non-zero components of the Minkowski metric and epsilon tensor are
\be
\eta_{+-} = \e_{-+} =-2,\qquad\qquad \eta^{+-}=\e^{+-} =-\hlf.
\ee
Finally, for performing Fourier transforms, we note that
\be
\tilde f(p)\equiv \int d^2x\ e^{-ip\cdot x}f(x),\quad f(x) \equiv \int {d^2p\over(2\pi)^2}\, e^{ip\cdot x} \tilde f(p),\quad \int d^2x\ e^{-ip\cdot x} = (2\pi)^2\dd^2(p).
\ee
This corresponds to the replacements
\be
-i\del_\pm \rightarrow p_\pm,\qquad D_+ \rightarrow \del_{\th^+} +\thbar^+p_+,\qquad \bar D_+ \rightarrow -\del_{\thbar^+} - \th^+ p_+.
\ee

\subsection{Loop integrals with IR cutoffs}\label{sec:Loops}

Here we compile a list of the various loop integrals needed for computing the Wilsonian effective action. We follow the prescription of~\cite{Bilal:2007ne}, which requires us to impose the IR cutoff, $\mu$, on the shifted loop momenta. That is, we use Feynman parameters to combine denominators in the usual manner, and shift the integration variables to put the integrals in the form
\be
\I_{p,q}(M^2) = \int_{\ell_E^2\geq\m^2} {d^2\ell\over(2\pi)^2}\, {\left(\ell^2\right)^p \over \left(\ell^2+M^2\right)^q}. \label{integral}
\ee
After Wick-rotating ($\ell^0\rightarrow i\ell^0_E$) we integrate over the (shifted) Euclidean momenta with $\ell_E^2\geq\mu^2$. When $q\geq p+2$ the integrals are convergent; for example,
\be
\I_{0,n+2}(M^2) = {i\over4\pi} {1\over (n+1)} {1\over\left(\mu^2+M^2\right)^{n+1}},\qquad \forall n\geq0.
\ee
More generally,
\be
\I_{m,n+m+2}(M^2) = {i\over4\pi} {m!\, n! \over (n+m+1)!} \sum_{k=0}^m \left(\begin{array}{c} n+m+1 \\ k\end{array}\right) {M^{2(m-k)} \m^{2k} \over \left(\m^2+M^2\right)^{m+n+1}},\  \forall n,m\geq0.
\ee

When $q=p+1$, the integrals diverge and a UV regulator is required.\footnote{Of course, $\I_{p,q}$ diverges for $q<p+1$ as well, but those cases will not concern us.} Following~\cite{Bilal:2007ne}, we use dimensional reduction: carrying out all $D_+$-algebra in $d=2$, but continuing loop momenta to $d=2-2\e$ in order to evaluate divergent integrals. For $p=0$ we note that
\be
\I_{0,1}(M^2) = {i\over4\pi}\left(\Gamma(\e) -\log\left(\m^2+M^2\over4\pi\right) + O(\e)\right),
\ee
while for $p>0$ we have
\be
\I_{m,m+1}(M^2) ={i\over4\pi}\left(\Gamma(\e) -\log\left(\m^2+M^2\over4\pi\right) + P_m\left(M^2\over\m^2+M^2\right) + O(\e)\right),
\ee
where $P_m(x)$ is an $m$-th order polynomial given by
\be\label{P(x)}
P_m(x) = \sum_{k=1}^m \left(\begin{array}{c} m \\ k\end{array}\right) {(-x)^k\over k}.
\ee
In particular,
\be
P_1(x) = -x, \qquad P_2(x) = \hlf x^2 -2 x.
\ee

\subsection{The action}\label{sec:action}

In general, we are interested in $N$ charged triplets of chiral superfields $(\S^a,P^a,\G^a)$, where $\G^a$ are fermionic, coupled by a superpotential:
\be
S_J = -\sum_{a=1}^n \left\{ {m_a\over\sqrt{2}} \int d^2x d\th^+\, \G^a \S^a P^a +c.c. \right\}.
\ee
The charges of $(\S^a,P^a,\G^a)$ are only constrained by  gauge invariance of the superpotential: $Q_{\S^a} + Q_{P^a}+Q_{\G^a}=0$. In general there must be other charged fields $\left( \Phi^i,\G^\a \right)$ such that the total gauge anomaly vanishes. These additional fields will not concern us here.

If we restrict to loops of $(P^a,\G^a)$ there is no need to fix the gauge, though we may choose a unitary gauge by setting, say, $\S^1=1$. Next we expand the fields about a generic point in moduli space $(A_0,\S^a_0)$, which together with the expectation values $\Phi^i_0$ ensures that the $V_-$ tadpole vanishes. For simplicity, we will usually include $\S^1_0$ along with the rest of $\S^a_0$, even though $\S^1_0\equiv1$ when we fix unitary gauge. With this in mind, the terms in the action which contain $(P^a,\G^a)$ take the form,
\bea
S[P^a,\G^a] &=& -\hlf\sum_a \int d^2xd^2\th^+\left[i \bar{P}^a e^{2Q_{P^a}(A_0+A)}\nabla_-P^a + \bar{\G}^a e^{2Q_{\G^a}(A_0+A)}\G^a\right] \\
&&-{1\over\sqrt{2}}\sum_a \int d^2x d\th^+\ m_a\left( \S^a_0 +\S^a \right) \G^a P^a  +c.c., \non
\eea
where $\nabla_- = \del_- + Q_{P^a}(\del_-A + iV_-)$.

\subsection{Deriving the free field propagators}\label{sec:deriving}

There is a well known difficulty in deriving the Feynman rules for chiral superfields because they satisfy a differential constraint:
\be \bar D_+ P^a = \bar D_+\G^a=0.\ee
This is similar to the case of electromagnetism, where the field strength satisfies $dF=0$. In this latter case, the well-known solution is to introduce a potential, $A$, such that $F=dA$, which can then be quantized easily. The penalty is, of course, that $A$ is not unique, but is instead a member of an equivalence class: $A\sim A+df$ for any real-valued function $f$. Associated with this redundancy is the fact that the kinetic operator for $A$, denote it $K$, has a kernel: $K(df)=0$. In order to find the propagator for $A$, we must invert $K$ on the orthogonal complement to this kernel. We will follow an analogous approach to derive the $(P^a,\G^a)$ propagator.

We begin by introducing (unconstrained) potential fields $(\Pi^a,G^a)$, such that\footnote{Let $A$ be a superfield with fermion number $F$; then $\overline{D_+ A} = (-)^{F} \Dbar_+ \bar A$ and $\overline{D_+\Dbar_+ A}= -\Dbar_+D_+ \bar A$.}
\be
P^a = \bar D_+ \Pi^a,\qquad \G^a = \bar D_+ G^a,\qquad \bar P^a=-D_+\bar \Pi^a,\qquad \bar\G^a = + D_+\bar G^a.
\ee
Note that the potential fields have the opposite statistics of their corresponding field strengths. The case where $\bar D_+ \G^a = E^a$ is easily adapted to this construction, though we will not pursue it here. These potential fields are not unique, since $\Pi^a\sim \Pi^a + \bar D_+ F_-^a$ for some bosonic superfields $F_-^a$, and similarly for $G^a$. The free part of the $(P^a,\G^a)$ action can be written succinctly as
\be
S_{free} = \int d^2x d^2\th^+\, \left(X^a\right)^\dagger K^{ab} X^b,
\ee
with
\be
X^a = \begin{pmatrix} \Pi^a \\ \bar G^a \end{pmatrix}, \qquad K^{ab} = \hlf\begin{pmatrix} ie^{2Q_{P^a}A_0} D_+\bar D_+\del_- & \sqrt{2}m_a \bar{\S}^a_0 D_+ \\ -\sqrt{2}m_a\S^a_0 \bar D_+ & -e^{2Q_{\G^a}A_0} \bar D_+  D_+ \end{pmatrix}\dd^{ab}.
\ee
Notice that introducing the potential fields $(\Pi^a,G^a)$ allows us to write the $F$-term mass as an integral over all of superspace.

At this point, one should expect that $K^{ab}$ has a non-trivial kernel. Indeed $\textrm{ker}(K^{ab})=\Im(L_+^{ab})$, where
\be
L_+^{ab} = \begin{pmatrix} \bar D_+ & 0 \\ 0 & D_+\end{pmatrix}\dd^{ab}.
\ee
$K^{ab}$ can only be inverted on the orthogonal complement of its kernel. To implement this restriction, consider the following dimension zero operator:
\be
\hat\Pi = {1\over2i\del_+}\begin{pmatrix} D_+\bar D_+ & 0 \\ 0 & \bar D_+ D_+ \end{pmatrix}.
\ee
It is not difficult to verify that $\hat\Pi$ defines a self-adjoint projection operator with $\hat\Pi L_+^{ab}=0$ and $\textrm{ker}(\hat\Pi) = \Im(L_+^{ab})$. Thus $\hat\Pi$ is the projection operator we need in order to invert $K^{ab}$.

The propagator for $X^a$ then satisfies the defining relation:
\be
K^{ab}(z_1) \D^{bc}(z_{12}) = \hat\Pi(z_1)\dd^{ab}\dd^4(z_{12}).
\ee
The desired solution turns out to be
\be
\D^{ab}(z_{12}) = -\begin{pmatrix} e^{-2Q_{P^a}A_0} & {\bar M_a \over\sqrt{2}i\del_+} e^{Q_{\S^a}A_0}D_+ \\ -{M_a \over\sqrt{2}i\del_+} e^{Q_{\S^a}A_0} \bar D_+ & -ie^{-2Q_{\G^a}A_0}\del_- \end{pmatrix} {\dd^4(z_{12}) \over \del_+\del_- + M_a^2}\ \dd^{ab},
\ee
where $M_a \equiv m_a \S^a_0 e^{Q_{\S^a}A_0}$. Equivalently, transforming to momentum space gives
\be
\D^{ab}(p) = -\begin{pmatrix} e^{-2Q_{P^a}A_0} & -{\bar M_a \over\sqrt{2}p_+} e^{Q_{\S^a}A_0}D_+ \\ {M_a \over\sqrt{2}p_+} e^{Q_{\S^a}A_0} \bar D_+ & e^{-2Q_{\G^a}A_0}p_- \end{pmatrix} {\dd^2(\th_{12}) \over p^2 + M_a^2 -i\e}\ \dd^{ab},
\ee
with an appropriate $i\e$ prescription. Note that by $D_+$ we mean $D_{1+}\equiv D_+(p,\th_1)$, although we can easily convert it to $D_{2+} \equiv D_+(-p,\th_2)$ by the relation
\be
D_{1+}\dd^2(\th_{12}) = -D_{2+}\dd^2(\th_{12}).
\ee

This defines the propagator for the potential fields $(\Pi^a,G^a)$. To obtain the propagator for the chiral fields $(P^a,\G^a)$, we should act on the left and right by $\bar D_+$ ($D_+$) for (anti-)chiral legs.

\subsection{Interactions}\label{sec:inter}
The vertices of the theory can be read off directly from the interaction Lagrangian:
\bea
S_{int} &=& \hlf\sum_a \int d^2xd^2\th^+\left[ {i\over2} e^{2Q_{P^a}A_0} \left(e^{2Q_{P^a} A}-1\right) \left(P^a \del_-\bar P^a - \bar P^a\del_- P^a\right)\right. \non\\
&&\left.\frac{}{} + Q_{P^a} e^{2Q_{P^a}A_0} e^{2Q_{P^a} A} V_- |P^a|^2 - \bar\G^ae^{2Q_{\G^a}A_0}\left(e^{2Q_{\G^a} A}-1\right)\G^a\right] \\
&&- {1\over\sqrt{2}}\sum_{a} \int d^2xd\th^+\, m_a \S^a\G^a P^a +c.c. .\non
\eea
Each interaction vertex is accompanied by $i\int d^2\th^+$ except for $F$-term interactions, which only require $i\int d\th^+$ or $i\int d\thbar^+$. To make things more symmetric, we use one of the $\bar D_+$ or $D_+$ operators that act on an internal $(P^a,\G^a)$ propagator to convert the chiral measure into a full $\int d^2\th^+$ integral. We will follow the convention that the $\bar D_+$ or $D_+$ is pulled off from $\G^a$ or $\bar\G^a$.

\subsection{The rules}\label{sec:rules}

We now summarize the rules for computing each term in the quantum effective action:
\begin{enumerate}
\item[(1)] The various propagators are given by $i\D^{ab}(p)$ with
\bea
\left(\begin{array}{cc} \langle P^a_1 \bar P^b_2 \rangle & \langle P^a_1 \G^b_2 \rangle \\ \langle\bar\G^a_1 \bar P^b_2\rangle & \langle \bar\G^a_1 \G^b_2\rangle \end{array}\right) =  -i\begin{pmatrix} e^{-2Q_{P^a}A_0} & -{\bar M_a \over\sqrt{2}p_+} e^{Q_{\S^a}A_0}D_+ \\ {M_a \over\sqrt{2}p_+} e^{Q_{\S^a}A_0} \bar D_+ & e^{-2Q_{\G^a}A_0}p_- \end{pmatrix}   {\dd^2(\th_{12}) \over p^2 + M_a^2 -i\e}\ \dd^{ab}, \non
\eea
where,
\be \langle X_1 Y_2 \rangle \equiv \langle X(p,\th_1) Y(-p,\th_2)\rangle, \qquad M_a\equiv m_a \S^a_0 e^{Q_{\S^a}A_0}, \qquad D_+\equiv D_+(p,\th_1).\ee
\item[(2)] The vertex factors are
\bea
\langle P^a\bar P^b V_- A\ldots A\rangle &=& {i\over2}Q_{P^a} \left(2Q_{P^a}\right)^n e^{2Q_{P^a}A_0}\dd^{ab}, \\
\langle P^a(p) \bar P^b(p') A \ldots A \rangle &=& {i\over4}\left(2Q_{P^a}\right)^n e^{2Q_{P^a}A_0}(p - p')_- \dd^{ab},\\
\langle \bar\G^a\G^b A\ldots A\rangle &=& -{i\over2}\left(2Q_{\G^a}\right)^n e^{2Q_{\G^a}A_0}\dd^{ab}, \\
\langle \S^a \G^b P^c\rangle = {i\over\sqrt{2}}m_a \dd^{ab} \dd^{ac},&&\quad \langle \bar\S^a \bar\G^b \bar P^c\rangle = {i\over\sqrt{2}}m_a \dd^{ab} \dd^{ac},\quad a\neq0.
\eea
Here $n$ denotes the number of $A$ legs. For each internal (anti-)chiral line, include a $\bar{D}_+$ $(D_+)$ acting on the associated propagator, \textit{except} for $\G^a$ ($\bar\G^a$) connected to a $\S\G P$ ($\bar\S\bar\G\bar P$) vertex.
\item[(3)] For each vertex, include an integral $\int d^2\th^+_{vert}$.
\item[(4)] For each loop, include an integral $\int {d^2 p\over(2\pi)^2}$.
\item[(5)] For each loop of Fermi fields $\G$, include a factor of $(-1)$.
\item[(6)] For a term in the effective action with $n$ field insertions, denoted collectively by $X(p_i)$, include an overall
\be
\prod_{i=1}^n\left( \int {d^2 p_i \over(2\pi)^2} X(p_i)\right)(2\pi)^2\dd^2\left(\sum_{i=1}^n p_i\right).
\ee
\item[(7)] Divide by the usual combinatoric factor.
\end{enumerate}
Note that we take all momenta in the vertex factors as incoming.

\subsection{Tips and tricks}\label{sec:tricks}
In computing the effective action, it is always possible (by integration by parts) to move all the $D_+$ and $\bar D_+$ operators so that they act on either external fields or on $\dd^2(\th_{ij})$ of a single propagator. In doing so, it is helpful to convert all of the $D_+$ and $\bar D_+$ operators to be of the same ``type", by using the identity
\be
D_+(p,\th_i)\dd^2(\th_{ij}) = - D_+(-p,\th_j)\dd^2(\th_{ij}).
\ee
Note that this yields the following rule for converting products of $D_+$ and $\bar D_+$:
\be
\bar D_{i+} D_{i+}\dd^2(\th_{ij}) = -\bar D_{i+} D_{j+}\dd^2(\th_{ij}) = + D_{j+} \bar D_{i+}\dd^2(\th_{ij}) = - D_{j+} \bar D_{j+}\dd^2(\th_{ij}).
\ee
So for a product, the order is reversed and an overall sign is introduced. After these manipulations are performed, the resulting expression can be further simplified using the identities:
\bea
\dd^2(\th_{ij}) D_{i+} \bar D_{i+} \dd^2(\th_{ij}) = + \dd^2(\th_{ij}),&\qquad& \dd^2(\th_{ij}) \bar D_{i+} D_{i+} \dd^2(\th_{ij}) = - \dd^2(\th_{ij}), \\
\dd^2(\th_{ij}) D_{i+}\dd^2(\th_{ij}) = 0, &\qquad& \dd^2(\th_{ij}) \bar D_{i+} \dd^2(\th_{ij}) =0.
\eea
In the end, one is left with enough ``bare" $\dd^2(\th_{ij})$ to trivially carry all but one of the fermionic integrals. In this way, every term in the effective action can be reduced to a single $\int d^2\th^+$ and is therefore local in the $\th^+$ coordinates, even though the 1PI effective action may be non-local in $x$.

Computing a one point function requires some care since it can involve derivatives of $\dd^2(\th_{11})\equiv0$. We define these propagators, from one point to itself, as a limit of a standard propagator between two points. Thus,
\be
D_{1+}\bar D_{1+} \dd^2(\th_{11}) \equiv \lim_{2\rightarrow1} D_{1+}\bar D_{1+} \dd^2(\th_{12}) = 1.
\ee

\section{Feynman rules for other superfields}\label{ap:others}

Although the effective action, $W$, is determined solely by integrating out $(P^a,\G^a)$, for consistency we will also carry out the path integral over the high-energy modes of the other light fields. As one might expect, we will see that this leads only to a renormalization of the dimensionless FI parameter.

\subsection{Light chiral superfields}\label{sec:light}

The action for the light chiral superfields $\Phi^i$ and $\G^\a$ is identical to that of $P^a$ and $\G^a$, except there are no superpotential couplings that give rise to mass terms $(m=0)$. The derivation of the propagator is nearly identical to the discussion in Section~\ref{sec:deriving}. The result is
\be
\langle \Phi^i_1 \bar\Phi^j_2 \rangle = -i e^{-2Q_{\Phi^i} A_0}\, {  \dd^2(\th_{12}) \over p^2-i\epsilon}\,\dd^{ij},\qquad \langle \G^\a_1 \bar\G^\beta_2 \rangle = -i e^{-2Q_{\G^\a} A_0}\, {p_-\dd^2(\th_{12}) \over p^2-i\epsilon}\,\dd^{\a\beta},
\ee
and vertices
\bea
\langle \Phi^i\bar \Phi^j V_- A\ldots A\rangle &=& {i\over2}Q_{\Phi^i} \left(2Q_{\Phi^i}\right)^n e^{2Q_{\Phi^i}A_0}\dd^{ij}, \\
\langle \Phi^i(p) \bar \Phi^j(p') A \ldots A \rangle &=& {i\over4}\left(2Q_{\Phi^i}\right)^n e^{2Q_{\Phi^i}A_0}(p - p')_- \dd^{ij},\\
\langle \bar\G^\a\G^\beta A\ldots A\rangle &=& -{i\over2}\left(2Q_{\G^\a}\right)^n e^{2Q_{\G^\a}A_0}\dd^{\a\beta}.
\eea
Since there are no $F$-term interactions, we always act on the $\Phi^i$ and $\G^\a$ propagators with $-\bar D_{1+}D_{2+}\sim \bar D_{1+}D_{1+}$.

The computation of $f_V$ from a loop of $\Phi^i$ fields is exactly the same as~\C{fV}, except we set $m=0$. This leaves
\be
f_V = -{Q_{\Phi^i}\over8\pi}\log\left(\m^2\over\m_r^2\right).
\ee
This correction is field-independent, and so will not affect $W$, though it is important for understanding the beta function for the FI parameter $r$. The light chiral fields do not contribute to $f_A$ or $f_\S$ either, because they do not have any classical coupling to $\S$.

\subsection{The Higgs and vector multiplets}\label{sec:higgsandvec}

Because of the spontaneous symmetry breaking that occurs, it is best to examine the Higgs and gauge sectors simultaneously. For simplicity, we will only consider a single Higgs field, $\S$. Their combined action, expanded about $(A_0,\S_0)$, is
\bea
S[\S,A,V_-] &=& \int d^2xd^2\th^+\left[-{i\over4}\left(\bar\S_0+\bar\S\right)e^{2Q_\S(A_0+A)}\nabla_-\left(\S_0+\S\right) +c.c.\right] \non\\
&&- {1\over8e^2}\int d^2xd^2\th^+\left[\bar\U_-\U_- +{1\over\xi}\bar F F\right] ,\label{higgsvector}
\eea
where we have included a gauge-fixing term $|F|^2$. $F$ must be a fermionic function; the choice $F=D_+(\del_-A+iV_-)$ leads to $-{1\over2\xi}(\del\cdot A)^2$ in the component action. However, the non-zero value of $\S_0$ gives a mixing between $\S$ and $(A,V_-)$ in the quadratic action.

What we need is a $(0,2)$ version of $R_\xi$ gauge, where the propagators are diagonal. Supersymmetric $R_\xi$ gauges for four-dimensional gauge theories were introduced in~\cite{Ovrut:1981wa}. It turns out that the correct choice for our purposes is
\be
F = D_+(\del_-A +iV_-) -i\xi {2M_A^2S\over Q_\S \S_0}, \label{gaugefix}
\ee
where
\be
M_A^2 = 2e^2 Q_\S^2 |\S_0|^2 e^{2Q_\S A_0},\qquad \textrm{and}\qquad\bar D_+S=\S_0+\S.
\ee
It will be important in the next section to notice that the potential field $S$ is defined for the total field $\S'\equiv\S_0+\S$, not just the shifted part $\S$.

With this choice of gauge-fixing, the quadratic part of the action becomes
\bea
S_{quad}[\S,A,V_-] &=& \hlf\int d^2xd^2\th^+ e^{2Q_\S A_0} \bar S\left(iD_+\bar D_+\del_- -2\xi M_A^2\right)S \\
&-&{1\over 2\xi e^2}\int d^2xd^2\th^+ \left[ A\left(\del^2-\xi M_A^2\right)V_- +\left({\xi-1\over4}\right)\bar\U_-\U_-\right].\non
\eea
For the $\S$ field, we find the propagator
\be
\langle \S_1 \bar\S_2 \rangle = ie^{-2Q_\S A_0}{D_+\bar D_+\dd^2(\th_{12}) \over 2p_+ \left(p^2+\xi M_A^2\right)} \sim -i e^{-2Q_\S A_0} {\dd^2(\th_{12})\over p^2 + \xi M_A^2 -i\e},
\ee
where in the last step we have used the fact that every internal $\S\bar\S$ propagator will be acted on by $-\bar D_{1+}D_{2+}$ to write an equivalent propagator with the pole at $p_+=0$ removed.\footnote{Note that $-\bar D_{1+} D_{2+}$ still acts on this equivalent form of the propagator.} We can recover unitary gauge by sending $\xi\rightarrow\infty$. In this limit, $\S$ does not propagate and it is effectively eliminated from the spectrum, as expected. However for the vector multiplet, $\xi=1$ is a much more natural choice since the kinetic terms simplify tremendously. This is a natural generalization of Feynman-'t Hooft gauge. We will henceforth work only in $\xi=1$ gauge, where the vector field propagator reduces to
\be
\langle A_1 V_2 \rangle = (2ie^2){\dd^2(\th_{12})\over p^2 + M_A^2 -i\e}.
\ee
The coefficient $2i=-i\eta_{+-}$ is exactly as one would expect for $\langle A_+A_-\rangle$ in Feynman gauge.

Interaction vertices can be read off directly from
\bea
S_{int} &=& \int d^2xd^2\th^+ \left[-{i\over4}e^{2Q_\S A_0}\left(e^{2Q_\S A}-1\right)\left(\bar\S\del_-\S - \S\del_-\bar\S\right) + {Q_\S\over2}e^{2Q_\S A_0}|\S|^2 e^{2Q_\S A} V_-\right.\non\\
&&\qquad\qquad+{Q_\S\over2}e^{2Q_\S A_0}\left(e^{2Q_\S A}-1\right)\left[V_-\left(\bar\S_0\S + \S_0\bar\S\right) +i\del_-A\left(\bar\S_0\S -\S_0\bar\S\right)\right] \non\\
&&\left.\qquad\qquad+{Q_\S\over2}|\S_0|^2e^{2Q_\S A_0} \left(e^{2Q_\S A}-2Q_\S A -1\right)V_- \right].
\eea
The terms in the first line are exactly the same as in the $\S_0=0$ case, while the terms in the second line are related to the first by replacing $\S$ or $\bar\S$ by its vev. Finally the terms of the third line, which come from setting $|\S|^2$ to its vev, give rise to a set of couplings between $V_-$ and $A$ only. In particular, the vertices are
\bea
\langle \S\bar \S V_- A\ldots A\rangle &=& {i\over2}Q_{\S} \left(2Q_{\S}\right)^n e^{2Q_{\S}A_0}, \label{1}\\
\langle \S(p) \bar \S(p') A \ldots A \rangle &=& {i\over4}\left(2Q_{\S}\right)^n e^{2Q_{\S}A_0}(p - p')_-,\label{2}\\
\langle \S V_- A\ldots A\rangle &=& {i\over2}Q_{\S} \left(2Q_{\S}\right)^n e^{2Q_{\S}A_0}\bar\S_0, \label{3}\\
\langle \bar \S V_- A\ldots A\rangle &=& {i\over2}Q_{\S} \left(2Q_{\S}\right)^n e^{2Q_{\S}A_0}\S_0, \label{4}\\
\langle \S(p) A \ldots A \rangle &=& {i\over4}\left(2Q_{\S}\right)^n e^{2Q_{\S}A_0}\bar\S_0 p_-,\label{5}\\
\langle \bar \S(p') A \ldots A \rangle &=& -{i\over4}\left(2Q_{\S}\right)^n e^{2Q_{\S}A_0}\S_0 p'_-,\label{7}\\
\langle V_- A\ldots A\rangle &=& {i\over2}Q_{\S} \left(2Q_{\S}\right)^n e^{2Q_{\S}A_0}|\S_0|^2.\label{self}
\eea
In~\C{1}\ $n\geq0$, while in~\C{2}-\C{4}\ we require $n\geq1$, and in~\C{5}-\C{self}\ $n\geq2$.

An important point to note in computing loops with these Feynman rules is that the $AV_-$ propagator is \textit{not} acted on by $\Dbar_+ D_+$, and so these propagators contribute ``bare" $\dd^2(\th_{12})$ to the loop integrals. To get a non-zero result, a loop with an internal vector line must also contain a line of chiral fields, otherwise it will be proportional to $\left(\dd^2(\th_{12})\right)^2=0$ in the case of two-point vertices, or $\dd^2(\th_{11})=0$ in the case of a single vertex.

\begin{figure}[h]
\centering
\subfloat[][]{
\includegraphics[width=0.475\textwidth]{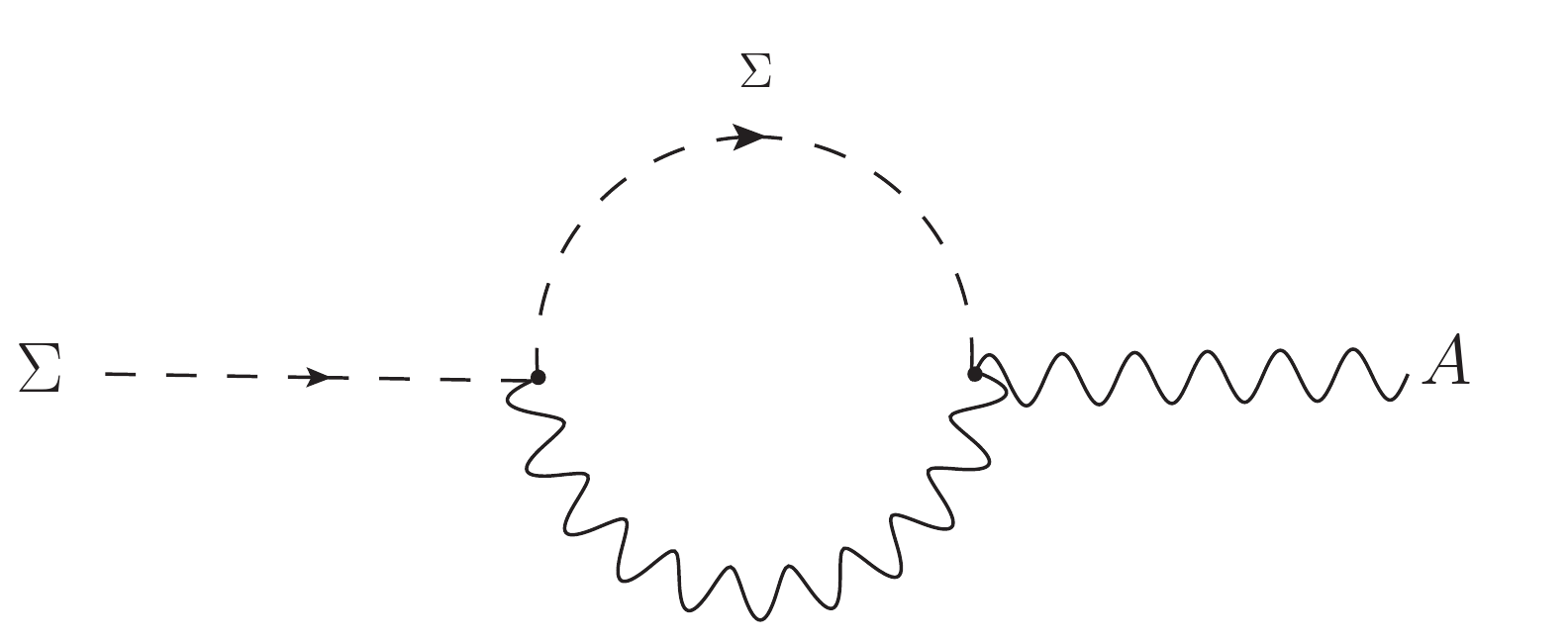}
\label{ASigma2}}
\
\subfloat[][]{
\includegraphics[width=0.475\textwidth]{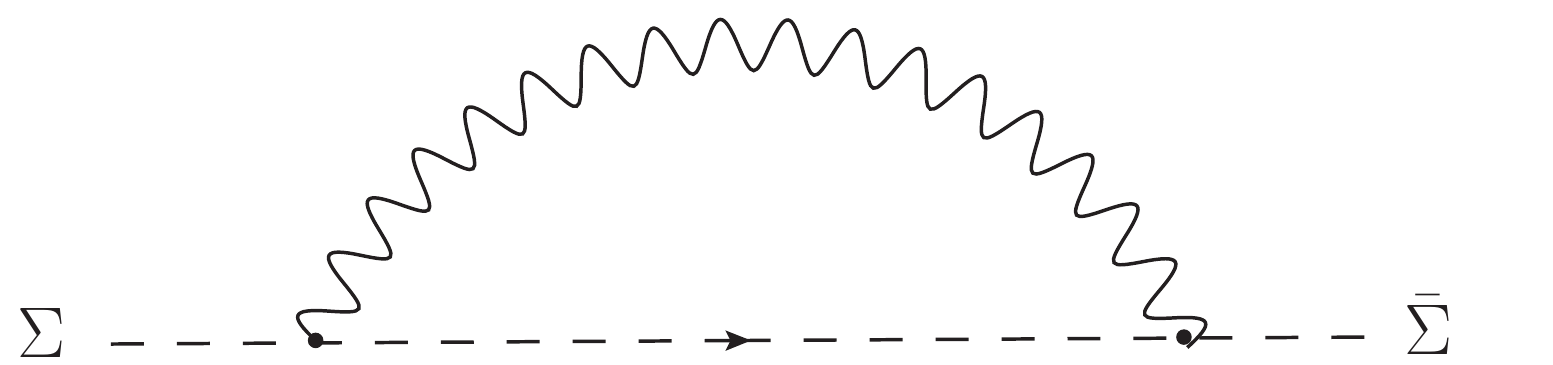}
\label{SSbar2}}
\caption{\textit{The remaining diagrams that contribute to the effective action in a general gauge.}}
\label{sigmaloops}
\end{figure}

Integrating down to a scale $\m\gg M_A$ where the gauge theory is still perturbative, the contribution to $f_V$ coming from a $\S$ loop is
\be
f_V =-{Q_{\S}\over8\pi} \log\left(\m^2+M_A^2 \over \m_r^2\right) =-{Q_{\S}\over8\pi} \log\left(\m^2\over \m_r^2\right) +\ldots.
\ee
This is a field-independent renormalization of $t$, which we can ignore.  The only diagrams which could contribute to $f_A$ and $f_\S$, and do not vanish identically, are shown in Figure~\ref{sigmaloops}.
However, it is easy to check that these diagrams are suppressed by $(M_A^2/\m^2)$, and can therefore be neglected.

\subsection{Ghosts}\label{sec:ghosts}

Even though we are dealing with an abelian gauge theory, fixing $R_\xi$ gauge leads to $F$-term interactions between both the chiral and anti-chiral ghosts, and the Higgs field $\S$. However, the ghost and matter sectors still turn out to decouple from one another; the ghosts could only possibly renormalize  $F$-terms, but we know that cannot happen. So we will find that ghosts cannot modify the effective action in this theory, despite coupling to the Higgs.

To demonstrate this claim we begin from the gauge fixed action, which naturally splits into three pieces:
\be
S = S_0[X] + S_F[\Om,F(X)] + S_{gh}[B,C,F(X)],\label{Stot}
\ee
where $S_0$ is the GLSM action, and we denote all the gauge and matter fields collectively by $X$. The second piece,
\be
S_F = \hlf\int d^2xd^2\th^+\left(\Om F(X) + \bar F(X)\bar\Om -4e^2\xi\,\bar\Om\Om\right),
\ee
is the gauge-fixing term with $\Om$ an auxiliary chiral Fermi field.\footnote{That $\Om$ is chiral follows from the fact that $F(X)$ is (essentially) anti-chiral. This is certainly true when $\xi=0$, and by exploiting the ``gauge symmetry" $S\sim S +\bar D_+T$ we can force $F$ to be anti-chiral for finite $\xi$ as well.} When $\xi=0$, $\Om$ acts as a Lagrange multiplier enforcing our gauge condition: $F(X)=0$. For non-zero $\xi$ we can solve for $\Om$ to recover the standard Gaussian average over gauge choices, as in~\C{higgsvector}. Finally, the third piece of~\C{Stot} gives the ghost action
\be
S_{gh} = -\hlf\int d^2xd^2\th^+\left[B\left(\dd_\La F\right)C - \bar B\left(\dd_{\bar\La}\bar F\right)\right],
\ee
where $B$ is a chiral commuting left-moving Fermi supermultiplet\footnote{Not to be confused with the $B$-field of a target space sigma model. Since we only discuss ghosts in this section, we hope the reader will forgive out abuse of notation.}, and $C$ is a chiral anti-commuting scalar supermultiplet:
\be
B = \beta + \th^+ b -i\th^+\thbar^+\del_+\beta,\qquad C = c + \th^+\gamma -i\th^+\thbar^+\del_+ c.
\ee
If we denote the gauge transformation of $X$ by $X^\La = X +\La\dd_\La X + \ldots,$  then
\be
\dd_\La F\equiv \left.{\dd F(X^\La) \over \dd\La}\right|_{\La=0}.
\ee
Rather than derive~\C{Stot} directly by a  $(0,2)$ version of the standard Faddeev-Poppov procedure, which can be done but has its own subtleties stemming from the fermionic nature of the gauge-fixing condition~\C{gaugefix}, we will instead offer the evidence that~\C{Stot}\ is invariant under the super-BRST symmetry:
\bea
\dd X = \e C\dd_\La X + \bar\e\bar C \dd_{\bar\La} X, &\qquad& \dd B = \e\Om,\qquad \dd \bar B = \bar\e\bar\Om, \\
\dd \Om = 0,\qquad \dd \bar \Om = 0,&\qquad& \dd C = 0,\qquad \dd \bar C = 0.
\eea
Verifying this symmetry is particularly straightforward, since $\dd C$ vanishes for an abelian gauge group.

For the gauge-fixing function~\C{gaugefix}, the ghost action is given by
\bea
S_{gh}[B,C,\S] &=& {i\over2}\int d^2xd^2\th^+\ B D_+\del_-C - \xi M_A^2 \int d^2xd\th^+ \left(1+{\S\over\S_0}\right) BC \label{ghost} \\
&&+  {i\over2}\int d^2xd^2\th^+\ \bar B \bar D_+\del_-\bar C - \xi M_A^2 \int d^2xd\bar\th^+ \left(1+{\bar\S\over\bar\S_0}\right) \bar B\bar C, \non
\eea
where we have used part of the Grassmann measure to convert the interaction with $S$ into an $F$-term interaction with $\S'=\S_0+\S$. We should stress that it is the field $\S'$ which transforms linearly under the gauge symmetry: $\S'^\La=e^{iQ_\S\La}\S'$. Since the $(B,C)$ and $(\bar B,\bar C)$ sectors decouple, it is clear that they cannot renormalize the effective action which must be a $D$-term. It should be pointed out that if $F$ were chosen so that $\dd_{\bar\La} F\neq0$ then the two sectors would be coupled and could  combine into a $D$-term.

To see this non-renormalization in greater detail, we write $C=\bar D_+\gamma$, but leave $B$ alone, giving the Feynman rules:
\bea
\langle C_1 B_2 \rangle = \langle \bar C_1 \bar B_2 \rangle = {i\dd^2(\th_{12}) \over p^2+\xi M_A^2 -i\e}, \quad \langle \S BC \rangle = -i{\xi M_A^2\over\S_0},\quad \langle \bar\S\bar B\bar C\rangle = -i {\xi M_A^2\over\bar\S_0}.
\eea
Note that internal $CB$ propagators are not acted on by  $D_+$ or $\bar D_+$. The reason is that the $\bar D_{+}$, which converts $\gamma$ to $C$ and would usually act on a $CB$ propagator, gets absorbed by the vertex factor in order to write the interaction as a $D$-term. It is then easy to see that there are no possible diagrams with these interactions that renormalize the effective action.

\section{The full quadratic effective action} \label{ap:quad}

In this appendix, we will compute the full momentum-dependence of the effective action to quadratic order in gauge fields. Aside from general interest, there are several reasons we consider this a useful exercise. First, we wish to confirm the coefficient $-\cA/8\pi$ of~\C{Gammafinal}\ by directly computing $\langle A V_-\rangle$ without relying on the background field trick. Second, we want to demonstrate that when computing the Wilsonian effective action at a scale $\m$, the non-local term in the anomaly is smoothed out, as we claim in~\C{smoothing}. Finally, along the way we will deepen our understanding of how the local counter-term~\C{counterterm}\ arises in perturbation theory.

Let us begin by considering the light chiral fields $(\Phi^i,\G^\a)$. There are three diagrams
\begin{figure}[h]
\centering
\subfloat[][]{
\includegraphics[width=0.475\textwidth]{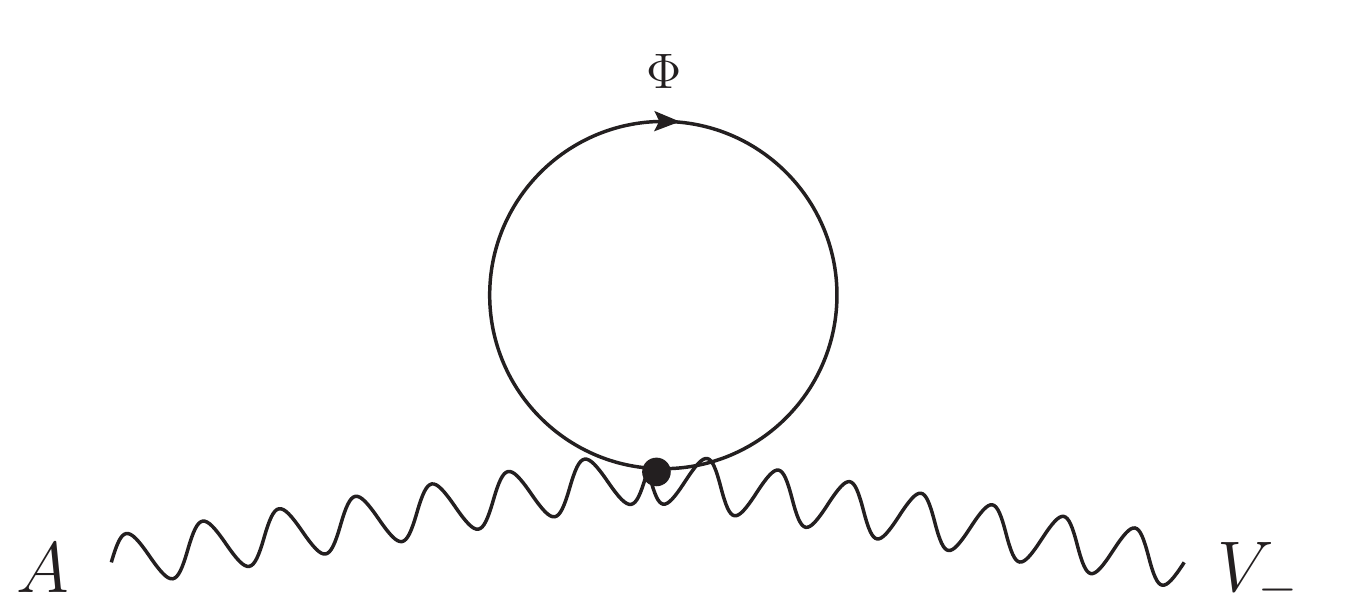}
\label{AV1}}
\
\subfloat[][]{
\includegraphics[width=0.475\textwidth]{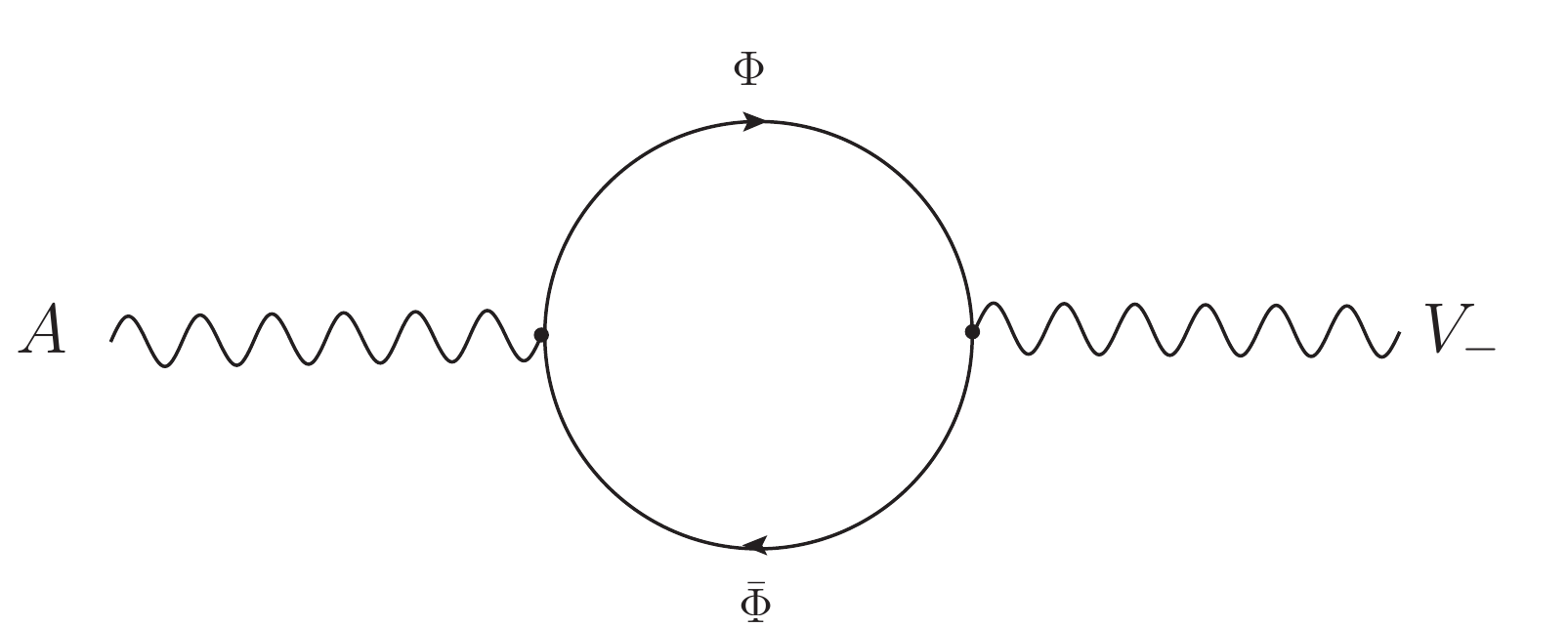}
\label{AV2}}
\\
\subfloat[][]{
\includegraphics[width=0.475\textwidth]{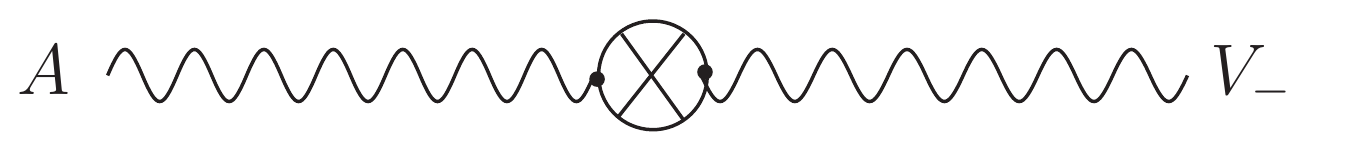}
\label{CT}}
\caption{\textit{The three loops contributions to $\langle AV_-\rangle$ coming loops of massless chiral fields.}}
\label{AVloops}
\end{figure}
that contribute to $\langle A V_- \rangle$, shown in Figure~\ref{AVloops}, where Figure~\ref{CT}\ corresponds to the counter-term~\C{counterterm}:
\be
W_{ct} = {1\over8\pi}\left(\sum_i Q_i^2 + \sum_\a Q_\a^2\right) \int d^2xd^2\th\ A V_- .\label{counterterm2}
\ee
To understand the origin of this counter-term, consider the computation of $\langle A_+ A_-\rangle$ in components. The integrand for the fermionic loop contains terms proportional to
\be
\sum_i Q_i^2\, \Tr\left[P_+\gamma^+ p\!\!/ P_+ \gamma^- p\!\!/\right] + \sum_\a Q_\a^2\, \Tr\left[P_-\gamma^+ p\!\!/ P_- \gamma^- p\!\!/\right] \equiv0,
\ee
where $P_\pm =\hlf(1\pm\g^5)$ and $\g^\pm =\hlf(\g^0\pm\g^1)$. The reason these terms vanish identically is that $P_\pm\g^\mp=0$. However if we consider the more general amplitude $\langle A_\m A_\n\rangle$ and work in $d=2-2\e$,  we find
\bea
&& \sum_i Q_i^2\, \Tr\left[P_+\gamma^\m p\!\!/ P_+ \gamma^\n p\!\!/\right] + \sum_\a Q_\a^2\, \Tr\left[P_-\gamma^\m p\!\!/ P_- \gamma^\n p\!\!/\right] \\
&&= \left(\sum_i Q_i^2 + \sum_\a Q_\a^2\right)  \left(d-2\over d\right)\eta^{\m\n} p^2 .\non
\eea
These terms appear inside divergent integrals so we end up with a net finite result for $\langle A_+ A_-\rangle$. This discrepancy arises when we carry out the gamma-matrix algebra in $d=2$ as opposed to $d=2-2\e$. This is the basic distinction between dimensional reduction and dimensional regularization.

It is well known that neither regularization scheme preserves supersymmetry, though in dimensional reduction the breakdown is only believed to occur at high loop order, at least for four-dimensional theories. Here we find a discrepancy already at one-loop that we can trace back to the inherently chiral structure of $(0,2)$ superspace, which cannot be continued away from $d=2$. We have already motivated the necessity of this counter-term in Section~\ref{ss:partf}, and now we have pinpointed its origin.

The loops appearing in figures~\ref{AV1}\ and~\ref{AV2}\ are separately divergent, but together they yield the finite result
\be
\int {d^2q\over(2\pi)^2}d^2\th^+\, A(q)q^2V_-(-q) \left[\sum_i{Q_i^2\over4}\int_0^1 dx\, \I_{0,2}\left(x(1-x)q^2\right) \right],\label{remainder}
\ee
which requires use of the identities
\be
\int_0^1  {(2x-1)\,dx \over M^2+x(1-x)q^2} =0, \label{identity2}
\ee
and
\be\label{identity1}
\int_0^1 dx\, \log\left(M^2 +x(1-x)q^2 \over M^2\right) = \int_0^1 dx\, {x(2x-1)q^2\over M^2+x(1-x)q^2}.
\ee

\begin{figure}[h]
\centering
\subfloat[][]{
\includegraphics[width=0.475\textwidth]{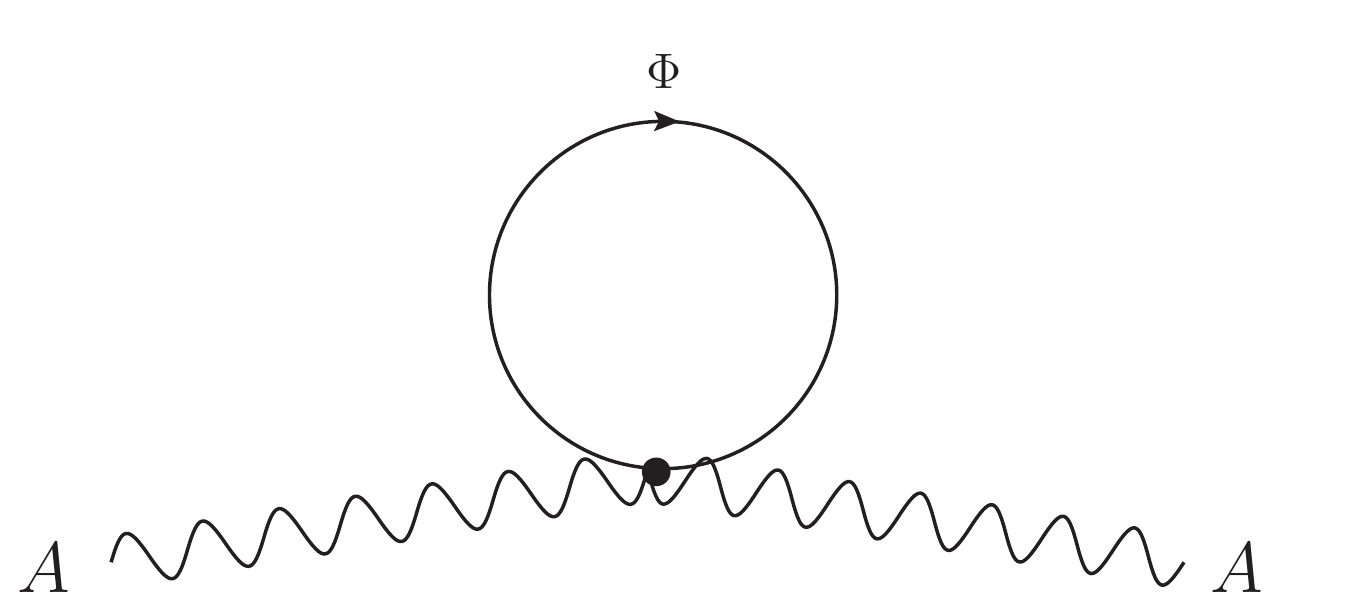}
\label{AA1}}
\
\subfloat[][]{
\includegraphics[width=0.475\textwidth]{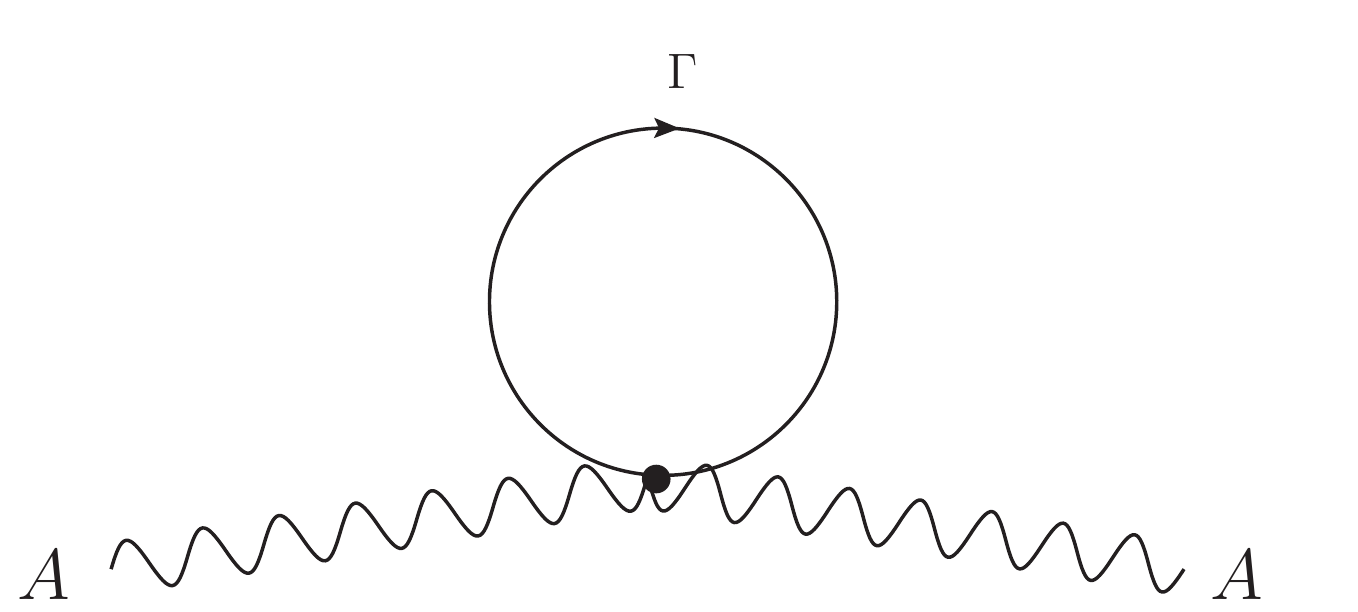}
\label{AA2}}
\\
\subfloat[][]{
\includegraphics[width=0.475\textwidth]{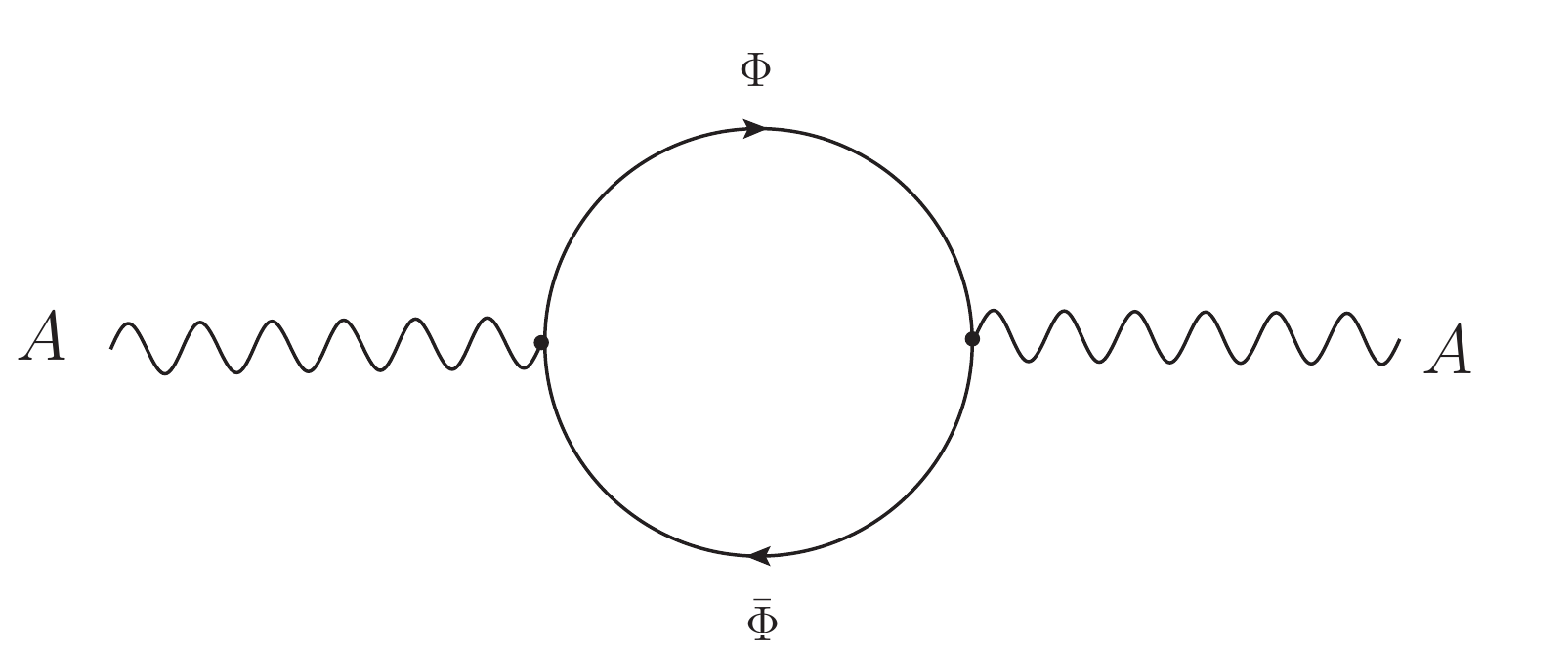}
\label{AA3}}
\
\subfloat[][]{
\includegraphics[width=0.475\textwidth]{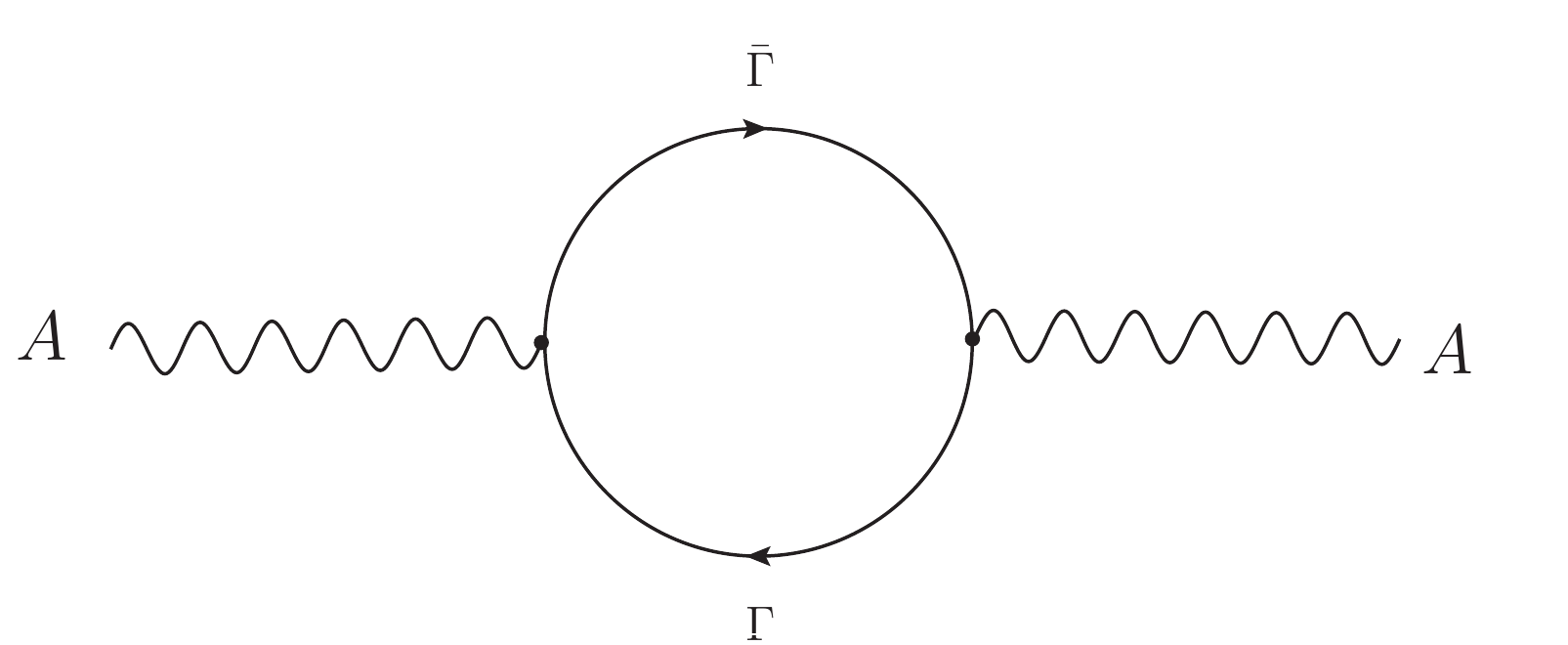}
\label{AA4}}
\caption{\textit{The diagrams contributing to $\langle AA\rangle$.}}
\label{AAloops}
\end{figure}

\begin{figure}[h]
\centering
\includegraphics[width=0.5\textwidth]{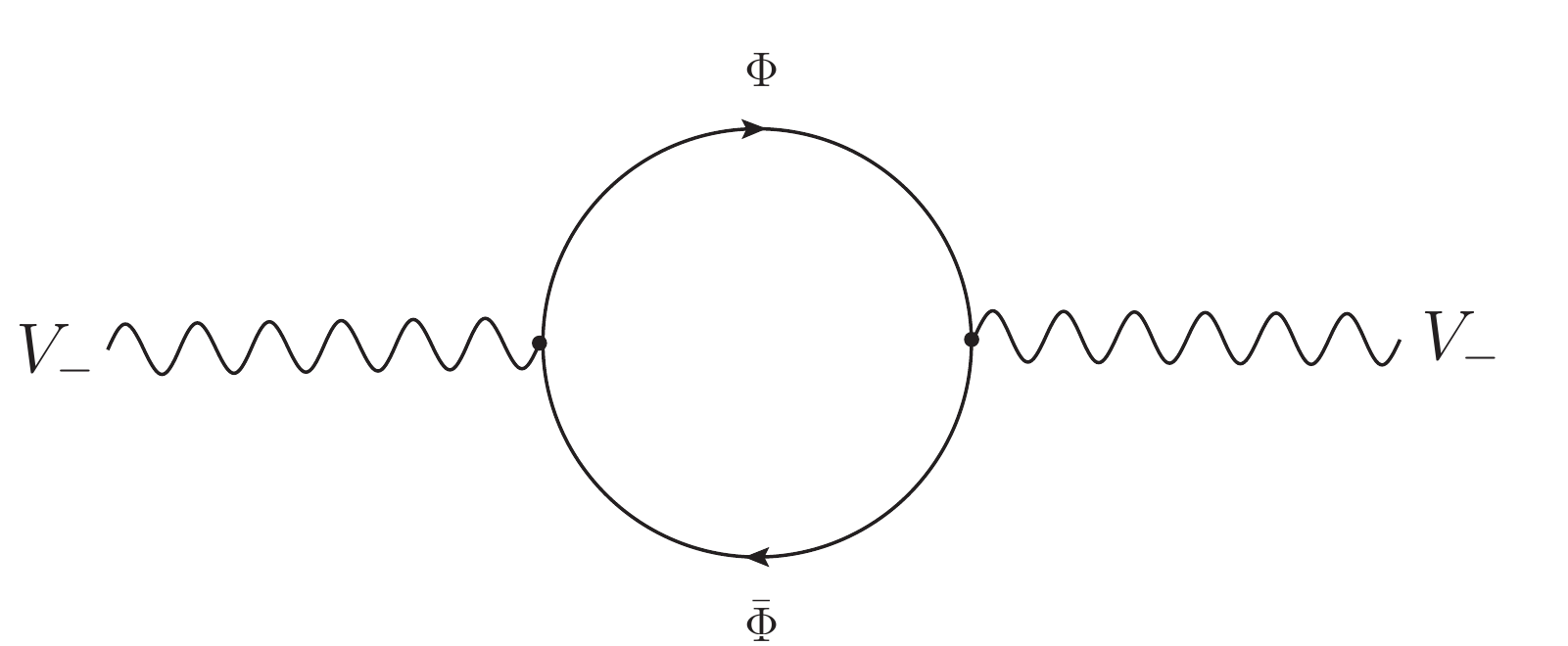}
\caption{\textit{The lone contribution to $\langle V_-V_-\rangle$.}}
\label{VV}
\end{figure}

The $\langle A A \rangle$ correlator receives contributions from the four diagrams of Figure~\ref{AAloops},
though diagrams~\subref{AA1}\ and~\subref{AA2}\ are easily shown to vanish. Only the single diagram of Figure~\ref{VV}\
contributes to $\langle V_- V_- \rangle$.  Together, these eight diagrams yield the following quadratic effective action:
\bea
W_{quad} &=& {1\over8\pi}\int d^2xd^2\th^+ \int_0^1 dx \left\{\sum_i{Q_i^2\over4}\bar\U_-\left(1\over\m^2-x(1-x)\del^2\right)\U_- \right. \\
&&\qquad \qquad\qquad \qquad \left. + \cA\left[\bar D_+\del_-A\left(x(1-x) \over \m^2 -x(1-x)\del^2 \right) D_+\del_-A - A V_-\right]\right\} \non,
\eea
where $\cA = \sum_i Q_i^2 - \sum_\a Q_\a^2$. Notice that when $\m=0$, $W_{quad}$ has a gauge-invariant term and a non-invariant term that produces the correct $(0,2)$ gauge anomaly. This would not have worked had we not included the counter-term~\C{counterterm2}. The non-locality of $W_{quad}$ that emerges at $\m=0$ signals that we have integrated out massless degrees of freedom. For $\m>0$ we see that the effective action has a perfectly local expansion in ${\del^2\over \m^2}$, and the non-local term in the anomaly gets smoothed out, as claimed in~\C{smoothing}.

\begin{figure}[t]
\centering
\subfloat[][]{
\includegraphics[width=0.475\textwidth]{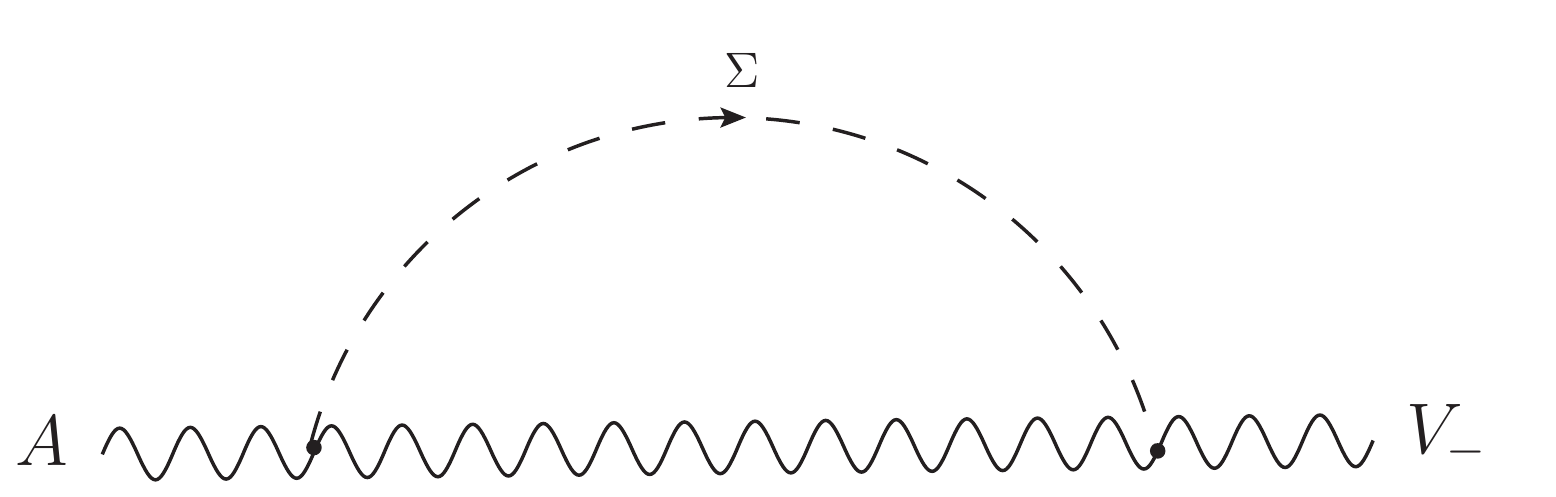}
\label{AV4}}
\
\subfloat[][]{
\includegraphics[width=0.475\textwidth]{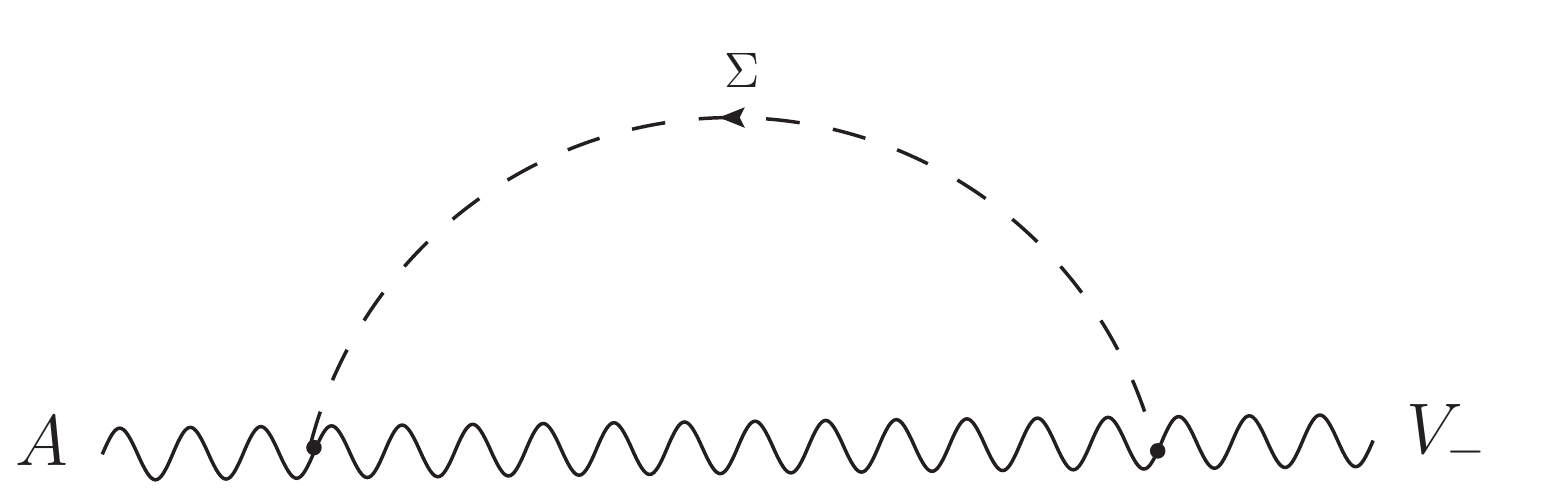}
\label{AV5}}
\caption{\textit{Novel contributions to $\langle AV_-\rangle$ from $\S$ and gauge multiplet loops.}}
\label{AVsigmaloops}
\end{figure}

Despite the self-interactions of the gauge multiplets, listed in~\C{self}, these couplings  do not give rise to any corrections to $W_{quad}$. Loops of $\S$ proceed exactly as in the case of the massless $\Phi$ fields discussed above, with two exceptions. The non-zero mass of $\S$ means we should replace $\m^2$ everywhere above with $\m^2+M_A^2$, and there are two additional diagrams to consider, shown in Figure~\ref{AVsigmaloops}.
These are easy enough to work out; they give,
\be
\int {d^2q\over(2\pi)^2}d^2\th^+\, A(q)V_-(-q) \left[2M_A^2 Q_\S^2\int_0^1 dx\, \I_{0,2}\left(\D_A\right) \right],
\ee
where $\D_A = M_A^2 + x(1-x)q^2$. In the limit $M_A^2\ll \m^2$ we can neglect the mass of $\S$, so these new contributions vanish and $\S$ behaves like any of the massless chiral fields discussed above. Thus we can just replace $\sum_i Q_i^2$ with $Q_\S^2 + \sum_i Q_i^2$ in the expressions above.

\begin{figure}[h]
\centering
\subfloat[][]{
\includegraphics[width=0.475\textwidth]{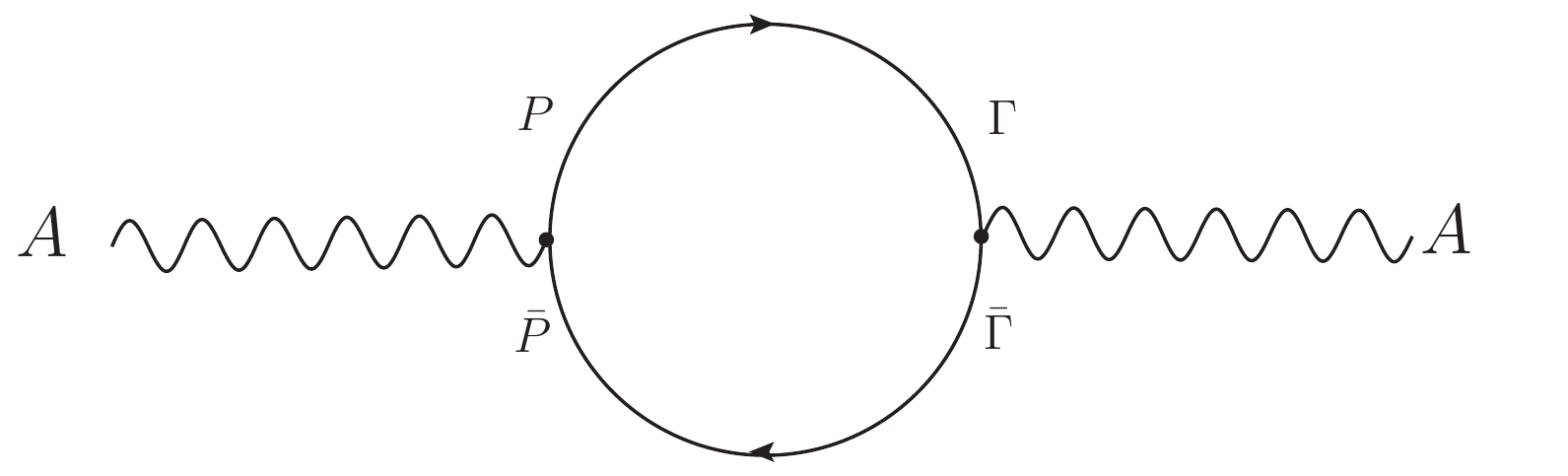}
\label{AA5}}
\
\subfloat[][]{
\includegraphics[width=0.475\textwidth]{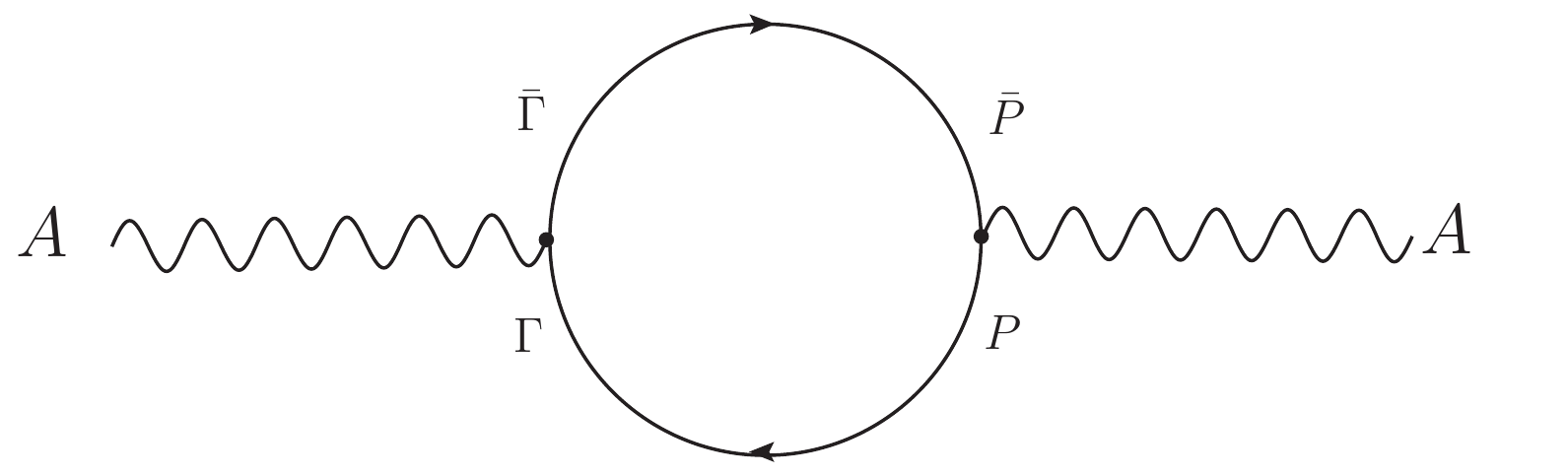}
\label{AA6}}
\
\subfloat[][]{
\includegraphics[width=0.475\textwidth]{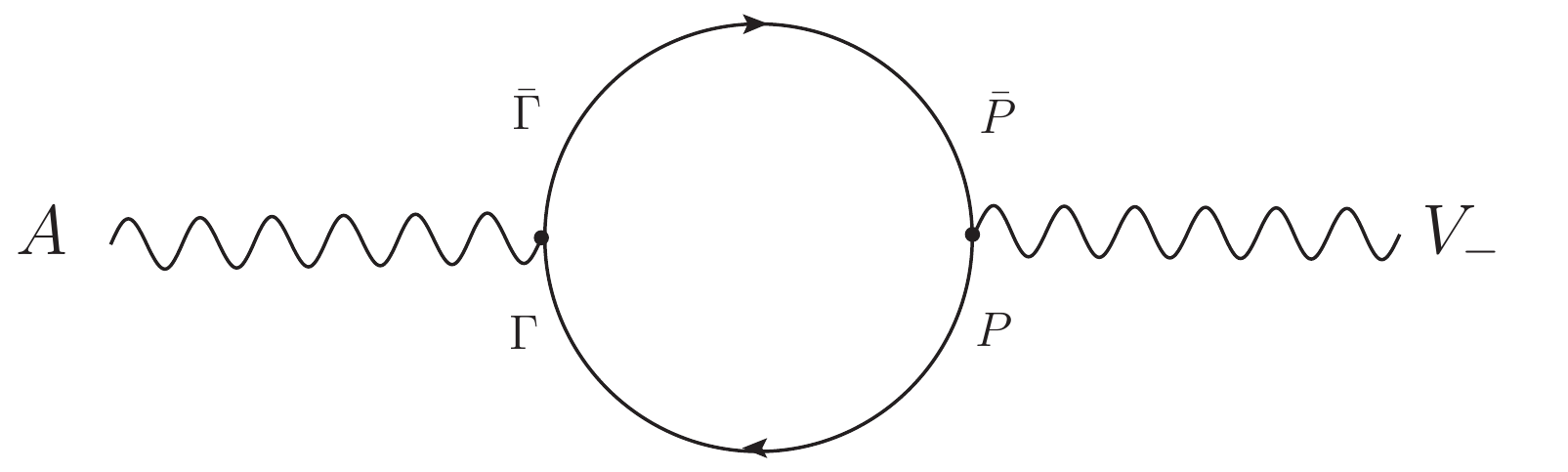}
\label{AV6}}
\caption{\textit{Novel contributions to $\langle AA\rangle$ and $\langle AV_-\rangle$ from $P$ and $\Gamma$ loops.}}
\label{sigmaloopssecond}
\end{figure}

Finally, we consider the massive fields $(P,\G)$. For simplicity we consider only one such pair with mass $M^2 = m^2|\S_0|^2 e^{2Q_\S A_0}$. Again, the main change from the massless case is the substitution $\m^2\rightarrow \m^2+ M^2$. There are also the three additional diagrams of Figure~\ref{sigmaloopssecond}\ to evaluate,
which can only appear when the $P\G$ propagator is non-vanishing. In fact,~\ref{AA5}\ and~\ref{AA6}\ sum to zero. Therefore, the novel contributions coming from the $(P,\G)$ sector are
\be
\int {d^2q\over(2\pi)^2}d^2\th^+\, A(q)V_-(-q) \left[2 M^2 Q_{P}\left(Q_{P}+Q_{\G}\right)\int_0^1 dx\, \I_{0,2}\left(\D\right) \right],
\ee
where $\D= M^2 + x(1-x)q^2$. The $Q_{P} Q_{\G}$ term clearly originates from Figure~\ref{AV6}, while the $Q_{P}^2$ term comes from the divergent graphs~\ref{AV1}\ and~\ref{AV2}. The latter contribution did not show up earlier in~\C{remainder}\ because it is proportional to $M^2$. In the limit $M^2\gg\m^2$,  only
\be
{Q_{\S}^2\over8\pi}  \int d^2x d^2\th^+\, A V_-,
\ee
survives from this sector, where we have used the fact that $Q_{\S} + Q_{P} + Q_{\G} =0$. This is the $AV_-$ term that arises in the one-loop correction to the $\S$ metric.

Putting everything together, we find that in the limit $M_A^2\ll \m^2 \ll M^2$ the coefficient of $AV_-$ in the effective action is
\be {1\over 8\pi} \left(Q_\S^2 -Q_\S^2 -\cA\right) = - {\cA\over 8\pi}, \ee
exactly as we found in~\C{Gammafinal}.


\newpage
\addcontentsline{toc}{chapter}{References}

\ifx\undefined\bysame
\newcommand{\bysame}{\leavevmode\hbox to3em{\hrulefill}\,}
\fi

\end{document}